\documentclass[apj]{emulateapj}

\usepackage{graphicx,fancyhdr,natbib,subfigure}
\usepackage{epsfig,psfig,epsf}
\usepackage{amsmath,cancel}
\usepackage{amssymb}
\usepackage{lscape}
\usepackage{dcolumn}% Align table columns on decimal point
\usepackage{bm}% bold math
\usepackage{hyperref,ifthen}
\usepackage{verbatim}

 \def\apjl{ApJ~Lett.} 
\def\apjs{ApJS}

 \def\aap{Astron.~\&~Astrophys.} 
\def\jcap{\ref@jnl{J. Cosmology Astropart. Phys.}}

\def \kms {{\rm ~km~s}^{-1}}

\def \hmpc{~\;h^{-1}~{\rm Mpc}}

\def \mi {M_{\rm i}}

\def \gsim { \lower .75ex \hbox{$\sim$} \llap{\raise .27ex \hbox{$>$}} }
\def \lsim { \lower .75ex \hbox{$\sim$} \llap{\raise .27ex \hbox{$<$}} }

\newcommand{\ciii}{C\,{\sc iii}]\ }

\newcommand{\feii}{Fe\,{\sc ii}\ }

\begin{document}

\shorttitle{The SDSS-III BOSS: Quasar Luminosity Function from DR9}
\shortauthors{N.~P.~Ross {\it et al.}}

\title{The SDSS-III Baryon Oscillation Spectroscopic Survey: \\
The Quasar Luminosity Function from Data Release Nine}

\author{
  Nicholas~P.~Ross\altaffilmark{1}, 
  Ian~D.~McGreer\altaffilmark{2}, 
  Martin~White\altaffilmark{1,3}, 
  Gordon~T.~Richards\altaffilmark{4},
  Adam~D.~Myers\altaffilmark{5}, 
  Nathalie~Palanque-Delabrouille\altaffilmark{6}, 
  Michael~A.~Strauss\altaffilmark{7},  
  Scott~F.~Anderson\altaffilmark{8}, 
  Yue~Shen\altaffilmark{9}, 
  W.~N.~ Brandt\altaffilmark{10,11}, 
  Christophe~Y{\`e}che\altaffilmark{6}, 
  Molly~E.~C.~Swanson\altaffilmark{9},
%% Below is the alphabetical group
  \'Eric~Aubourg\altaffilmark{12}, 
  Stephen~Bailey\altaffilmark{1}, 
  Dmitry~Bizyaev\altaffilmark{13},
  Jo~Bovy\altaffilmark{14,15},
  Howard~Brewington\altaffilmark{13},    
  J.~Brinkmann\altaffilmark{13},
  Colin~DeGraf\altaffilmark{16},
  Tiziana~Di~Matteo\altaffilmark{16},
  Garrett~Ebelke\altaffilmark{13},    
  Xiaohui~Fan\altaffilmark{2},  
  Jian~Ge\altaffilmark{17},
  Elena~Malanushenko\altaffilmark{13},      
  Viktor~Malanushenko\altaffilmark{13},   
  Rachel~Mandelbaum\altaffilmark{16},
  Claudia~Maraston\altaffilmark{18}
  Demitri~Muna\altaffilmark{19},
  Daniel~Oravetz\altaffilmark{13},    
  Kaike~Pan\altaffilmark{13}, 
  Isabelle~P\^aris\altaffilmark{20,21}, 
  Patrick~Petitjean\altaffilmark{20}, 
  Kevin~Schawinski\altaffilmark{22}, 
  David~J.~Schlegel\altaffilmark{1}, 
  Donald~P.~Schneider\altaffilmark{10,11}, 
  John~D.~Silverman\altaffilmark{23},  
  Audrey~Simmons\altaffilmark{13},      
  Stephanie~Snedden\altaffilmark{13},     
  Alina~Streblyanska\altaffilmark{24},
  Nao~Suzuki\altaffilmark{1},  
  David~H.~Weinberg\altaffilmark{25}, 
  Donald~York\altaffilmark{26} 
}
\altaffiltext{1}{Lawrence Berkeley National Laboratory, 1 Cyclotron Road, Berkeley, CA 92420, U.S.A.} 
\email{npross@lbl.gov}
\altaffiltext{2}{Steward Observatory, 933 North Cherry Avenue, Tucson, AZ 85721, U.S.A.}
\altaffiltext{3}{Department of Physics, 366 LeConte Hall, University of California, Berkeley, CA 94720, U.S.A.}
\altaffiltext{4}{Department of Physics, Drexel University, 3141 Chestnut Street, Philadelphia, PA 19104, U.S.A}
\altaffiltext{5}{Department of Physics and Astronomy, University of Wyoming, Laramie, WY 82071, U.S.A.} 
\altaffiltext{6}{CEA, Centre de Saclay, IRFU, 91191 Gif-sur-Yvette, France}
\altaffiltext{7}{Department of Astrophysical Sciences, Princeton University, Princeton, NJ 08544, U.S.A.}
\altaffiltext{8}{Department of Astronomy, University of Washington, Box 351580, Seattle, WA 98195, U.S.A.}
\altaffiltext{9}{Harvard-Smithsonian Center for Astrophysics, 60 Garden Street, Cambridge, MA 02138, U.S.A.}
\altaffiltext{10}{Department of Astronomy and Astrophysics, The Pennsylvania State University, 525 Davey Laboratory, University Park, PA 16802, U.S.A.}
\altaffiltext{11}{Institute for Gravitation and the Cosmos, The Pennsylvania State University, 104 Davey Laboratory, University Park, PA 16802, U.S.A.}
\altaffiltext{12}{APC, University of Paris Diderot, CNRS/IN2P3, CEA/IRFU, Observatoire de Paris, Sorbonne Paris Cite, France.}
\altaffiltext{13}{Apache Point Observatory, P.O. Box 59, Sunspot, NM 88349-0059, U.S.A.}
\altaffiltext{14}{Institute for Advanced Study, Einstein Drive, Princeton, NJ 08540, U.S.A.}
\altaffiltext{15}{Hubble Fellow}
\altaffiltext{16}{McWilliams Center for Cosmology, Carnegie Mellon University, 5000 Forbes Avenue, Pittsburgh, PA 15213, U.S.A.}
%\altaffiltext{16}{Department of Physics, Carnegie Mellon University, 5000 Forbes Avenue, Pittsburgh, PA 15213, U.S.A.}
\altaffiltext{17}{Dept. of Astronomy, University of Florida, 211 Bryant Space Science Center, Gainesville, FL, 32611, U.S.A.}
\altaffiltext{18}{Institute of Cosmology \& Gravitation, Dennis Sciama Building, University of Portsmouth, Portsmouth, PO1 3FX, U.K.}
\altaffiltext{19}{Center for Cosmology and Particle Physics, Department of Physics, New York University, 4 Washington Place, New York, NY 10003, U.S.A.}
\altaffiltext{20}{UPMC-CNRS, UMR7095, Institut d'Astrophysique de Paris, F-75014, Paris, France}  
\altaffiltext{21}{Departamento de Astronom\'ia, Universidad de Chile, Casilla 36-D, Santiago, Chile}
\altaffiltext{22}{Department of Physics, Yale University, New Haven, CT 06511, U.S.A.}
\altaffiltext{23}{Institute for the Physics and Mathematics of the Universe (IPMU), University of Tokyo, Kashiwanoha 5-1-5, Kashiwa, Chiba 277-8568, Japan}
\altaffiltext{24}{Instituto de Astrofsica de Canarias (IAC), E-38200 La Laguna, Tenerife, Spain}
\altaffiltext{25}{Astronomy Department and Center for Cosmology and AstroParticle Physics, Ohio State University, 140 West 18th Avenue, Columbus, OH 43210, U.S.A.}
\altaffiltext{26}{Department of Astronomy and Astrophysics and the Fermi Institute, The University of Chicago, Chicago, IL 60637, U.S.A.}

\date{\today}

\begin{abstract}
We present a new measurement of the optical Quasar Luminosity Function
(QLF), using data from the Sloan Digital Sky Survey-III: Baryon
Oscillation Spectroscopic Survey (SDSS-III: BOSS).  From the SDSS-III
Data Release Nine (DR9), we select a uniform sample of \hbox{22,301}
$i\lesssim21.8$ quasars over an area of 2236~deg$^{2}$ with confirmed
spectroscopic redshifts between $2.2<z<3.5$, filling in a key part of
the luminosity-redshift plane for optical quasar studies.
We derive the completeness of the survey through simulated quasar
photometry, and check this completeness estimate using a sample of
quasars selected by their photometric variability within the BOSS
footprint. We investigate the level of systematics associated with
our quasar sample using the simulations, in the process generating
color-redshift relations and a new quasar $k$-correction.
We probe the faint end of the QLF to $M_{i}(z=2.2)\approx-24.5$ and
see a clear break in the QLF at all redshifts up to $z=3.5$. We find
that a log-linear relation (in $\log\Phi^{*}-M^{*}$) for a luminosity
and density evolution (LEDE) model adequately describes our data
within the range $2.2<z<3.5$; across this interval the break
luminosity increases by a factor of $\sim$2.3 while $\Phi^{*}$
declines by a factor of $\sim$6.  At $z\lesssim2.2$ our data is
reasonably well fit by a pure luminosity evolution (PLE) model.
We see only a weak signature of ``AGN downsizing'', in line with recent
studies of the hard X-ray luminosity function.
We compare our measured QLF to a number of theoretical models and find that
models making a variety of assumptions about quasar triggering and halo
occupation can fit our data over a wide range of redshifts and luminosities.
\end{abstract}


\maketitle

%\clearpage

%%%%%%%%%%%%%%%%%%%%%%%%%%%%%%%%%%%%%%%%%%%%%%%%%%%%%%%%%%%%%%
%%%%%%%%%%%%%%%%%%%%%%%%%%%%%%%%%%%%%%%%%%%%%%%%%%%%%%%%%%%%%%
%%
%%   SECTION 1  SECTION 1  SECTION 1  SECTION 1  SECTION 1  SECTION 1  
%%   SECTION 1  SECTION 1  SECTION 1  SECTION 1  SECTION 1  SECTION 1  
%%   SECTION 1  SECTION 1  SECTION 1  SECTION 1  SECTION 1  SECTION 1  
%%
%%%%%%%%%%%%%%%%%%%%%%%%%%%%%%%%%%%%%%%%%%%%%%%%%%%%%%%%%%%%%%
%%%%%%%%%%%%%%%%%%%%%%%%%%%%%%%%%%%%%%%%%%%%%%%%%%%%%%%%%%%%%%
\section{Introduction}\label{sec:intro}
Quasars, i.e. luminous active galactic nuclei (AGN), represent a
fascinating and unique population of objects at the intersection of
cosmology and astrophysics. The cosmological evolution of the quasar
luminosity function (QLF) has been of interest since quasars were
first identified a half-century ago \citep{Sandage61, Hazard63,
Schmidt63, Oke63, Greenstein63, Burbidge67}.

Measuring the QLF, and its evolution with redshift, is important for
several reasons. It is generally believed that present-day
supermassive black holes (SMBHs) gained most of their mass via gas
accretion during an active nuclear phase, potentially at quasar
luminosities \citep[$L_{\rm Bol}\gtrsim10^{45}$ erg
s$^{-1}$;][]{Salpeter64, ZelNov65, LyndenBell69, Sol82}, so an
accurate description of the QLF allows us to place constraints on the
formation history of supermassive black holes \citep[e.g., ][]{Rees84,
MadauRees01, Volonteri03, Volonteri06, Netzer07, Haiman12} and to map
the black hole accretion history of the Universe via the black hole
mass function \citep{Shankar09, Shankar10, Shen09b, Shen_Kelly12}, as
well as constrain the effect of black hole spin on the central engine
\citep{Volonteri05, Fanidakis11}.

Measurements of the QLF also place constraints on the intensities and
nature of various cosmic backgrounds, including the buildup of the
cosmic X-ray \citep{Shanks91, Comastri95, Ueda03, Brandt05, Hickox06},
ultraviolet \citep{Henry91} and infrared (IR) \citep{Hauser_Dwek01,
Dole06} backgrounds. Knowledge of the UV background is relevant for
calculations that involve the contribution of quasar UV photons to the
epoch of H reionization \citep[e.g.,][]{Fan06review} at
$z\gtrsim6$. At lower ($z\lesssim6$) redshift, quasars contribute
towards a fraction of the ionizing photons that keep most of the 
H ionized, allowing studies of the Ly-$\alpha$ forest 
\citep[Ly$\alpha$F; e.g. ][]{Lynds71,Meiksin09}. Helium reionization 
(\ion{He}{2}$\rightarrow$\ion{He}{3})
can be measured by its effect on the Ly$\alpha$F \citep{Jakobsen94,
Reimers97, Smette02, Reimers05, Syphers11, Worseck11He}. This second
epoch of reionization occurs at $z\sim3$, and may be driven by
UV photons from quasars, so an accurate determination of the QLF at
this epoch is a key consistency check on the He reionization
measurements.

Furthermore, the co-evolution of galaxies and AGN is a crucial
ingredient in, and test of, modern theories of galaxy formation. The
energy feedback from AGN is thought to impact their host galaxies, and
thus influencing their present-day properties \citep[e.g.,
][]{Cattaneo09, Fabian12}. Observations of the evolution of quasar
properties over cosmic time can inform such models and therefore our
understanding of the galaxy-black hole connection.

Recent large quasar surveys have allowed us to study the properties of
the quasar population with unprecedented statistical precision. The
number of known quasars has increased nearly 100-fold since the late
1990s, \citep[for photometrically identified quasars, see
][]{Richards09} and since that time, there has been a large effort to
measure the QLF in the UV/optical \citep[][]{Boyle00, Fan01, Wolf03,
Hunt04, Fan04, Croom04, Hao05b, Richards05, Richards06, Fan06,
Jiang06, Fontanot07, Bongiorno07, Reyes08, Jiang08, Jiang09, Croom09b,
Glikman10, Willott10, Glikman11, Ikeda11, Ikeda12, Masters12},
mid-infrared \citep{Brown06a, Siana08, Assef11} and the soft and hard
X-ray \citep{Cowie03, Ueda03, Hasinger05, Barger05, Silverman05,
Silverman08, Aird08, Treister09, Aird10, Fiore12}. An overview of
recent determinations of the optical QLF is given in
Table~\ref{tab:previous_surveys}.

\begin{table*}
  \begin{center}
    \caption{Selected optical quasar luminosity function measurements.\\ 
      $^{a}$Cosmic Evolution Survey \citep{Scoville07}. \\
      $^{b}$No Type-1 quasars were identified, though a low-luminosity $z\sim5.07$ Type-2 quasar was discovered. \\
      $^{c}$NOAO Deep Wide-Field Survey \citep{JD99} and the Deep Lens Survey \citep{Wittman02}. \\
      $^{d}$SDSS Faint Quasar Survey. \\
      $^{e}$The ``boss21'' area on the SDSS Stripe 82 field. \\
      $^{f}$2dF-SDSS LRG And QSO Survey \citep{Croom09a}. \\
      $^{g}$Photometric sample from SDSS; spectroscopic confirmation from SDSS and other telescopes.\\
      $^{h}$Canada-France High-$z$ Quasar Survey \citep{Willott09}\\
      $^{i}$2dF Quasar Redshift Survey \citep{Croom04}. \\
      $^{j}$From our ``uniform'' sample defined in Section~\ref{sec:data_uniform} \\
      $^{k}$From a catalog of $>$1,000,000 photometrically classified quasar candidates. \\}
    \begin{tabular}{lrrccl}
      \hline
      \hline
      Survey                      & Area (deg$^{2}$) & N$_{\rm Q}$ & Magnitude Range & $z$-range & Reference  \\
      \hline 
      GOODS(+SDSS)        &   0.1+(4200)  &   13(+656)    &  $22.25 < z_{850} < 25.25$  & $3.5<z<5.2$  & \citet{Fontanot07} \\
      VVDS                       &        0.62    & 130      &  $17.5 < I_{\rm AB} < 24.0$ &  $0<z<5$    &  \citet{Bongiorno07} \\
      COMBO-17              &       0.8     & \hbox{  192} & $R<24$                       &  $1.2 < z < 4.8$  & \citet{Wolf03} \\
      COSMOS$^{a}$         &       1.64   & 8     & $22 < i' < 24$              & $3.7 \lesssim z \lesssim 4.7$   & \citet{Ikeda11} \\      
      COSMOS                  &       1.64    & $^{b}$0         & $22 < i' < 24$     & $4.5 \lesssim z \lesssim 5.5$   & \citet{Ikeda12} \\   
      COSMOS                  &       1.64    & 155              & $16 \leq  I_{\rm AB} \leq 25$      & $3<z<5$   & \citet{Masters12} \\   
      NDWFS+DFS$^{c}$   &       4        &                24  & $R \leq 24$                    & $3.7<z<5.1$       & \citet{Glikman11} \\
      SFQS$^{d}$               &       4        & \hbox{  414} & $g<22.5$                          & $z<5$                 & \citet{Jiang06} \\
      BOSS$^{e}$+MMT     & 14.5+3.92 & \hbox{1 877} & $g\lesssim 23$  & $0.7<z<4.0$ & \citet{Palanque-Delabrouille12} \\ 
      2SLAQ$^{f}$            &     105     & \hbox{ 5 645} & $18.00 < g < 21.85$         & $z\leq2.1$               & \citet{Richards05}     \\
      SDSS$^{g}$               &    182     &                  39  &  $i\leq 20$                        &  $3.6<z<5.0$    & \citet{Fan01} \\
      SDSS+2SLAQ           &    192     & \hbox{10 637} & $18.00 < g < 21.85$        & $0.4<z<2.6 $    & \citet{Croom09b}     \\
      SDSS Main+Deep     &    195     &                     6 & $z_{\rm AB} < 21.80$        & $z\sim6 $         & \citet{Jiang09}     \\
      {\bf BOSS Stripe 82} & {\bf 220}  &{\bf \hbox{5 476}} & $i>${\bf 18.0 and} $g< ${\bf 22.3}  & {\bf 2.2}$<z<${\bf 3.5}  &\citet{Palanque-Delabrouille11} \\ 
      CFHQS$^{h}$             & 500      &  \hbox{ 19}                  & $z'<22.63$           &  $5.74 < z < 6.42. $ & \citet{Willott10} \\
      2QZ$^{i}$                 &   700     & \hbox{23 338} & $18.25<b_{\rm J}<20.85$   & $0.4<z<2.1$     & \citet{Boyle00, Croom04}     \\
      SDSS DR3                 & 1622     & \hbox{15 343} &  $i\leq 19.1$ and $i\leq 20.2$ &  $0.3<z<5.0$ & \citet{Richards06} \\
     {\bf BOSS DR9}  &  {\bf 2236}  & $^{j}${\bf \hbox{23 201}} & $g<${\bf 22.00 or }$r<${\bf 21.85}  & {\bf 2.2}$<z<${\bf 3.5}  & {\bf this paper}           \\
     SDSS DR7                 &  6248 & \hbox{57 959}  &  $i\leq 19.1$ and $i\leq 20.2$ &  $0.3<z<5.0$ & \citet{Shen_Kelly12} \\
      SDSS Type 2            &  6293 & \hbox{   887}  & $L_{\rm O III} \geq 10^{8.3} L_\odot$ &  $z<0.83$ & \citet{Reyes08} \\
      SDSS DR6$^{k}$        &  8417 & $\gtrsim\hbox{850,000}$ &  $i<21.3 $        &  $z\sim2$ and $z\sim4.25$ & \citet{Richards09}
  \\
      \hline
      \hline
      \label{tab:previous_surveys}
    \end{tabular}
  \end{center}
\end{table*}

Quasar number density evolves strongly with redshift
\citep[][]{Schmidt70, Osmer82, Schmidt95, Fan01, Richards06,
Croom09a}, and one of the key goals of quasar studies is to understand
what drives this strong evolution.  A caveat here is that the
evolution of the optical QLF is a composite of intrinsic quasar
evolution and the evolution of the obscuring medium in quasar
hosts. In this study, we concentrate on the unobscured quasar
population, defined as objects that were selected via their UV/optical
rest-frame continuum and the presence of broad, $\gtrsim$1000 $\kms$,
emission lines. We leave investigations of the obscured AGN population
to other studies, e.g., in the mid-infrared
\citep[e.g.,][]{Lacy04, Richards06b, Stern12, Yan12}, and X-ray
\citep{Tueller10, Luo11, Corral11, Brightman12, Lehmer12}. As the QLF
is observed to have a broken power-law form, it is
necessary to probe below the luminosity at which the power-law breaks
in order to distinguish {\it luminosity evolution} (where the
luminosity of AGN changes with time, but their number density remains
constant) from {\it density evolution} (where the number density of
AGN changes, but the luminosities of individual objects remains
constant), or a combination of the two.

The QLF is defined as the number density of quasars per unit
luminosity. It is often described by a double power-law
\citep[][hereafter, R06]{Boyle00,Croom04,Richards06} of the form
\begin{equation}
  \Phi(L, z) = \frac{ \phi_{*}^{(L)} }
                            {  (L/L^{*})^{\alpha}    +  (L/L^{*})^{\beta}  }
 \ \,
 \label{eq:double_powerlaw}
\end{equation}
with a characteristic, or break, luminosity $L_{*}$.  An alternative
definition of this form of the QLF gives the number density of quasars
per unit magnitude,
\begin{equation}
  \Phi(M, z) = \frac{ \phi_{*}^{(M)} }
       { 10^{0.4{(\alpha +1)[M-M^{*}(z)]}}+10^{0.4{(\beta +1)[M-M^{*}(z) ]}} }
 \label{eq:double_powerlaw_mag}
\end{equation}
The dimensions of $\Phi$ differ in the two conventions.  We have
followed R06 such that $\alpha$ describes the faint end QLF slope, and
$\beta$ the bright end slope.  The $\alpha$/$\beta$ convention in some
other works \citep[e.g.,][]{Croom09b} is in the opposite sense from our 
definition. Evolution of the QLF can be encoded in the redshift dependence 
of the break luminosity, $\phi_{*}$, and also potentially in the evolution of
the power-law slopes.

\citet{Boyle00} and \citet{Croom04} found that the QLF measured in the
2dF Quasar Redshift Survey \citep[2QZ, ][]{Croom04} was well fit by a
pure luminosity evolution model where $\Phi_{*}^{(M)}$ was constant
but $M_*$ evolved with redshift. In this model, $M_*$ changed from 
$\approx -26.0$ to $\approx -22.0$ between
$z\sim2.5$ and $z\sim0$. However, this paradigm is challenged using
recent, deeper data. \citet{Croom09b} measured the optical quasar
luminosity function at $z\leq2.6$ from the combination of the 2dF-SDSS
LRG And QSO survey \citep[2SLAQ; ][]{Croom09a}, which probes down to a
magnitude limit of $g=21.85$, and the SDSS-I/II Quasar survey
\citep{Richards02, Schneider10} to $i=19.1$ ($z\lesssim3$) and
$i=20.2$ ($z\gtrsim3$). Here, the double power-law form with pure
luminosity evolution provides a reasonable fit to the observed QLF
from low $z$ up to $z\simeq 2$, but it appears to break down at higher
redshift.  However, the 2SLAQ sample has few objects above $z\sim2$,
and SDSS does not probe down to $L_{*}$ at higher redshifts, making it
difficult to constrain the faint end of the QLF at high $z$.

At $z\gtrsim2$, the constraints on the QLF are less clear-cut, as the
selection of luminous quasars becomes less efficient.  This situation
arises because the broad-band colors of $z\approx2.7$ and
$z\approx3.5$ quasars are very similar to those of A and F stars
\citep{Fan99, Fan01, Richards02, Ross12} in the Sloan Digital Sky
Survey color system \citep{Fukugita96}.  Although we have good
constraining power at the bright end at $z>2$,
\citep[e.g.][]{Richards06, Jiang09}, there is uncertainty in the form,
and evolution of the QLF at $z>2$, especially at the faint end. The
redshift range $z\sim2-3$ is of particular importance since the
luminous quasar number density peaks here; this is often referred to
as the ``quasar epoch'' \citep{Osmer82, Warren94, Schmidt95, Fan01,
Richards06, Croom09b}.

For our study, we use data from the SDSS-III: Baryon Oscillation
Spectroscopic Survey \citep[BOSS; ][]{Dawson12} that is specifically
designed to target faint, $g\lesssim22$, quasars in the redshift range
$z=2.2-3.5$ \citep{Ross12}. The first two phases of the SDSS
\citep[``SDSS-I/II'', hereafter simply SDSS;][]{York00} have been
completed \citep{Abazajian09}, with a sample of $\approx$100,000
spectroscopically confirmed quasars at $0<z\lesssim5$
\citep{Schneider10}. The third incarnation of the Sloan Digital Sky
Survey \citep[SDSS-III; ][]{Eisenstein11} is taking spectra of 150,000
$z>2.2$ quasars as part of the BOSS. The main scientific motivation
for the SDSS-III BOSS Quasar survey is to measure the baryon acoustic
oscillation feature (BAO) in the Lyman-$\alpha$ forest
\citep[Ly$\alpha$F; ][]{Slosar11}. This sample is designed to select
quasars with $2.2<z<3.5$, and will have an order of magnitude more
objects at $z>2$ than SDSS, sampling the quasar luminosity function
$\sim2$ magnitudes deeper at each redshift. Combining the BOSS and
SDSS observations gives a dynamic range of $\sim$5 magnitudes at a
given redshift, and a primary motivation for our study is to extend
the work presented in \citet{Richards06}, both in dynamic range in
luminosity, and concentrating on the redshift range $z=2.2-3.5$, where
the original SDSS selection was sparse-sampled in an attempt to
minimize the contamination by stars \citep{Richards02}.

In this paper we present the optical quasar luminosity function (QLF)
from the first two years of BOSS spectroscopy, data included in SDSS
Data Release Nine \citep[DR9;
][]{DR9}\footnote{\href{http://www.sdss3.org/surveys/}{http://www.sdss3.org/surveys/}}. We
use data from the 3671 deg$^{2}$ observed over the DR9 footprint, and
supplement this with deeper data over a smaller area (14.6 deg$^{2}$),
in order to probe the redshift range $0.7<z<2.2$, also observed as
part of the BOSS
\citep[Table~\ref{tab:previous_surveys};][]{Palanque-Delabrouille12}. Table~\ref{tab:previous_surveys}
places the BOSS DR9 survey in context as a wide-field, medium-depth
survey, and we will return to the surveys that match BOSS in redshift.

The outline of this paper is as follows. In Section 2 we describe our
data sets, which includes both color and variability selected AGN
samples.  In Section 3, we qualitatively compare our different quasar
samples, and quantify our selection function using both empirical
data, and new, updated template quasar spectra. In Section 4 we
present the SDSS+BOSS quasar number counts and a new quasar
$k$-correction based on our simulations. In Section 5, we present the
combined SDSS+BOSS QLF, sampling the range $-24.5 < M_{i} < -30$ in
absolute magnitude across redshifts $0.7<z<3.5$ and compare to
previous measurements. In Section 6, various models of the
double-power law form are fit to our data, we compare our results to
recent theoretical predictions in the literature, and place our new
results in a broader context. We present our conclusions in Section
7. In Appendix~\ref{appndx:spec_models} we investigate further the
selection function models introduced in Section 3 and in
Appendix~\ref{appndx:tables} provide tables of our measured QLFs.  For
direct comparison with, and extension of, R06, we assume a flat
cosmology with $\Omega_{\Lambda}$=0.70 and $H_{0} = 70\hmpc$. Our
magnitudes are based on the AB zero-point system \citep{Oke83} and are
PSF magnitudes \citep{Stoughton02}, corrected for Galactic extinction
following \citet{Schlegel98}. Absolute magnitudes ($M$) are determined
using luminosity distances for this cosmology \citep{Peebles80book,
Peebles93book, Hogg02, Wright06}.

%%%%%%%%%%%%%%%%%%%%%%%%%%%%%%%%%%%%%%%%%%%%%%%%%%%%%%%%%%%%%%
%%%%%%%%%%%%%%%%%%%%%%%%%%%%%%%%%%%%%%%%%%%%%%%%%%%%%%%%%%%%%%
%%
%%   SECTION 2     SECTION 2     SECTION 2     SECTION 2     SECTION 2     SECTION 2     
%%   SECTION 2     SECTION 2     SECTION 2     SECTION 2     SECTION 2     SECTION 2     
%%   SECTION 2     SECTION 2     SECTION 2     SECTION 2     SECTION 2     SECTION 2     
%%
%%%%%%%%%%%%%%%%%%%%%%%%%%%%%%%%%%%%%%%%%%%%%%%%%%%%%%%%%%%%%%
%%%%%%%%%%%%%%%%%%%%%%%%%%%%%%%%%%%%%%%%%%%%%%%%%%%%%%%%%%%%%%   
\begin{figure}
  \begin{center}
    \includegraphics[height=10.0cm, width=8.0cm]
    {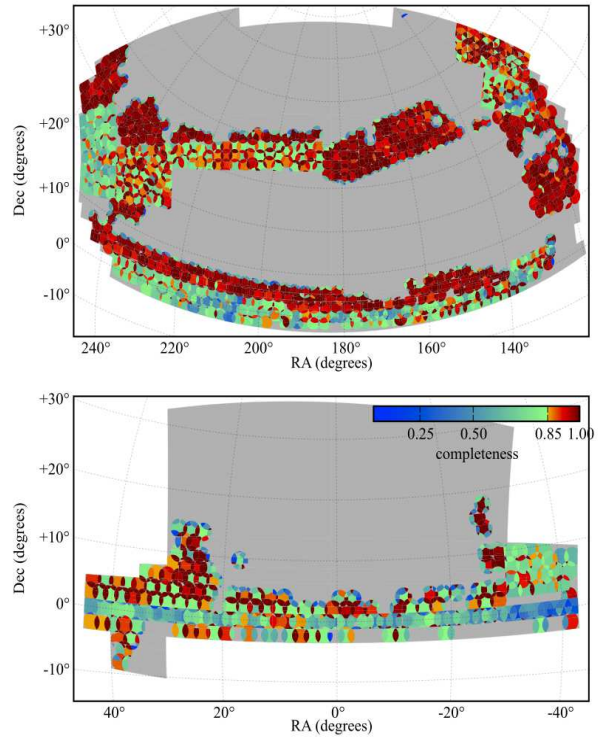}
    \caption{
      The sky coverage of the SDSS-III: BOSS DR9 quasar dataset
      (colored regions) overlaid on the final expected footprint of BOSS
      (gray). These areas are 3671 and 10 269 deg$^{2}$, respectively. The
      upper panel shows the coverage in the NGC, and the lower one is the
      SGC. Each sector (covered by a unique combination of spectroscopic
      tiles) is colored according to the fraction of quasar targets,
      selected with the uniform XDQSO algorithm, which have successful
      redshifts. The sectors which contribute data to the QLF analysis have
      $>85\%$ spectroscopic completeness (yellow-orange-red regions). This
      area is 2236 deg$^{2}$. The Stripe 82 field runs from $-43^{\circ}<$
      R.A. $<45^{\circ}$ at Decl.$=\pm1.25^{\circ}$ and generally has
      $f_{\rm sc}<0.85$ (where $f_{\rm sc}$ is defined in
      Sec.~\ref{sec:spec_completeness}).  However, since Stripe 82 had
      quasar targets that were variability selected
      (see Sec.~\ref{sec:variability}), the {\it true} number of quasars in this
      field is actually very high.}
    \label{fig:DR9_coverage}
  \end{center}
\end{figure}

\section{Data}\label{sec:data}
We use imaging data that are part of Data Release Eight
\citep[DR8;][]{DR8} in order to select spectroscopic targets that form
the Data Release Nine \citep[][]{DR9} dataset. \citet{Ross12}
describes the BOSS quasar target selection algorithms used to identify
objects for spectroscopy.  In summary, we use the subset of the DR9
data that employs the ``Extreme Deconvolution'' (XDQSO) algorithm of
\citet{Bovy11} to select quasars based on their optical fluxes and
colors to define a uniform sample.  The XDQSO procedure is
supplemented by a selection using optical variability, where we have
repeat imaging data within the DR8 footprint. The DR9 data are the
first two years of BOSS spectroscopic data, and the full DR9 Quasar
Catalog (DR9Q) is detailed in \citet{Paris12}. Fig~\ref{fig:DR9_coverage}
shows the sky coverage of the DR9 quasar dataset. {\it However, the
XDQSO selection was not implemented in the first year}, leading to
effects on completeness that we will address below in order to perform
a QLF measurement.

    \subsection{Imaging and Target Selection}\label{sec:data_imaging}
    The SDSS-III:BOSS uses the imaging data gathered by a dedicated
    2.5m wide-field telescope \citep{Gunn06}, which collected light for a
    camera with 30 2k$\times$2k CCDs \citep{Gunn98} over five broad bands
    - {\it ugriz} \citep{Fukugita96} - in order to image 14,555 unique
    deg$^{2}$ of the sky. This area includes \hbox{7,500 deg$^{2}$} in the
    North Galactic Cap (NGC) and \hbox{3,100} deg$^{2}$ in the South
    Galactic Cap (SGC). The imaging data are taken on dark photometric
    nights of good seeing \citep{Hogg01} and are calibrated
    photometrically \citep{Smith02, Ivezic04, Tucker06, Padmanabhan08a},
    and astrometrically \citep{Pier03} before object parameters are
    measured \citep{Lupton01, Stoughton02}.

    Using the imaging data, BOSS quasar target candidates are selected
    for spectroscopic observation based on their fluxes and colors in SDSS
    bands. However, selection of quasars for BOSS spectroscopy is
    complicated by two facts: {\it(i)} The optical colors of $z\sim2.7$ quasars
    resemble faint A and F stars \citep{Fan99, Richards01, Ross12} and {\it(ii)} 
    to maximize the number density of quasars for Ly$\alpha$F cosmology,
    we are required to work close to the magnitude limit of the
    (single-epoch) imaging data, leading to larger photometric errors,
    expansion of the stellar locus and higher stellar contamination. All
    objects classified as point-like and having magnitudes of $g\leq22$ {\it
      or} $r<21.85$ are passed to the quasar target selection code.
    
    As was the case for the original SDSS Quasar
    survey, radio data was used to select quasars. Specifically, optical
    stellar objects with $g \leq 22$ or $r \leq 21.85$ which have
    matches within 1$''$ to radio sources apparent in the Faint Radio
    Sources at Twenty cm (FIRST) survey \citep{Becker95} are considered as
    potential quasar targets, irrespective of their radio
    morphology. Approximately 2\% of targets, and $\approx$1.3\% of our
    uniform quasar sample (defined in~\S~\ref{sec:data_uniform} below), satisfy
    the radio selection criteria.

    As the main science goal of the BOSS quasar sample is to probe the
    foreground hydrogen in the IGM, priority was placed on maximizing the
    surface density of $z>2$ quasars \citep{McDonald07, McQuinn11}, rather
    than creating a homogeneous dataset. The target selection is
    consequently a complicated heterogenous combination of several methods
    \citep{Ross12}. However, a uniform subsample \citep[called ``CORE'' in
    ][]{Ross12} was defined to allow statistical studies of quasar
    demographics to be performed. The spectroscopic observations, and
    creation of this uniform subset of objects, are described in the next two sections.
    
    \subsection{Spectroscopy}\label{sec:data_spectra}
    The BOSS spectrographs and their SDSS predecessors are described
    in detail by \citet{Smee12}. In brief, there are two double-armed
    spectrographs that are significantly upgraded from those used by
    SDSS-I/II. They cover the wavelength range $3600\,$\AA\ to
    $10,400\,$\AA\ with a resolving power of 1500 to 2600 \citep{Smee12}.
    In addition, the throughputs have been increased with new CCDs,
    gratings, and improved optical elements, and the 640-fibre cartridges
    with $3"$ apertures have been replaced with 1000-fibre cartridges with
    $2"$ apertures. Each observation is performed in a series of
    900-second exposures, integrating until a minimum signal-to-noise
    ratio is achieved at a fiducial magnitude for the given spectroscopic
    plate \citep{Dawson12}.
    
    Once target selection is completed, the spectroscopic targets are
    assigned to tiles of diameter $3^\circ$ using an algorithm that is
    adaptive to the density of targets on the sky \citep{Blanton03}. Of
    the 1000 available fibers on each tile, a maximum of 900 fibers are
    allocated for science targets, of which $\sim$160-200 are
    allocated to quasar targets, while 560-630 fibers are assigned to
    galaxy targets, and 20-90 to ancillary science targets
    \citep{Dawson12}. Because of the 62$''$ diameter of the cladding around
    each optical fiber, two targets with a separation smaller than that
    angle cannot both be observed on a given spectroscopic plate, and
    different classes of targets are assigned priorities when such a
    collision arises. CORE quasars are assigned higher tiling priority
    than the galaxy targets \citep[Appendix B of][]{Ross12}. To cover the
    survey footprint without leaving gaps, adjacent tiles overlap,
    alleviating the fiber collisions problem somewhat. This leads to the
    definition of a ``sector'' - a region covered by a unique set of tiles
    \citep[see ][]{Blanton03, Tegmark04, Swanson08, White11}.  As in
    previous SDSS analyses, we work on a sector-by-sector basis to define
    our various completenessess.

    \begin{figure}
      \begin{center}
        \includegraphics[height=8.0cm, width=8.6cm]
        {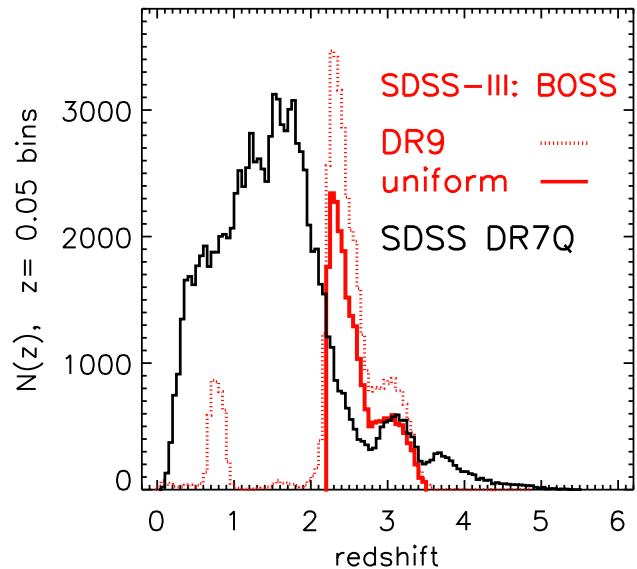}
        \caption{Quasar $N(z)$ redshift distributions. The dotted red
          histogram shows the redshift distribution for the full SDSS-III: BOSS
          DR9 quasar dataset, while the solid red line shows those objects
          uniformly selected by the ``XDQSO'' method across $2.2<z<3.5$.  The
          black histogram is the final distribution from the DR7Q catalog of
          \citet{Schneider10}.}
        \label{fig:Nofz}
      \end{center}
    \end{figure}

    The DR9 footprint is \hbox{3 671} deg$^{2}$, and is given in
    Fig.~\ref{fig:DR9_coverage}.  In total, we obtained \hbox{182 973}
    spectra of objects that were selected as BOSS quasar targets, and
    \citet[][]{Bolton12} describes the automated spectral classification,
    redshift determination, and parameter measurement pipeline used for
    the BOSS. A total of \hbox{167 331} of these had the {\tt specPrimary}
    flag set to 1, indicating that this was the best spectroscopic
    observation of an object; this cut, by definition, removes objects
    with duplicate spectra. As described in \citet{Adelman-McCarthy08} and
    \citet{Bolton12}, each redshift is accompanied by a flag, {\tt
      zWarning}, which is set when the automatically derived
    (a.k.a. pipeline) redshift and classification are not reliable;
    \hbox{132 290} of these objects do {\it not} have this flag set. Of these,
    \hbox{54 019} have pipeline redshifts between 2.2 and 3.5. A summary
    of these numbers is given in Table~\ref{tab:DR9_key_numbers}.
    
    Each of the quasar target spectra has also been visually
    inspected, and the redshift corrected where necessary. In total there
    are \hbox{87 822} spectroscopically confirmed quasars in the DR9Q,
    while approximately half the quasar candidates were stars
    \citep{Paris12}. If there was confidence in a secure redshift from the
    visual inspection of the spectrum, the flag {\tt z\_conf\_person} was
    set to be $\geq3$. Most of the {\tt zWarning} $\neq0$ objects have
    secure redshifts after visual inspection; over 97\% of the {\tt specPrimary}
    BOSS quasar target spectra have secure redshifts
    (Table~\ref{tab:DR9_key_numbers}). Among these objects, \hbox{ 54 593}
    are confirmed, by visual inspection, to be at $2.2<z<3.5$. For
    comparison the DR7Q \citep{Schneider10} has \hbox{14 063} objects in
    this redshift range.

\begin{table}
  \begin{center}
    \begin{tabular}{lrr}
      \hline
      \hline
      Description & \multicolumn{2}{c}{No. of Objects}  \\
                         & Pipeline$^{a}$         &  Visually                    \\
                         &                               &  Inspected                    \\
      \hline 
      All DR9 {\tt boss\_target1} quasar targets$^{b}$            & \hbox{182 973} &  ---\\
      \,\,\,\,  {\tt specPrimary = 1}                                        & \hbox{167 331} &  ---\\
      \,\,\,\,\,\,\,\,\, " and reliable redshift$^{c}$               & \hbox{132 290} & \hbox{163 128}  \\ 
      \,\,\,\,\,\,\,\,\,  \,\,\,\,\,\,\,\,\, '' and  $2.2<z<3.5$  & \hbox{ 54 019} & \hbox{ 54 593}  \\  
      & &  \\
      XDQSO DR9 quasar targets                                          & \hbox{ 74 607} & ---\\
      \,\,\,\,\,\,\,\,  '' with spectra$^{d}$                             & \hbox{ 63 061} & ---\\
      \,\,\,\,\,\,\,\,\,\,\,\,\  " and reliable redshift$^{c}$   & \hbox{  54 416} & \hbox{62 048} \\
      \,\,\,\,\,\,\,\,\,\,\,\,\,\,\,\,\,\, " and $2.2<z<3.5$   & \hbox{  34 803} & \hbox{35 099} \\
\,\,\,\,\,\,\,\,\,\,\,\,\,\,\,\,\,\,\,\,\,\,\,\,\,\,\, " and $f_{\rm sc}\geq 0.85$ & & \hbox{{\bf 23 301}} \\
      \hline
      \hline
    \end{tabular}
    \caption{Properties of the SDSS-III BOSS DR9 QLF dataset. \\
      $^{a}$The automated redshift determination algorithms described in \citet{Bolton12}. 
      $^{b}$The DR9 quasar {\tt boss\_target1} target flag is defined in \citet{Ross12}.
      $^{c}$Totals include stars. 
      $^{d}$All  XDQSO DR9 quasar target spectra have {\tt specprimary}=1 by design.
     The uniform sample defined in Section 2 is based upon the XDQSO selection.}
    \label{tab:DR9_key_numbers}
 \end{center}
\end{table}

    \subsection{DR9 Uniform Sample}\label{sec:data_uniform}
    We now define a uniform subsample from the parent
    DR9 quasar dataset. XDQSO models the distribution in SDSS flux space
    of stars and quasars as a function of redshift, as a sum of Gaussians
    convolved with photometric measurement errors, allowing the Bayesian
    probability that any given object is a quasar to be calculated. XDQSO
    is specifically trained and designed to select quasars in the redshift
    range $2.2<z<3.5$ down to the BOSS limiting magnitude.
    
    XDQSO was only chosen as the algorithm to define the uniform
    sample after the first year of BOSS spectroscopic observations. Each
    object is assigned a probability, P(QSOMIDZ), that it is a quasar with
    $2.2<z<3.5$. Objects with P(QSOMIDZ) $>0.424$ \citep{Bovy11} are
    targeted as part of the uniform (CORE) sample. Knowing this threshold,
    we are able to say which targets XDQSO {\it would have targeted} in
    the first year of observations, many of which BOSS did obtain spectra
    for. There are \hbox{74 607} quasar targets selected by XDQSO over the
    DR9 footprint, \hbox{63 061} of which have spectroscopic observations,
    and of these, over half (\hbox{35 099}) are confirmed
    $2.20<z<3.50$ quasars by visual inspection \citep{Paris12}.
    
    This sample of \hbox{35 099} quasars is over an order of magnitude
    more objects in this redshift range than in the study of
    \citet{Richards06} from DR3 \citep{Abazajian05}. Fig.~\ref{fig:Nofz}
    shows the redshift distribution of this sample.  Although we plot the
    full redshift range of the quasars, we do not use data from quasars
    which have a redshift below 2.2 or above 3.5, where the mid-$z$ XDQSO
    selection, by design, is quite incomplete. The BOSS DR9 uniform quasar
    sample has a mean (median) redshift of $\langle z\rangle=2.59$ (2.49).
     
    This uniform dataset is our primary basis for the QLF
    measurement. We supplement these data with a complementary dataset,
    selected by photometric variability criteria.
    
    \subsection{Variability selection: Stripe 82}\label{sec:variability}
    The $\sim$300 deg$^{2}$ area centered on the Celestial Equator in
    the Southern Galactic Cap, commonly referred to as ``Stripe 82'', was
    imaged repeatedly by the SDSS over 10 years, generating up to 80
    epochs \citep{Abazajian09}, due in large part to the SDSS Supernova
    Survey \citep{Frieman08}. These data are beneficial for quasar target
    selection for two reasons: {\it (i)} the improved photometry of the
    deeper data better defines the stellar locus \citep{Ivezic07} and {\it
      (ii)} quasars can be selected based on their variability
    \citep[][]{Sesar07, Bramich08, Schmidt10, MacLeod11,
      Palanque-Delabrouille11, Palanque-Delabrouille12}.
    
    \citet{Palanque-Delabrouille11} describe the spectroscopic quasar
    target selection for BOSS on 220 deg$^{2}$ of Stripe 82 based on
    variability. This variability selection was designed to select quasars
    with $i > 18.0$ and $g < 22.3$ mag and redshift $z>2.15$
    \citep[][Sec. 3.2]{Palanque-Delabrouille11}.  This dataset --- which
    we shall refer to as the Stripe 82 (S82) data in what follows ---
    includes $\sim$6000 $z>2$ quasars, roughly half of which would have been
    selected by XDQSO (as seen in Fig.~\ref{fig:DR9_coverage}). Since
    the completeness of the variability selection is only very weakly
    dependent on redshift \citep[e.g., Fig. 11
    of][]{Palanque-Delabrouille11} these data are subject to different,
    and arguably much weaker, selection biases than a color-based
    selection, as we show in Section~\ref{sec:NNVar_vs_XDQSO}.

    \begin{table*}
      \begin{center}
        \begin{tabular}{rr ccccc c rr c}
          \hline
          \hline
          R.A.       & Decl.   &  {\it u}-band &  {\it g}-band &  {\it r}-band &  {\it i}-band & {\it z}-band &  {\it i}-band &$z_{\rm pipe}$ & $z_{\rm vis}$  & $f_{\rm sc}$\\
          (J2000) & (J2000) &                     &                      &                    &                      &                    &  extinction    &  &   & \\
          \hline
          0.031620 &    0.495352 &  20.845$\pm$0.060 &  20.319$\pm$0.027 &  20.377$\pm$0.028 &  20.206$\pm$0.035 &  19.922$\pm$0.092 &  0.0527 &   2.260 &   2.254 &   0.4615 \\ 
          0.058656 &    1.497665 &  22.591$\pm$0.295 &  20.455$\pm$0.029 &  20.013$\pm$0.026 &  19.686$\pm$0.030 &  19.650$\pm$0.081 &  0.0509 &   3.228 &   3.228 &   0.8947 \\ 
          0.063211 &    0.809249 &  22.357$\pm$0.190 &  19.852$\pm$0.024 &  19.240$\pm$0.017 &  19.129$\pm$0.018 &  18.898$\pm$0.035 &  0.0585 &   3.028 &   3.028 &   0.8333 \\ 
          0.074886 &    0.407500 &  21.434$\pm$0.139 &  20.876$\pm$0.030 &  20.805$\pm$0.038 &  20.648$\pm$0.042 &  20.214$\pm$0.132 &  0.0540 &   2.282 &   2.281 &   0.4615 \\ 
          0.075538 &    1.610326 &  21.568$\pm$0.144 &  20.848$\pm$0.036 &  20.903$\pm$0.044 &  20.811$\pm$0.059 &  20.417$\pm$0.151 &  0.0485 &   2.400 &   2.400 &   0.8947 \\ 
          0.077683 &    3.548377 &  21.097$\pm$0.108 &  20.439$\pm$0.027 &  20.349$\pm$0.032 &  20.216$\pm$0.038 &  19.772$\pm$0.090 &  0.0563 &   2.237 &   2.238 &   0.7143 \\ 
          0.085803 &    3.399193 &  24.714$\pm$0.886 &  21.823$\pm$0.056 &  21.257$\pm$0.046 &  21.499$\pm$0.080 &  20.850$\pm$0.179 &  0.0569 &   2.904 &   2.903 &   0.7143 \\ 
          0.112584 &    3.120975 &  19.307$\pm$0.028 &  18.788$\pm$0.020 &  18.736$\pm$0.018 &  18.747$\pm$0.022 &  18.571$\pm$0.034 &  0.0451 &   2.353 &   2.343 &   1.0000 \\ 
          0.113820 &    1.523919 &  21.532$\pm$0.141 &  21.006$\pm$0.040 &  20.889$\pm$0.044 &  20.910$\pm$0.065 &  20.428$\pm$0.156 &  0.0500 &   2.589 &   2.589 &   0.8947 \\ 
          0.132704 &    1.685750 &  22.735$\pm$0.319 &  21.830$\pm$0.057 &  21.918$\pm$0.082 &  21.769$\pm$0.096 &  21.639$\pm$0.275 &  0.0476 &   2.526 &   2.526 &   0.8947 \\ 
          \hline
          \hline
        \end{tabular}
        \caption{
          The BOSS DR9 statistical quasar dataset. The first ten
          lines are shown here for guidance regarding its format and
          content. The full table is published in the electronic edition of {\it
            The Astrophysical Journal}. More details of the pipeline and visual
          inspection redshifts are documented in \citet{Bolton12} and
          \citet{Paris12}.}
        \label{tab:uniform_DR9_quasars}
      \end{center}
    \end{table*}
%%%%%%%%%%%%%%%%%%%%%%%%%%%%%%%%%%%%%%%%%%%%%%%%%%%%%%%%%%%%%% 
%%%%%%%%%%%%%%%%%%%%%%%%%%%%%%%%%%%%%%%%%%%%%%%%%%%%%%%%%%%%%% 
%% 
%%   SECTION 3     SECTION 3     SECTION 3     SECTION 3     SECTION 3     SECTION 3     
%%   SECTION 3     SECTION 3     SECTION 3     SECTION 3     SECTION 3     SECTION 3     
%%   SECTION 3     SECTION 3     SECTION 3     SECTION 3     SECTION 3     SECTION 3     
%%
%%%%%%%%%%%%%%%%%%%%%%%%%%%%%%%%%%%%%%%%%%%%%%%%%%%%%%%%%%%%%%
%%%%%%%%%%%%%%%%%%%%%%%%%%%%%%%%%%%%%%%%%%%%%%%%%%%%%%%%%%%%%%   
    \begin{figure*}[!th]
      \begin{center}
        %% 5.9cm works, 6.0cm doesn't... :-) 
        \includegraphics[height=5.9cm, width=5.9cm]{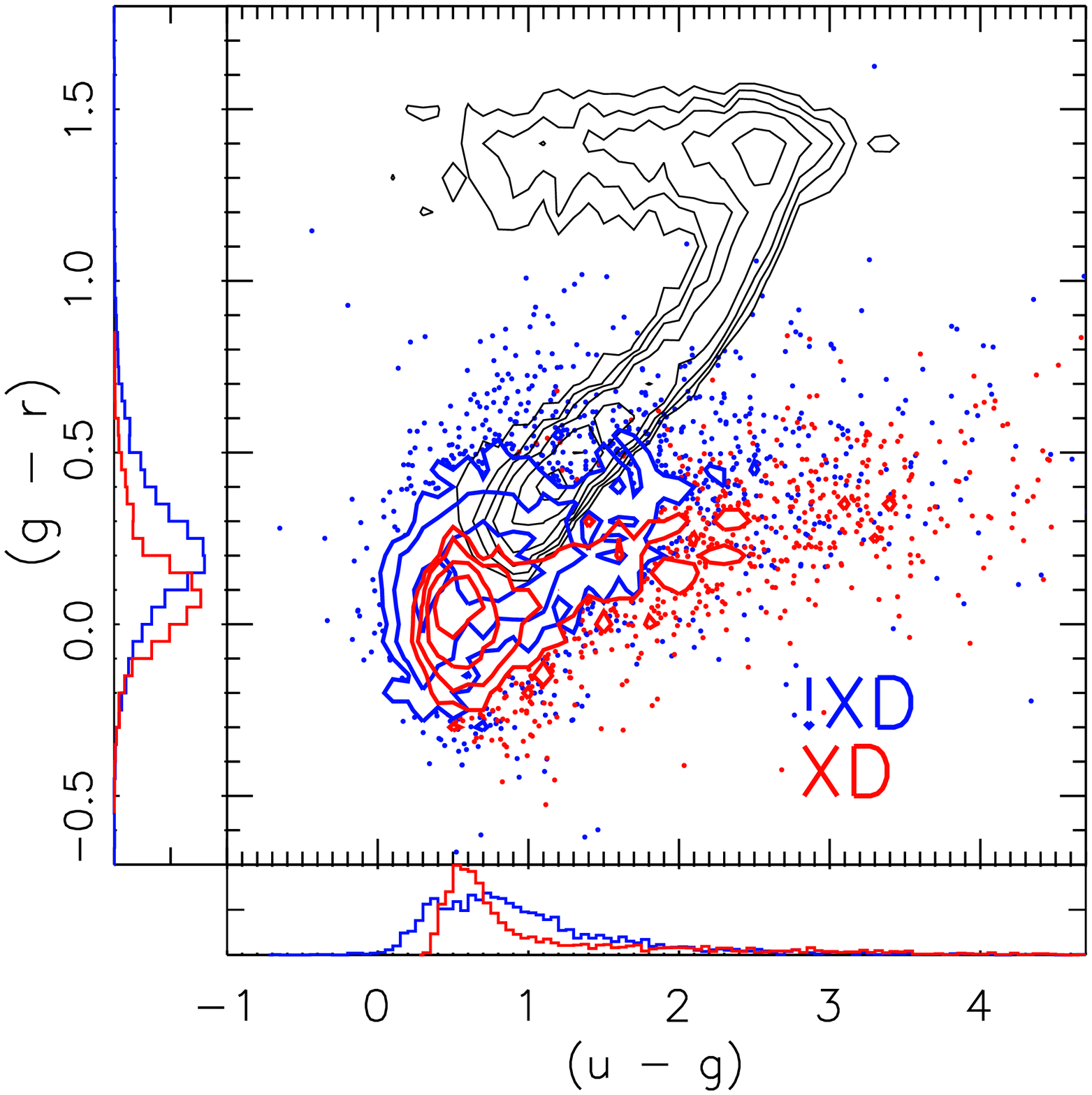}
        \includegraphics[height=5.9cm, width=5.9cm]{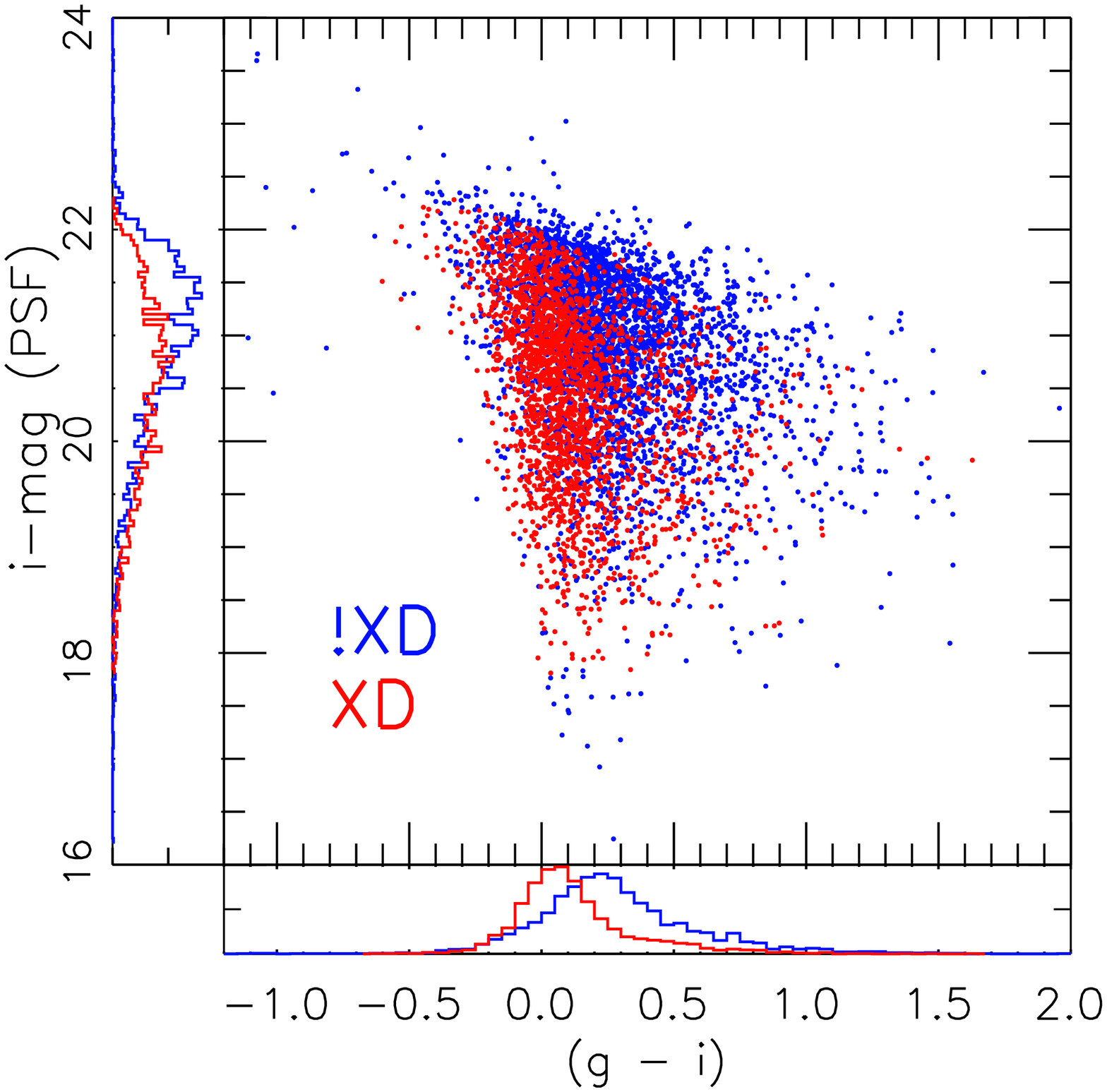}
        \includegraphics[height=5.9cm, width=5.9cm]{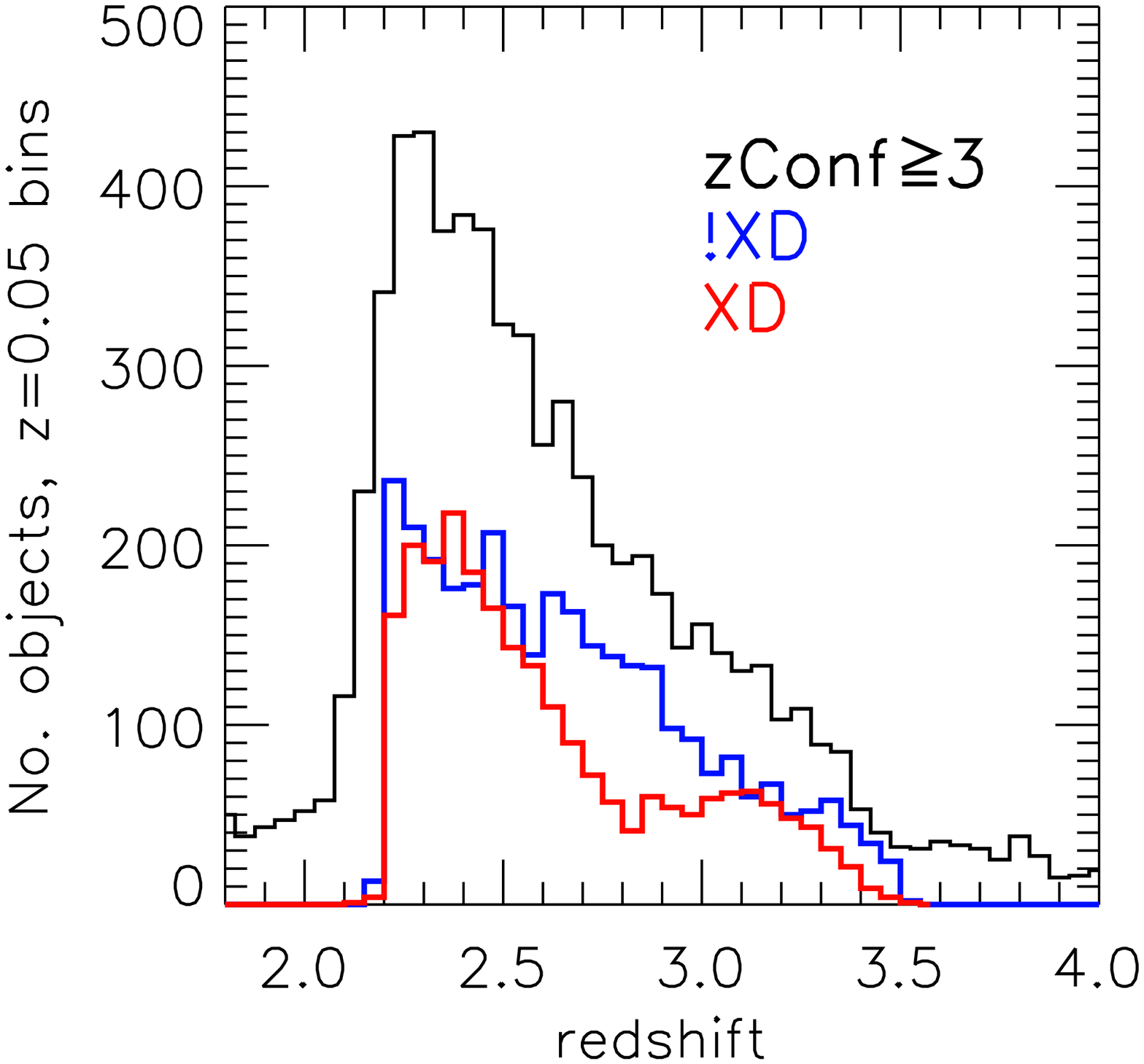}
        \caption{
          Splitting the sample of variability selected $2.2<z<3.5$
          Stripe 82 quasars into the \hbox{2 333} that are selected by the XDQSO
          algorithm (red) and those (\hbox{3 143}) that are not (blue).  {\it
            (Left):} The distributions in the $(u-g)$ vs. $(g-r)$ color-color
          plane, (the stellar locus is given by the black contours); {\it
            (Center):} the $(g-i)$ vs. $i$-band and {\it (Right):} the resulting
          $N(z)$ histogram, (with all {\tt z\_conf\_person$\geq$3} objects indicating a
          secure, visually inspected, redshift) plotted by the black line.        }
        \label{fig:Stripe82_XDnotXD}
      \end{center}
    \end{figure*}

\section{Survey Completeness}\label{sec:completeness}
To measure the QLF we must quantify the probability, $P(z,M,{\rm
SED})$, of spectroscopically confirming a quasar of a given redshift $z$, 
absolute magnitude $M$ and Spectral Energy Distribution (SED) shape. 
In this section, we describe the sources of incompleteness in the sample, 
and our checks of our completeness corrections. Our focus in this section
will be on the DR9 data; discussion of the completeness for the S82
sample can be found in \citet{Palanque-Delabrouille11} and
\citet{Ross12}.

    \subsection{Incompleteness Descriptions}
    We follow the approaches of \citet{Croom04} and \citet{Croom09a} to
    quantify four avenues of potential sample incompleteness:
    Morphological, Targeting, Coverage and Spectroscopic, and give
    descriptions of each type.
    
        \subsubsection{Morphological completeness} 
        The input catalog to the BOSS quasar targeting algorithm is
        restricted to objects with stellar morphologies in the single-epoch
        SDSS imaging. Host galaxies of $z>2$ quasars are highly unlikely to be
        detected in this imaging, thus viable quasar targets should be
        unresolved. However, any true quasars not targeted because they are
        erroneously classified as resolved in the photometry will contribute
        to the survey incompleteness; this is referred to as morphological
        completeness ($f_{\rm m}$).
        
        We checked the assumption that $z>2$ host galaxies are
        undetected in SDSS imaging, and tested the reliability of the
        star/galaxy classifier from the SDSS photometric pipeline {\tt photo}
        \citep{Lupton01, Scranton02}, to the BOSS target selection magnitude
        limit. To do this, we compare to the {\it Hubble Space Telescope}
        (HST) observations of the COSMOS field \citep{Scoville07HST}, which
        has been observed by the SDSS imaging camera. Objects classified as
        extended by SDSS that are actually unresolved in the COSMOS imaging,
        could be true high-$z$ quasars that we fail to target in BOSS. We
        found that at $r\leq21.0$, $\lesssim$3\% of objects classified
        morphologically as galaxies by SDSS are unresolved in COSMOS; this
        fraction rises to $\approx$8\% at $r=22.0$. We also found that all
        BOSS quasars at $z>2$ lying within the COSMOS field are unresolved by
        {\it HST}. Thus we conclude that host galaxy contribution to
        morphological incompleteness is minimal, and we do not account for the
        misclassification rate of stellar objects by {\tt photo} in our QLF
        calculations.

        \subsubsection{Targeting completeness}
        Targeting completeness, $f_{t}$, accounts for any true quasars
        which are not targeted by our selection algorithm. We use the XDQSO
        method to select our high-$z$ quasar targets, but, as we demonstrate
        below, the completeness of XDQSO is a strong function of color,
        redshift and magnitude.  Also, as we have noted, the XDQSO
        method was not used to select a uniform sample until the end of Year
        One.
        
        The area targeted with Year One target selection was 1661
        deg$^{2}$, although the 220 deg$^{2}$ of Stripe 82 was re-targeted and
        re-observed in Year Two.  \citet{Ross12} found that apart from over
        the Stripe 82 area, the fraction of objects selected by the XDQSO CORE
        algorithm {\it which actually were targeted}, was 87\% for the DR9
        footprint. This result is consistent with the numbers of XDQSO
        targets (\hbox{74,607}) that have spectra (\hbox{63,074}), as given in
        Table~\ref{tab:DR9_key_numbers}.
        
        In Stripe 82, this fraction declines to 65.4\%. This drop in
        targeting completeness is due to the deeper Stripe 82 photometry which
        eliminates many noisy stellar contaminants in the single-epoch XDQSO
        target list, while selecting nearly all of the true quasars selected
        by CORE.  The high targeting completeness fraction of XDQSO in the
        remainder of Year One is because many of the CORE quasar targets (and
        consequently true quasars) were selected by other target selection
        methods. There are in some sense the ``easiest'' quasars to
        discover. Indeed, \citet{Bovy11} demonstrate that XDQSO and the
        Likelihood method \citep{Kirkpatrick11}, used as CORE for Year One,
        select similar samples.

        \begin{figure}
          \begin{center}
            \includegraphics[height=3.9cm, width=5.9cm]{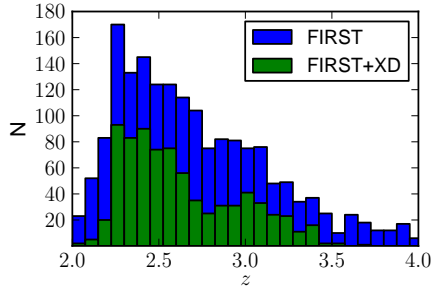}
            \caption{The redshift distribution for radio-selected objects (``FIRST'', blue
              histogram) and those that passed both the radio and XDQSO selection
              (``FIRST''+XD, green histogram).}
            \label{fig:radio_zhist_xdcompare}
          \end{center}
        \end{figure}

       \subsubsection{Coverage completeness}
        Coverage completeness, $f_{\rm c}(\theta)$, is defined as the
        fraction of BOSS quasar targets that have spectroscopic observations,
        is quantified on a sector-by-sector basis, and is thus a function of
        angular position, $\theta$. The main source of coverage incompleteness
        is fiber collisions, i.e. fibers cannot be placed closer than 62$''$
        to each other on a single plate. In Year One, the coverage
        completeness was $\gtrsim90\%$, and in Year Two CORE quasars were
        given highest tiling priority, and $f_{\rm c}(\theta)$ is $>98$\%.
        
        \subsubsection{Spectroscopic completeness}\label{sec:spec_completeness}
        Spectroscopic completeness, $f_{s}( {\rm mag}, z, \theta)$, is
        the fraction of BOSS quasar targets with spectra, from CORE, that have
        reliable redshifts. With the visual inspections of all the quasar
        target spectra, this fraction is $>90\%$. We define a
        ``spectro-coverage completeness'' as $f_{\rm sc} = f_{\rm c} \times
        f_{s}$, which is the fraction of XDQSO targets that were allocated
        fibers, and returned a reliable spectrum. Tests showed that the
        computation of the QLF is only very weakly sensitive to the value of
        $f_{\rm sc}$, and we choose a threshold of $f_{\rm sc} =0.85$ as a
        good compromise between high completeness and large sample
        size. Sectors with $f_{\rm sc}\geq85$\% are shown in red shades in
        Fig.~\ref{fig:DR9_coverage}; we limit our LF analysis to this
        area. This approach tends to exclude regions that have Year One
        spectroscopy, leaving an area of 2236 deg$^{2}$. {\bf There are
          \hbox{23 301} quasars in this area (all with visually confirmed
          redshifts) and this sample is given in
          Table~\ref{tab:uniform_DR9_quasars}.}
        
          \begin{table}
             \begin{center}
               \begin{tabular}{lrr}
                 \hline
                 \hline
                 Description & \multicolumn{2}{c}{No. of Objects}  \\
                 & Pipeline                   &  V.I.                    \\
                 \hline 
                 All Stripe 82 quasar targets$^{a}$                    & \hbox{ 15 576}  & --- \\ 
                 \,\,\,\,  with {\tt specPrimary = 1}                 & \hbox{ 12 576} & --- \\ 
                 \,\,\,\,\,\,\,\, AND reliable redshifts             & \hbox{ 10 506} & \hbox{ 11 990}\\
                 \,\,\,\,\,\,\,\,\,\,\,\,   AND $2.20<z<3.50$  & \hbox{  5 433} & \hbox{   5 476} \\
                 \,\,\,\,\,\,\,\,\,\,\,\,\,\,\,\,   AND w/ XDQSO seln.  & \hbox{ 2 318} & \hbox{   2 333} \\
                 \hline
                 \hline
               \end{tabular}
               \caption{Properties of the Stripe 82 BOSS QLF dataset,
                 described in \citet{Palanque-Delabrouille11} and with data from two of
                 the Ancillary Programs that targeted quasars due to their
                 near-infrared colors or radio properties.}
               \label{tab:S82_key_numbers}
               \end{center}
             \end{table}
             
   \subsection{Empirical checks using Variability selected Quasars}\label{sec:NNVar_vs_XDQSO}
    The 220 deg$^{2}$ of spectroscopy across the Stripe 82 field has
    targets selected via their optical variability \citep[][and
    \S~\ref{sec:variability}]{Palanque-Delabrouille11}. We concentrate on
    the \hbox{5 476} $2.20<z<3.50$ quasars in Stripe 82
    (Table~\ref{tab:S82_key_numbers}), including 122 quasars selected
    solely due to their near-infrared colors or radio properties
    \citep{Dawson12}. In this area, we find a higher surface density of
    high-$z$ quasars, 24.9 deg$^{-2}$, than across the full DR9 dataset
    (14.7 deg$^{-2}$) and the XDQSO uniform sample (9.6 deg$^{-2}$). Thus,
    this enhanced Stripe 82 dataset is more complete and less affected by
    color-induced selection biases, and we will use it to measure the targeting 
    completeness of our XDQSO uniform sample empirically. 
    
    We split the sample of \hbox{5 476} visually confirmed $2.2<z<3.5$
    quasars into the \hbox{2 333} that would have been selected by the
    XDQSO algorithm ({\tt XD}) and those (\hbox{3 143}) that would not
    have been ({\tt !XD}). Over 96\% of the {\tt !XD} sample was selected
    by a variability algorithm. The $(u-g)$ vs. $(g-r)$ color-color plane,
    their distribution in $(g-i)$ vs. $i$-band and the resulting $N(z)$
    redshift histograms of these two samples are given in
    Fig.~\ref{fig:Stripe82_XDnotXD}.

    The difference between the two selections is apparent and
    consistent with that in \citet[][their
    Fig. 18]{Palanque-Delabrouille11}.  The {\tt !XD} sample more heavily
    overlaps with the stellar locus in $(u-g)$ vs. $(g-r)$, and is
    generally redder than the {\tt XD} population in $(g-r)$. Thus, the
    variability selection is able to recover quasars from the stellar
    locus. The distribution of {\tt XD} and {\tt !XD} objects in the
    $(g-r)$ vs. $(r-i)$ color-color plane (not shown) is similar, in that
    the {\tt !XD} population overlaps with the stellar locus, but both
    populations have similar distributions in the $(r-i)$ color
    \citep[again in agreement with Fig.18 of
    ][]{Palanque-Delabrouille11}. However, the {\tt !XD} population is
    redder in $(g-i)$ (Fig.~\ref{fig:Stripe82_XDnotXD}, {\it center}).
   
    The $N(z)$ histograms for the two samples are also very much in
    line with previous studies \citep[e.g., ][]{Richards06,
      Palanque-Delabrouille11}. The decrement at $z\sim2.7-2.9$ for the
    XD selection is due to the overlap of such objects with the 
    stellar locus in color-space, a key issue in the original studies in SDSS. 
 
    The two samples are similar in their distributions across
    $z=3.0-3.3$, though at $z\approx3.4-3.5$, there are more {\tt !XD}
    objects, probably due to the efficient cut-off of the $z=2.2-3.5$
    ``mid-$z$'' XDQSO selection, and the fact that at $z\sim3.5$ quasar
    colors again approach the stellar locus.

    \begin{figure*}[!th]
      \begin{center}
        \includegraphics[scale=1.00]{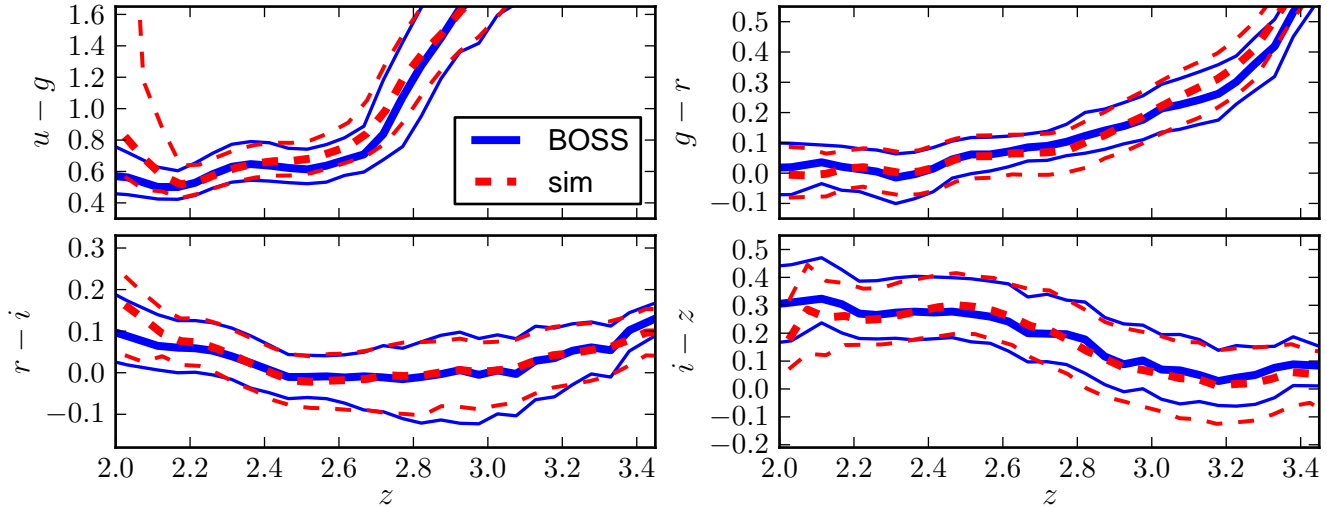}
        \caption{Color-redshift relation for BOSS quasars selected by XDQSO
          (solid, blue) and for simulated quasars (dashed, red). The thick lines
          show the median values, and the thin lines show the 25\%-75\% range,
          at each redshift. The agreement between the data and models is
          generally very impressive. At $z<2.2$ the colors are less well
          matched, but are very noisy due to the very low completeness of XDQSO
          at those redshifts.}
       \label{fig:colorz_simcompare}
      \end{center}
    \end{figure*}
      
   \subsection{Radio Selection versus Color Selection} 
    Figure~\ref{fig:radio_zhist_xdcompare} displays the redshift
    distribution for radio-selected objects (``FIRST'', blue histogram)
    and those that passed both the radio and XDQSO selection
    (``FIRST''+XD, green histogram). This graph can be compared directly
    to Figure 10 in R06, which compares the redshift distribution of
    radio-selected quasars to those that were both radio- and
    color-selected using the full DR3Q \citep{Schneider05} sample. The
    redshift distribution of the radio-only selected objects is smoother
    and has a smaller decrement of objects at $z=2.7-2.8$ than the
    radio$+$color selection. This was also seen in the R06 DR3Q
    investigation.
    
    However, we are wary of over-interpreting this for several
    reasons. First, only $\sim2$\% (3 348) of the DR9Q \citep{Paris12},
    and $\sim2$\% (747) of the XDQSO quasars are targeted via their radio
    properties, of which half are selected {\it only} via their radio
    properties. Second, BOSS is deeper than SDSS, whereas the FIRST
    detection limits are the same for the two optical surveys, so BOSS
    radio sources are more radio loud. If radio loudness correlates
    with redshift and/or luminosity \citep[e.g.,][]{Jiang07, Singal11,
      Singal12}, an attempt to correct the $N(z)$ distribution using
    radio-loud quasars would be incorrect (see also the discussion in
    \S~3.4 of R06). Finally, radio-loud quasars are not drawn from the
    same color distribution as radio-quiet quasars, with the radio-loud
    population tending to have redder colors \citep{RWhite03, McGreer09,
      Kimball11}. Thus objects in a radio-selected quasar sample do not
    share the same selection function as a purely color-selected sample.
    
   \subsection{Simulated Quasar Spectra and Completeness}\label{sec:simqso} 
    In Section~\ref{sec:NNVar_vs_XDQSO} we presented the sample
    established by the XDQSO targeting algorithm, and compared it to a
    dataset constructed from the Stripe 82 sample of quasars selected
    independently of that algorithm. We can use the results of
    \S~\ref{sec:NNVar_vs_XDQSO} to quantify the completeness of XDQSO only
    in the limit that the Stripe 82 sample is itself complete. Here we
    adopt another approach: we construct a model for the observed
    spectroscopic and photometric properties of quasars, generate a large
    sample of simulated quasars, and then test the targeting algorithm
    against this model using the simulated quasars
    \citep[e.g.,][]{Fan99,Richards06}.
    
    The broadband optical fluxes used by XDQSO are dominated by a
    featureless power-law continuum.  However, quasar selection is highly
    sensitive to the {\em colors} of quasars, which evolve strongly with
    redshift as the broad, high-equivalent-width emission lines move
    through the optical bandpasses \citep{Richards01_colors}. Hence, a
    complete prescription for quasar properties must capture both the
    smooth continuum and the emission lines.
    
    Past models have generally adopted a continuum power law index of
    $\alpha_{\nu} = -0.5$, where $\alpha_{\nu}$ is the frequency power law
    index (i.e. $F(\nu) \propto \nu^{\alpha_{\nu}}$),
    typically measured from quasar spectra \citep[e.g., ][]{Richstone80,
      Francis91, VdB01}. Often a break is added to the near UV where a softer
    spectrum is observed \citep[$\alpha_{\rm UV} \sim -1.7$, ][]{Telfer02,
      Shang05, Shang11}.  Emission line templates, including \feii
    complexes, are then generated from composite mean quasar spectra
    constructed from large samples \citep[e.g., ][]{Francis91, VdB01}. To
    these emission components, absorption from the Ly$\alpha$ forest is
    added, given a model for its redshift dependence. With these basic
    assumptions, models can be generated that broadly reproduce the mean
    colors of quasars as a function of redshift \citep{Richards01_colors}.

    For this work we have taken advantage of the many improvements in
    our understanding of quasar spectral properties in recent years,
    namely, improved measurements of absorption due to the Ly$\alpha$
    forest \citep[e.g.,][]{Worseck11}, templates for iron emission
    \citep[a significant contributor to quasar colors;][]{Vestergaard01}
    and finally, large samples of quasars to calibrate the models. We have
    simulated the full survey, by passing the model quasars
    through the target selection algorithm and comparing the resulting
    color distribution to observations. Under the common assumption that
    quasar spectral features do not evolve with redshift, the selection
    function provides a redshift-dependent window into the underlying
    color distribution. By comparing the colors of quasars that the
    model predicts are selected by the survey, to those actually observed,
    we can determine a best-fit model that not only recovers the selection
    function, but also provides insight into the intrinsic properties of
    quasars.

 \begin{figure}[!h]
      \begin{center}
        \includegraphics[scale=0.9]{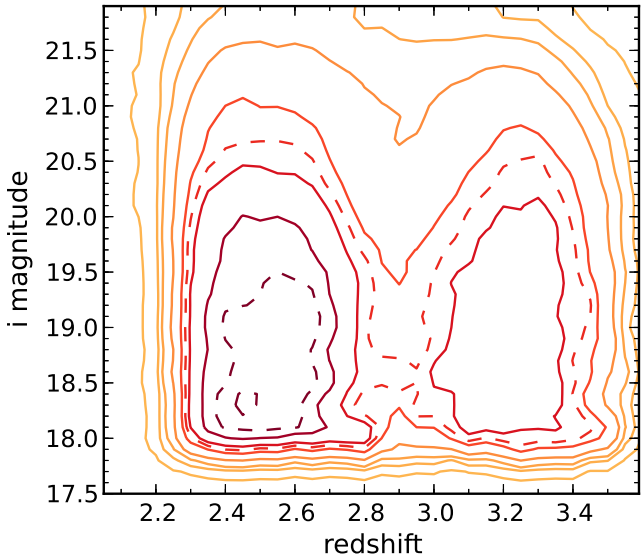}
        \caption{The selection function for the BOSS XDQSO sample via
          simulated quasar spectra and photometry. Contour levels are drawn at
          1, 5, 10, 20, 40, 50, 60, 80, and 90 percent completeness, as
          determined by the fraction of simulated quasars selected by XDQSO as a
          function of redshift and $i$-band magnitude. The 50\% and 90\% levels
          are drawn with dashed lines.}
        \label{fig:grid_newcont_i_z_PQSOMIDZmap}
      \end{center}
    \end{figure}

    \begin{figure*}[!ht]
      \begin{center}
       \includegraphics[scale=1.00]{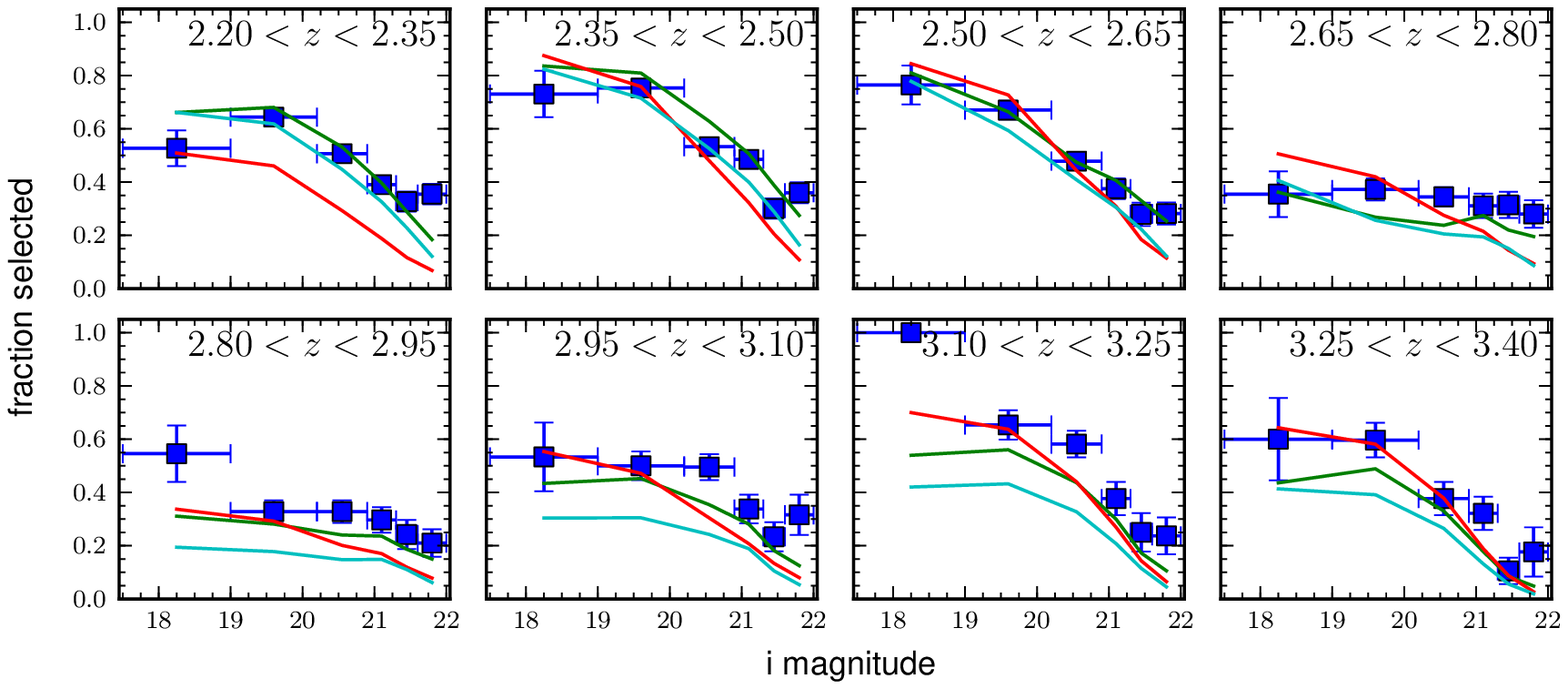}
        \caption{BOSS (XDQSO) quasar selection function, in discrete redshift
          bins covering the range $2.2 < z < 3.4$. The blue points with error
          bars show the empirical selection function derived from Stripe 82;
          specifically, they denote the fraction of Stripe 82 quasars selected
          by XDQSO within bins of magnitude and redshift (the $x$-error bar
          represents the $i$-magnitude bin width).  The green line shows our
          ``fiducial'' model selection function using simulated quasars as described in 
          \S~\ref{sec:simqso} , and the same binning as the
          empirical points.  The agreement at $z\la3$ shows that the two are
          consistent, as expected if both the model is a good representation of
          true quasars and the variability selection is highly complete.  The
          model predicts lower efficiency at $z>3$, suggesting that the
          completeness of the variability selection is lower at higher redshifts
          i.e., XDQSO recovers a higher fraction of the variability quasars than
          the model quasars, and variability is potentially missing a
          population of quasars. Note also that the efficiency
          predicted by the model is generally lower in the faintest magnitude
          bin ($i\sim22$), again suggesting that variability was less complete
          at the faint end.  Note that the model has a luminosity function prior
          \citep{HRH07} applied.  For comparison, we also plot the ``VdB lines''
          and ``exp dust'' model selection functions (red and cyan lines,
          respectively), both of which generally show poorer agreement with the
          empirical points from Stripe 82 than the fiducial model.}
        \label{fig:compare_stripe82_to_sims_selfn}
      \end{center}
    \end{figure*}

    This process will be detailed in a forthcoming work (McGreer et al., 
    in prep). Here we briefly outline the steps taken to generate a model for 
    the population of quasars observed (and not observed) by BOSS. 
    
    \begin{enumerate}
    \item{We construct a grid of model quasars in ($M$,$z$) space,
        using the luminosity function from \cite{HRH07}. For each quasar we
        randomly sample the following components:}
      
      \begin{enumerate}
      \item{A broken power-law continuum with a break at 1100\AA; at near-UV
          wavelengths the power law index is drawn from a Gaussian distribution
          with mean $\alpha_\nu = -1.7$ and scatter $\sigma(\alpha_\nu) = 0.3$;
          for $\lambda>1100$~\AA\ the distribution has a mean $\alpha_\nu = -0.5$
          and scatter $\sigma(\alpha_\nu) = 0.25$.}
      \item{Emission lines generated from composite spectra of BOSS
          quasars {\em binned in luminosity}, reproducing trends between
          emission line properties and continuum luminosity \cite[e.g., the
          Baldwin Effect:][]{Baldwin77, Wu09}.} The resulting emission line
          template provides the mean and scatter in line strength for prominent
          quasar emission lines as a function of luminosity; values for 
          individual objects are drawn from this distribution.
      \item{Fe emission from the template of \citet{Vestergaard01}. The
          template is divided into discrete wavelength segments 
          \citep[see ][]{Vestergaard01} that are scaled independently; the 
          scale values are determined during the fitting of the composite 
          spectra used for the emission line template.}
      \item{Ly$\alpha$ forest blanketing according to the prescription
          of \citet{Worseck11}. A population of absorbers is generated in a
          Monte Carlo fashion using the parameters given in
          \citet{Worseck11}. The lines are modeled as Voigt profiles using the
          approximation of \citet{TepperGarcia06}, and then applied to the
          forest regions of the simulated spectra. All Lyman series transitions
          up to $n=32$ are included.  A total of 5000 independent sightlines
          were generated and then randomly drawn for each of the simulated
          quasars.}
      \end{enumerate}
      
    \item{We generate spectra from this grid and calculate SDSS broadband fluxes
        from the spectra.}
    \item{The fluxes are transferred to observed values via empirical relations for
        the photometric uncertainties derived from single-epoch observations of
        stars on Stripe 82, using the coadded fluxes \citep{Annis11} as 
        the reference system.}
    \item{The XDQSO algorithm is used to calculate mid-$z$ quasar 
        probabilities for each model quasar in the same manner as for 
        BOSS selection, and a sample of model quasars is defined.}
    \end{enumerate}

    This describes our fiducial model. We further test two 
    modifications to the fiducial model. For comparison to previous
    work, we implement a second model where the emission line template is
    derived from a single composite spectrum and thus does not have any
    dependence on luminosity. This template comes from the SDSS composite
    spectrum presented in \citet{VdB01} and is referred to as ``VdB
    lines''. This model is closest to that of \citet{Richards06}. Finally,
    we test a third model that includes dust extinction from the host galaxy. 
	In this model, individual spectra are extincted using a SMC dust model
    \citep{Prevot84}, with values of $E(B-V)$ distributed exponentially
    around a peak of $0.03$ \citep[e.g.,][]{Hopkins04}. This model is
    referred to as ``exp dust''. We compare the three models in more detail
    in Appendix~\ref{appndx:spec_models}.

    We test the accuracy of the fiducial model by checking the
    simulated quasar colors against observed quasar colors. This prior is
    used to distribute the simulated quasars in flux and redshift space in
    a manner similar to the intrinsic distribution.  The simulated quasar
    photometry is passed through the XDQSO selection algorithm to mimic
    the observations, so that the final color relations for simulated
    quasars are derived only for objects that would have been targeted by
    the survey. We then construct the color-redshift relation
    \cite[e.g.,][]{Richards01_colors} of both simulated and observed
    quasars by dividing the samples into narrow redshift bins ($\Delta z =
    0.05$) and calculating both the median and scatter of the $u-g$, $g-r$,
    $r-i$, and $i-z$ colors within each redshift bin. The results for the
    fiducial model are shown in Figure~\ref{fig:colorz_simcompare}, 
    demonstrating that the model does an excellent job of reproducing the 
    observed quasar color distribution.

    Dust extinction is thought to produce the red tail of the color
    distribution often seen in quasar surveys. For example,
    \citet{Richards03}~and~\citet{Hopkins04} find that $\sim20$\% of SDSS
    quasars have colors consistent with reddening from dust with an
    SMC-like extinction curve with $E(B-V)\ga0.1$. We find that the
    exp dust model does not significantly improve the fit to the color 
    distribution of BOSS quasars compared to our fiducial model, and thus our 
    primary analysis does not include the effect of dust extinction. 
    Section~\ref{sec:lum_func} will explore how the differing assumptions of 
    the three models affect the calculation of the QLF.
    
    Table~\ref{tab:selec_func} and
    Fig.~\ref{fig:grid_newcont_i_z_PQSOMIDZmap} give the selection
    function, i.e. the fraction of selected quasars, in each bin of $M$
    and $z$, generated from the fiducial model outlined above.

    \begin{table}
      \begin{center}
        \begin{tabular}{lcccc}
          \hline
          \hline
          $i$\_start &    $i$\_end &  $z$\_start & $z$\_end  &  Selec. Func. \\
          \hline
          17.500  & 17.600  &  2.000 &   2.050 &  0.0000 \\
          17.500  & 17.600  &  2.050 &   2.100 &  0.0000 \\
          $\vdots$  & $\vdots$  &  $\vdots$ &  $\vdots$ &  $\vdots$ \\
          17.800 & 17.900  &  2.100  &  2.150  &  0.0000 \\
          17.800 & 17.900  &  2.150  &  2.200  &  0.0291 \\
          17.800 & 17.900  &  2.200  &  2.250  &  0.1710 \\
          17.800 & 17.900  &  2.250  &  2.300  &  0.4076 \\
          17.800 & 17.900  &  2.300  &  2.350  &  0.3365 \\ 
          17.800 & 17.900  &  2.350  &  2.400  &  0.5029 \\
          \hline
          \hline
        \end{tabular}
        \caption{The Quasar Selection Function for the fiducial model
          described in the text; see also
          Fig.~\ref{fig:grid_newcont_i_z_PQSOMIDZmap}. The final column gives
          the fraction of simulated quasars selected by
          XDQSO. Table~\ref{tab:selec_func} is published in its entirety in the
          electronic edition of {\it The Astrophysical Journal}; this excerpt
          here is shown here for guidance regarding its format and content.}
        \label{tab:selec_func}
      \end{center}
    \end{table}
    
    We compare the model selection function to an empirical relation
    from Stripe 82 in~Figure~\ref{fig:compare_stripe82_to_sims_selfn}. The
    green lines show the fraction of model quasars that are selected by
    XDQSO-CORE. This is compared to the fraction of Stripe 82 quasars ---
    predominantly selected by variability criteria --- that are recovered
    by XDQSO selection. This empirical relation is shown as blue squares
    with error bars (Poisson uncertainties). The agreement at $z\la3$
    shows that the two are consistent, as expected if both the model is a
    good representation of actual quasars, and the variability selection
    is highly complete.

    However, there is some disagreement between the two completeness
    estimates. For example, smaller fractions of model quasars are
    selected by XDQSO at $z>3$ than are selected by XDQSO from the Stripe
    82 sample in the same redshift and magnitude bins. This may indicate a
    deficiency of the models; however, we are encouraged by the excellent
    agreement between the colors predicted by the model and those observed
    (Figure~\ref{fig:colorz_simcompare}). Alternatively, our assumption
    that the variability-selected sample is both complete and unbiased may
    be invalid. Indeed, color criteria were applied to objects when
    constructing the variability sample \citep{Palanque-Delabrouille11};
    these criteria may exclude some populations of quasars, in particular, they
    may introduce bias against high redshift quasars. In that case, the
    disagreement in Figure~\ref{fig:compare_stripe82_to_sims_selfn}
    suggests that XDQSO recovers a higher fraction of the
    variability-selected quasars, but both XDQSO and variability are
    missing a population of objects. This effect may also explain why the
    model predicts lower completeness at the faintest magnitudes: i.e.,
    both XDQSO and variability have low completeness at $i\sim21.8$, but
    XDQSO recovers a higher fraction of the quasars that are also selected
    by variability.

    In what follows, we implement our fiducial selection function
    model to calculate the QLF from the DR9 uniform quasar sample. 
    Since the color selection incompleteness dominates over the other
    sources of incompleteness, we do not make any further corrections 
    during the QLF calculation.

%%%%%%%%%%%%%%%%%%%%%%%%%%%%%%%%%%%%%%%%%%%%%%%%%%%%%%%%%%%%%%
%%%%%%%%%%%%%%%%%%%%%%%%%%%%%%%%%%%%%%%%%%%%%%%%%%%%%%%%%%%%%%
%%
%%   SECTION 4     SECTION 4     SECTION 4     SECTION 4     SECTION 4     SECTION 4     
%%   SECTION 4     SECTION 4     SECTION 4     SECTION 4     SECTION 4     SECTION 4     
%%   SECTION 4     SECTION 4     SECTION 4     SECTION 4     SECTION 4     SECTION 4     
%%
%%%%%%%%%%%%%%%%%%%%%%%%%%%%%%%%%%%%%%%%%%%%%%%%%%%%%%%%%%%%%%
%%%%%%%%%%%%%%%%%%%%%%%%%%%%%%%%%%%%%%%%%%%%%%%%%%%%%%%%%%%%%%
\begin{figure}
  \begin{center}
    \includegraphics[width=8.0cm]
    {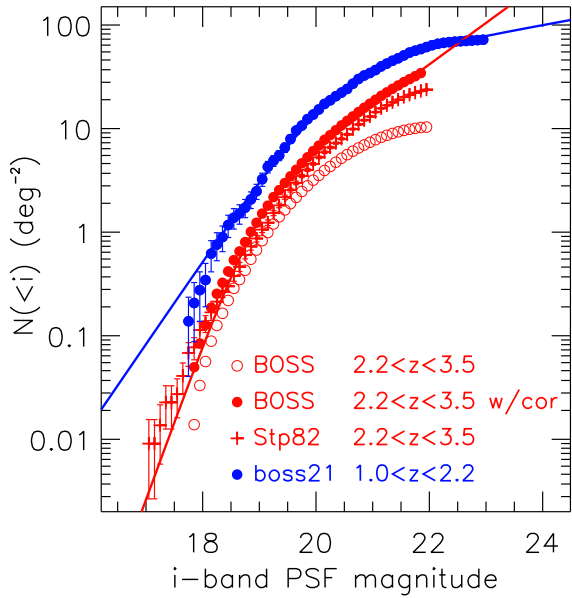}
    \caption{Cumulative $i$-band number counts. Here, the $2.2<z<3.5$ BOSS
      samples are in red, with the uncorrected BOSS uniform sample shown by
      the open circles, while the Stripe 82 data are given by the crosses.
      Also shown are the number counts from the deeper, $g\approx23$
      $1.0<z<2.2$ quasars selected from the ``boss21+MMT'' survey
      \citep{Palanque-Delabrouille12}. For clarity, we only plot errorbars
      at the bright end ($i<19$), since the errors are smaller than the
      points at the faint end. We also show the double power law fits to
      the data as described in the text.}
\label{fig:cuml_number_counts}
\end{center}
\end{figure}

\begin{figure}
  \begin{center}
    \includegraphics[width=8.0cm]
    {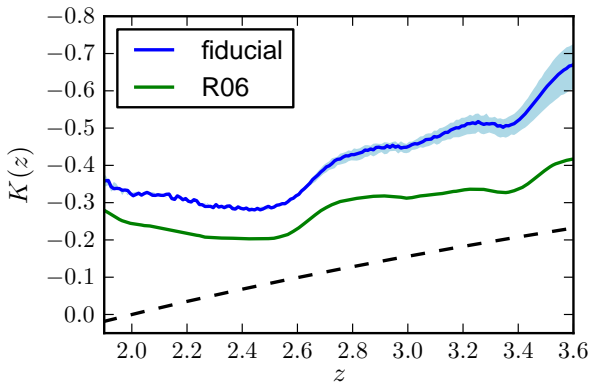}
    \caption{Comparison of the $k$-correction from our fiducial
    model (blue line) with the R06 $k$-correction (green line).
    Both are defined as the correction required to transfer the
    observed $i$-band flux to the $i$-band luminosity at $z=2$, 
    i.e., $M_i(z=2)$.
    The offset of $\sim$0.1 mag at $z<2.7$ is due to a different
    treatment of Fe emission, and grows somewhat larger at higher
    redshifts as the \ciii line enters the $i$-band.
    The blue line shows our quasar model for $M_i(z=2)=-26$, and the
    shaded regions show the variation of the $k$-correction with
    luminosity over the range $-27 < M_i(z=2) < -24.3$, covering
    most of the luminosity range of BOSS quasars.
    Finally, the dashed line shows a pure continuum $k$-correction
    for $\alpha_\nu=-0.5$.}
    \label{fig:kcorr_compare}
  \end{center}
\end{figure}

\section{Number Counts and $k$-corrections}\label{sec:no_counts}

    \subsection{Number Counts}
    In Fig.~\ref{fig:cuml_number_counts} we present the cumulative
    $i$-band number counts of the datasets described in Sec. 2: the XDQSO
    uniform sample of \hbox{23 301} quasars across 2236 deg$^{2}$ with
    $2.2<z<3.5$ (red circles) and the \hbox{5 470} quasars across 220
    deg$^{2}$ of Stripe 82 also with $2.2<z<3.5$ (red crosses).  Also
    shown are the number counts from $1.0<z<2.2$ quasars selected from
    deeper, $g\approx23$ spectroscopy, using data from the ``boss21+MMT''
    survey \citep[blue, filled circles;][]{Palanque-Delabrouille12}.

    For the uniform BOSS sample, the open red circles
    are for the raw, uncorrected number counts, whereas the filled circles
    use the correction derived in the previous section, integrated over
    our redshift range. These number counts can be compared to the
    $2.2<z<3.5$ quasars selected via their variability signature on Stripe
    82. The two are in reasonable agreement to $i\approx21.0$, with the
    corrected number counts being consistently higher.  Fainter than this,
    the variability number counts drop more noticeably below the corrected
    counts, suggesting that this dataset is incomplete at the faint end 
    (or that the incompleteness of the DR9 sample is overestimated). 
    Across the redshift range $2.2<z<3.5$ and down to $i=21.5$, the
    corrected BOSS DR9 uniform cumulative number counts reach 34.4
    deg$^{-2}$, whereas the Stripe 82 cumulative counts are 26.2
    deg$^{-2}$.

    Motivated by the double power-law form of the QLF
    (Eqn.~\ref{eq:double_powerlaw_mag}), and prior measurements
    \citep[e.g.][]{Myers03}, we also express the cumulative number counts
    as a double power-law,
    \begin{equation}
      \int dN = \frac{N_{0}}{10^{-\alpha_{d}(m-m_{0})} +  10^{-\beta_{d}(m-m_{0})} }
    \end{equation}
    and find best-fits to the (corrected) BOSS uniform and boss21+MMT
    counts.  For the BOSS sample we find slopes of $\alpha_{d} = 1.50$ and
    $\beta_{d} = 0.40$, while the ``break magnitude'' $m_{0}=$19.0 and
    the normalization, $N_{0} = 2.63$ deg$^{-2}$.  In comparison, the boss21+MMT 
    data has a less-steep bright end slope of $\alpha_{d} = 0.80$ and an almost
    flat faint end slope $\beta_{d} = 0.10$.  The break magnitude is
    fainter at $m_{0}=20.4$ and the normalization is significantly higher,
    $N_{0} = 43.6$ deg$^{-2}$. These power-law descriptions and surface densities
    will allow extrapolation for future Ly$\alpha$-forest cosmology
    experiments \citep[e.g., ][]{McQuinn11}.  For unobscured $1.0<z<2.2$
    quasars, there are 48 (78) objects deg$^{-2}$ down to $i\lesssim21.5$ (23.0),
    broadly consistent with the value of 99$\pm$4 quasars deg$^{-2}$ with
    $g_{\rm dered}<22.5$ from \citet[][their Table
    5]{Palanque-Delabrouille12} and a surface density similar to that
    selected by a shallow mid-infrared selection \citep{Stern12, Yan12}.

    \begin{table}
      \begin{center}
        \begin{tabular}{lc}
          \hline
          \hline
          $z_{\rm em}$ & $k$-correction \\ 
          \hline   
          0.105.............. &   0.323 \\
          0.115.............. &   0.250 \\
          0.125.............. &   0.317 \\
          0.135.............. &   0.332 \\
          0.145.............. &   0.335 \\
          0.155.............. &   0.334 \\
          0.165.............. &   0.285 \\
          0.175.............. &   0.291 \\
          0.185.............. &   0.334 \\
          0.195.............. &   0.249 \\
          \hline
          \hline
        \end{tabular}
        \caption{
          The $i$-band $k$-corrections. The $k$-correction is obtained
          using the fiducial quasar model described in Section~\ref{sec:simqso},
          and includes an updated treatment of the emission line template
          compared to \citet{Richards06}.  We define our $k$-correction to be
          our model $k$-correction at $M_i(z=2) = -26.0$ (see main text for
          details). The full table is published in the electronic edition of
          {\it The Astrophysical Journal}; the first ten lines are shown here
          for guidance regarding its format and content.       }
        \label{tab:kcorr}
      \end{center}
    \end{table}
    \subsection{$k$-corrections}\label{sec:k_corr}
    Following R06, we define the $k$-correction to determine the
    $i$-band luminosity at $z=2$; $M_i(z=2)$. This is $\sim$2700~\AA\ in
    the rest-frame, and is close to the median redshift of the BOSS
    quasars sample. R06 calculate the quasar $k$-correction as the sum of
    a component due to the underlying continuum, $k_{\rm cont}$, and a
    component due to the emission lines, $k_{\rm em}$. The sign convention
    of the $k$-correction, $k(z)$, is $m_{\rm intrinsic} = m_{\rm
      observed} - k(z)$ \citep{Oke_Sandage68, Hogg02}. 
    
    We obtain a $k$-correction from the fiducial quasar model defined
    in Section~\ref{sec:simqso}. This model is very similar to the one
    adopted by R06. The continuum model is identical: a power-law slope of
    $\alpha_{\nu}=-0.5$ at $\lambda>1100$\AA, and we set $k_{\rm
      cont}(z=2.0)=0.0$ by definition. On the other hand, our emission line
    template is not the same as the one used by R06. They defined their
    emission line template using a single composite spectrum derived from
    SDSS quasars \citep[similar to that of ][]{VdB01} with a power-law
    continuum removed. Our emission line template is similarly obtained
    from composite spectra; however, we use a suite of composite spectra
    binned in luminosity, and fit the continuum jointly with the Fe
    template of \citet{Vestergaard01}.
    
    One advantage of defining the $k$-correction to be in the $i$-band
    at $z=2$ is that the $i$-band is relatively free of strong emission
    lines at this redshift. At $z\la2.5$, the $i$-band samples rest-frame
    $\sim2200$\AA, where the only strong emission line features are from
    \ion{Fe}{2} and \ion{Fe}{3}. We find that our model for Fe emission
    introduces a shift of about $\sim$0.1 mag at $2<z<2.7$ relative to the
    R06 model; i.e., $k_{\rm fid} - k_{\rm R06} \approx -0.1$. At higher
    redshifts, the \ciii line enters the $i$-band and the offset between
    our $k$-correction and that of R06 grows somewhat larger, reaching
    $\approx -0.2$ mag at $z=3.5$. We compare our $k$-correction to that
    of R06 in Figure~\ref{fig:kcorr_compare}.
 
\begin{figure}
  \begin{center}
    \includegraphics[height=8.5cm,width=8.5cm]
    {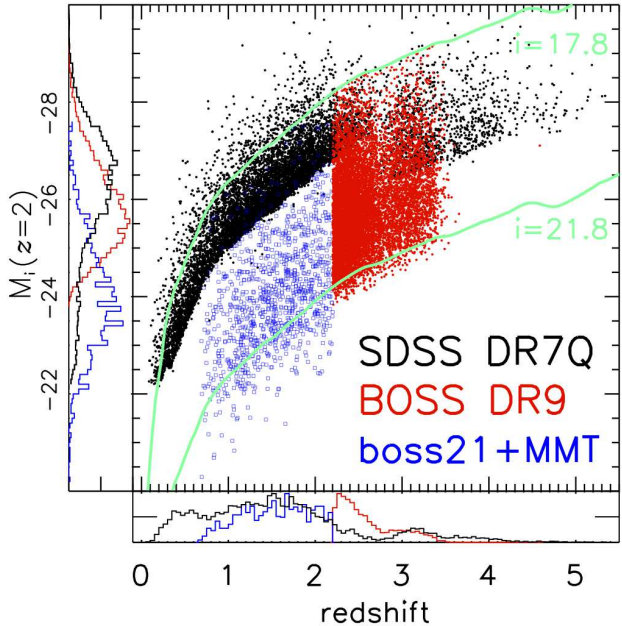}
    \caption{The Absolute Magnitude-Redshift distribution for SDSS DR7
      quasars (black points), BOSS DR9 (red points) and the fainter,
      boss21+MMT variability-selected dataset from Stripe 82 (blue squares).
      The normalized redshift distributions of the three datasets are shown in
      the bottom panel, while the normalized absolute {\it i}-band
      magnitudes are given in the side panel. The bright and faint magnitude
      limits of the BOSS DR9 sample are given by the solid turquoise lines; the
      wiggles are due to the redshift dependence of the $k$-correction.}
    \label{fig:L-z}
  \end{center}
\end{figure}
   
    Though our quasar model introduces a luminosity dependence to the
    $k$-correction due to the anticorrelation between emission line
    equivalent width and luminosity (the Baldwin Effect), we chose not to
    apply this further correction in this work. For the same reasons given
    above, at $z<2.7$ there is almost no variation in our $k$-correction
    with luminosity. At $z=3.5$, this effect only reaches $\sim5\%$ over
    the range of luminosities probed by BOSS (see
    Figure~\ref{fig:kcorr_compare}). This variation is much smaller than
    the intrinsic scatter in $k$-corrections at a given redshift; i.e.,
    even if we attempted to correct for the Baldwin Effect in the mean,
    the scatter in this correlation is far greater than the correction. We
    define our $k$-correction to be our model $k$-correction at $M_i(z=2)
    = -26.0$, near the median luminosity of the BOSS sample. Table~\ref{tab:kcorr}
    presents our new $k$-correction. 
   
\begin{table}
  \begin{center}
    \begin{tabular}{ccc ccc}
      \hline
      \hline
      $\bar{z}$ & $\langle M_{i}(z=2)\rangle$ & $M_{i}$ bin  & $N_{\rm Q}$ & $\log(\Phi)$   & $\sigma_{\Phi}/ 10^{-9}$  \\
         (1)         &            (2)                         &   (3)             &        (4)       & (5)                  &   (6)   \\
      \hline
      2.488 &   -28.297 &   -28.350 &      26 &   -7.892 &    1.994 \\
      2.386 &   -28.024 &   -28.050 &     105 &   -7.338 &    3.772 \\
      2.409 &   -27.737 &   -27.750 &     184 &   -7.125 &    4.821 \\
      2.399 &   -27.437 &   -27.450 &     306 &   -6.917 &    6.126 \\
      2.408 &   -27.143 &   -27.150 &     510 &   -6.699 &    7.874 \\
      2.397 &   -26.844 &   -26.850 &     825 &   -6.486 &   10.063 \\
      2.392 &   -26.544 &   -26.550 &    1037 &   -6.360 &   11.631 \\
      2.391 &   -26.246 &   -26.250 &    1382 &   -6.210 &   13.824 \\
      2.381 &   -25.948 &   -25.950 &    1604 &   -6.104 &   15.629 \\
      2.385 &   -25.653 &   -25.650 &    1778 &   -5.983 &   17.958 \\
      2.378 &   -25.351 &   -25.350 &    1878 &   -5.855 &   20.808 \\
      2.375 &   -25.051 &   -25.050 &    1768 &   -5.824 &   21.562 \\
      2.373 &   -24.758 &   -24.750 &    1484 &   -5.807 &   21.991 \\
      2.364 &   -24.457 &   -24.450 &    1059 &   -5.750 &   23.483 \\
      2.332 &   -24.187 &   -24.150 &     456 &   -6.143 &   14.936 \\
      2.296 &   -23.907 &   -23.850 &      58 &         --- &     --- \\
      2.216 &   -23.678 &   -23.550 &       2 &          --- &     --- \\
      \hline
      2.830 &   -28.566 &   -28.650 &       5 &   -8.127 &    1.527 \\
      2.761 &   -28.328 &   -28.350 &      46 &   -7.469 &    3.259 \\
      2.787 &   -28.032 &   -28.050 &      67 &   -7.304 &    3.939 \\
      2.796 &   -27.739 &   -27.750 &     120 &   -7.013 &    5.509 \\
      2.802 &   -27.440 &   -27.450 &     217 &   -6.750 &    7.452 \\
      2.782 &   -27.145 &   -27.150 &     291 &   -6.630 &    8.557 \\
      2.780 &   -26.845 &   -26.850 &     373 &   -6.483 &   10.131 \\
      2.777 &   -26.547 &   -26.550 &     484 &   -6.301 &   12.501 \\
      2.775 &   -26.239 &   -26.250 &     512 &   -6.210 &   13.876 \\
      2.775 &   -25.944 &   -25.950 &     536 &   -6.206 &   13.949 \\
      2.776 &   -25.651 &   -25.650 &     646 &   -6.109 &   15.595 \\
      2.773 &   -25.362 &   -25.350 &     669 &   -6.046 &   16.762 \\
      2.776 &   -25.056 &   -25.050 &     634 &   -5.975 &   18.188 \\
      2.764 &   -24.758 &   -24.750 &     382 &   -6.079 &   16.129 \\
      2.715 &   -24.487 &   -24.450 &     184 &   -6.552 &    9.362 \\
      2.681 &   -24.217 &   -24.150 &      25 &        --- &      --- \\
      \hline
      3.259 &   -28.864 &   -28.950 &       3 &   -8.588 &    0.815 \\
      3.283 &   -28.627 &   -28.650 &      25 &   -7.754 &    2.128 \\
      3.207 &   -28.327 &   -28.350 &      44 &   -7.543 &    2.711 \\
      3.219 &   -28.063 &   -28.050 &      72 &   -7.406 &    3.174 \\
      3.208 &   -27.746 &   -27.750 &     136 &   -7.131 &    4.356 \\
      3.206 &   -27.441 &   -27.450 &     218 &   -6.899 &    5.691 \\
      3.208 &   -27.141 &   -27.150 &     265 &   -6.803 &    6.358 \\
      3.190 &   -26.848 &   -26.850 &     371 &   -6.666 &    7.448 \\
      3.185 &   -26.539 &   -26.550 &     460 &   -6.551 &    8.494 \\
      3.181 &   -26.250 &   -26.250 &     515 &   -6.492 &    9.090 \\
      3.173 &   -25.954 &   -25.950 &     486 &   -6.426 &    9.817 \\
      3.170 &   -25.658 &   -25.650 &     402 &   -6.317 &   11.131 \\
      3.153 &   -25.359 &   -25.350 &     332 &   -6.174 &   13.114 \\
      3.138 &   -25.062 &   -25.050 &     218 &   -6.040 &   15.306 \\
      3.127 &   -24.800 &   -24.750 &      85 &   -6.329 &   10.966 \\
      3.097 &   -24.486 &   -24.450 &      16 &       --- &      --- \\
      \hline
      \hline
    \end{tabular}
    \caption{The binned BOSS DR9 Quasar Luminosity Function.  
      (1) The mean redshift of the bin; 
      (2) The mean $i$-band absolute magnitude of the bin; 
      (3) the absolute magnitude bin center; 
      (4) The number of quasars in each bin; 
      (5) $\Phi$ in units of Mpc$^{-3}$ mag$^{-1}$ and 
      (6) The (Poisson) error on $\Phi$, divided by $1\times10^{-9}$;
      The bins with no measured $\Phi$ are
      at the faint end limit where the selection function is rapidly
      approaching, or is equal to, 0.00, thus making our QLF estimation very
      uncertain. However, these bins are included so that
      $\sum N_{\rm Q}=23\,301$.
} 
    \label{tab:qlf_iband}
  \end{center}
\end{table}
%%%%%%%%%%%%%%%%%%%%%%%%%%%%%%%%%%%%%%%%%%%%%%%%%%%%%%%%%%%%%% 
%%%%%%%%%%%%%%%%%%%%%%%%%%%%%%%%%%%%%%%%%%%%%%%%%%%%%%%%%%%%%% 
%% 
%%   SECTION 5     SECTION 5     SECTION 5     SECTION 5     SECTION 5     SECTION 5     
%%   SECTION 5     SECTION 5     SECTION 5     SECTION 5     SECTION 5     SECTION 5     
%%   SECTION 5     SECTION 5     SECTION 5     SECTION 5     SECTION 5     SECTION 5     
%%
%%%%%%%%%%%%%%%%%%%%%%%%%%%%%%%%%%%%%%%%%%%%%%%%%%%%%%%%%%%%%%
%%%%%%%%%%%%%%%%%%%%%%%%%%%%%%%%%%%%%%%%%%%%%%%%%%%%%%%%%%%%%%
\section{Luminosity Functions}\label{sec:lum_func}
In Fig.~\ref{fig:L-z}, we show the coverage in the absolute
magnitude-redshift ($M_{i}-z$) plane for the three datasets of main
interest here: the SDSS \citep[black points;][]{Richards06,
Schneider10}; the XDQSO-selected $2.2\leq z \leq 3.5$ BOSS DR9 sample
(red points) and the fainter variability-selected dataset from the
boss21+MMT sample \citep[blue squares;
][]{Palanque-Delabrouille12}. We also analyze the Stripe 82 
variability-selected dataset of \citet{Palanque-Delabrouille11}, which has
a similar redshift distribution as the DR9 sample.
The bright and faint magnitude limits of
the BOSS DR9 sample, $i=17.8$ and $i=21.8$, respectively, are given
are given by the solid turquoise lines. Our binning is identical to
R06; the edges of the redshift bins in which we will calculated the
QLF are: 0.30, 0.68, 1.06, 1.44, 1.82, 2.20, 2.6, 3.0, 3.5, 4.0, 4.5,
and 5.0, and the $M_{i}$ bins start at -22.5 and are in increments of
0.30 mag\footnote{All the necessary data and code used here to produce
our results will be publicly available at
\href{http://www.sdss3.org/dr9/qlf}{{\tt www.sdss3.org/dr9/qlf}}.}.

   \begin{figure*}
      \begin{center}
        \includegraphics[height=14.0cm,width=18.0cm]
        {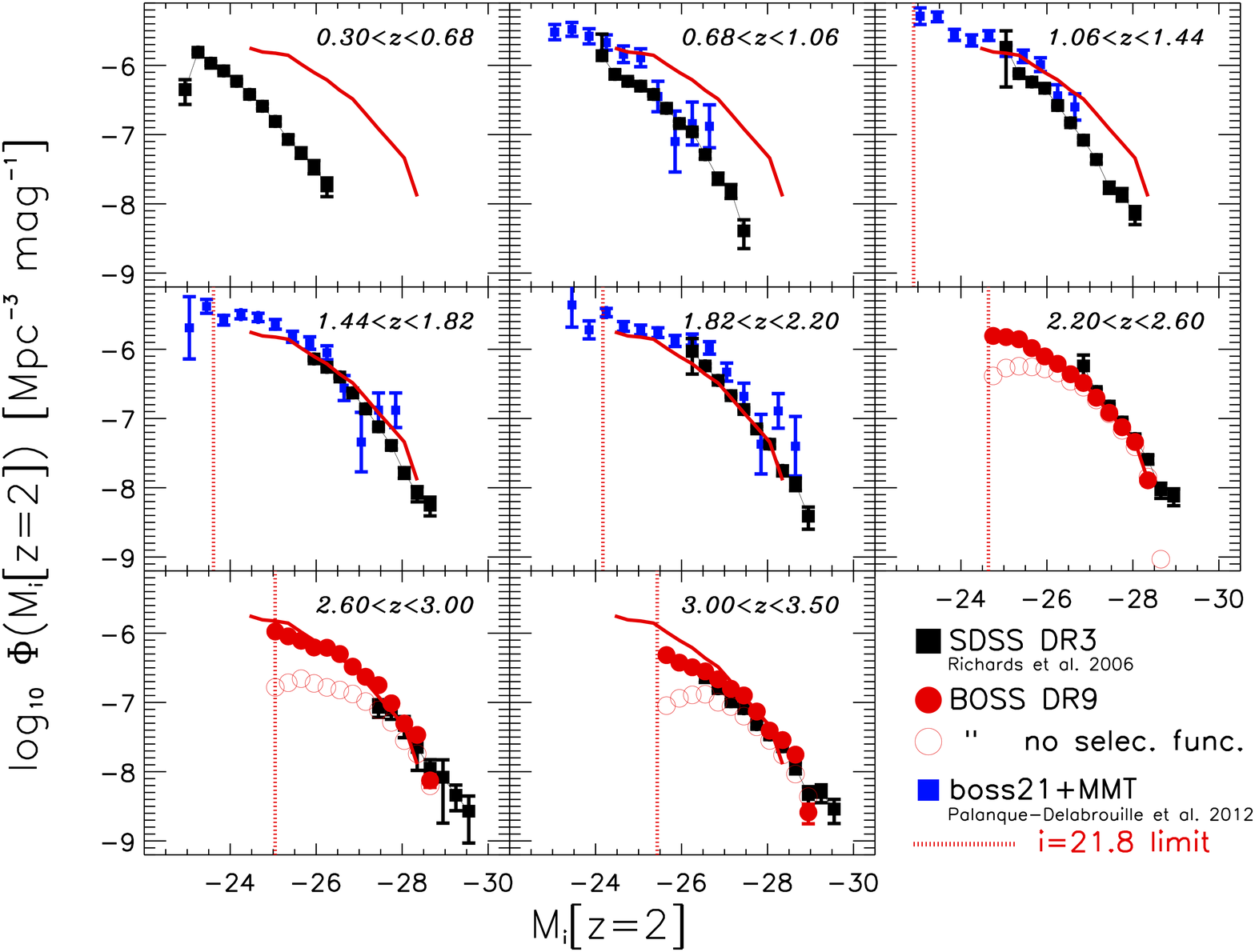}
        \vspace{-16pt}
        \caption{The $i$-band Quasar Luminosity Function. The red points are
          from our analysis of BOSS quasars from DR9, while the black squares
          are from the DR3 analysis of \citet{Richards06}. The boss21+MMT sample
          from \citet{Palanque-Delabrouille12} is also shown for $0.68<z<2.20$
          (blue filled squares). Over the redshift range $2.20\leq z \leq3.50$,
          we use the \hbox{23 301} DR9 quasars uniformly selected by XDQSO, and
          that are in sectors of spectro-completeness of 85\% or higher. The
          solid line in each panel is the BOSS DR9 QLF at $2.2 <z< 2.6$, to show
          how the luminosity function evolves. The open circles show the
          $2.2<z<3.5$ QLF without correcting for the (fiducial) selection
          function. There are no uniform DR9 measurements above $z=3.5$, since
          the XDQSO selection deliberately cuts off at this redshift.  The
          Poisson error bars for the BOSS measurements in the three panels
          spanning $2.2<z<3.5$ are the same size, or smaller, than the points
          shown.   }
        \label{fig:QLF_iband}
      \end{center}
    \end{figure*}

    \subsection{The Optical Luminosity Function to $i=21.8$}\label{sec:qlf}

    In Fig.~\ref{fig:QLF_iband} we show the $i$-band luminosity
    function from our BOSS DR9 uniform sample over $2.2<z<3.5$, as well as
    the fainter boss21+MMT sample of quasars covering $0.68<z<2.2$ from
    \citet{Palanque-Delabrouille12}.  We use the binned QLF
    estimator\footnote{\citet{Croom09b} show that the difference between
      the \citet{Page00} estimator and the ``model-weighted'' estimator of
      \citet{Miyaji01} is small, even at the bright end, for the $z\geq1$
      QLF.} of \citet{Page00},
    \begin{equation}
      \phi \approx \phi_{\rm est} = \frac{N_{q}}{\int^{L_{\rm max}}_{L_{\rm min}}
        \int^{z_{\rm max}(L)}_{z_{\rm min}}(dV/dz)\ dz\ dL}.
    \end{equation}
    This involves calculating the number of quasars, $N_{q}$ observed
    in a given ($M_{i}-z$) bin, correcting for our selection function, and
    dividing $N_{q}$ by the effective volume element $dV$ of that bin. The
    effective volume is calculated by using our fixed, flat
    $(\Omega_{\Lambda}, \Omega_{\rm m}, h)=(0.70, 0.30, 0.70)$ cosmology,
    and the area of our uniform DR9 sample (2236 deg$^{2}$). We check,
    and find that our redshift bins are sufficiently narrow to avoid
    complications due to evolution. The plotted error is estimated by
    \begin{equation}
      \delta \phi_{\rm est} = \frac{\delta N}{\int^{L_{\rm max}}_{L_{\rm min}}
        \int^{z_{\rm max}(L)}_{z_{\rm min}}(dV/dz)\ dz\ dL}
    \end{equation}
    and $\delta N$ is given by Poisson statistics including the
    up-weighting by the inverse of the completeness. We discuss the
    validity of this error estimate below. The binned QLF is also given in
    Table~\ref{tab:qlf_iband}, which gives the mean redshift of the
    quasars in each bin, the mean $i$-band magnitude of the quasars in the
    bin, the magnitude at the bin center, the raw number of quasars in the
    bin, the log of the space density, $\Phi$, in Mpc$^{-3}$ mag$^{-1}$,
    and the error $\times10^{9}$. The results from R06 using the SDSS DR3
    are given as the black squares in
    Fig.~\ref{fig:QLF_iband}. \citet{Shen_Kelly12} measured the QLF from
    the final DR7 SDSS quasar sample, and found excellent agreement with
    the DR3 results.
   
    \begin{figure*}
      \begin{center}
        \includegraphics[height=14.0cm,width=18.0cm]
        {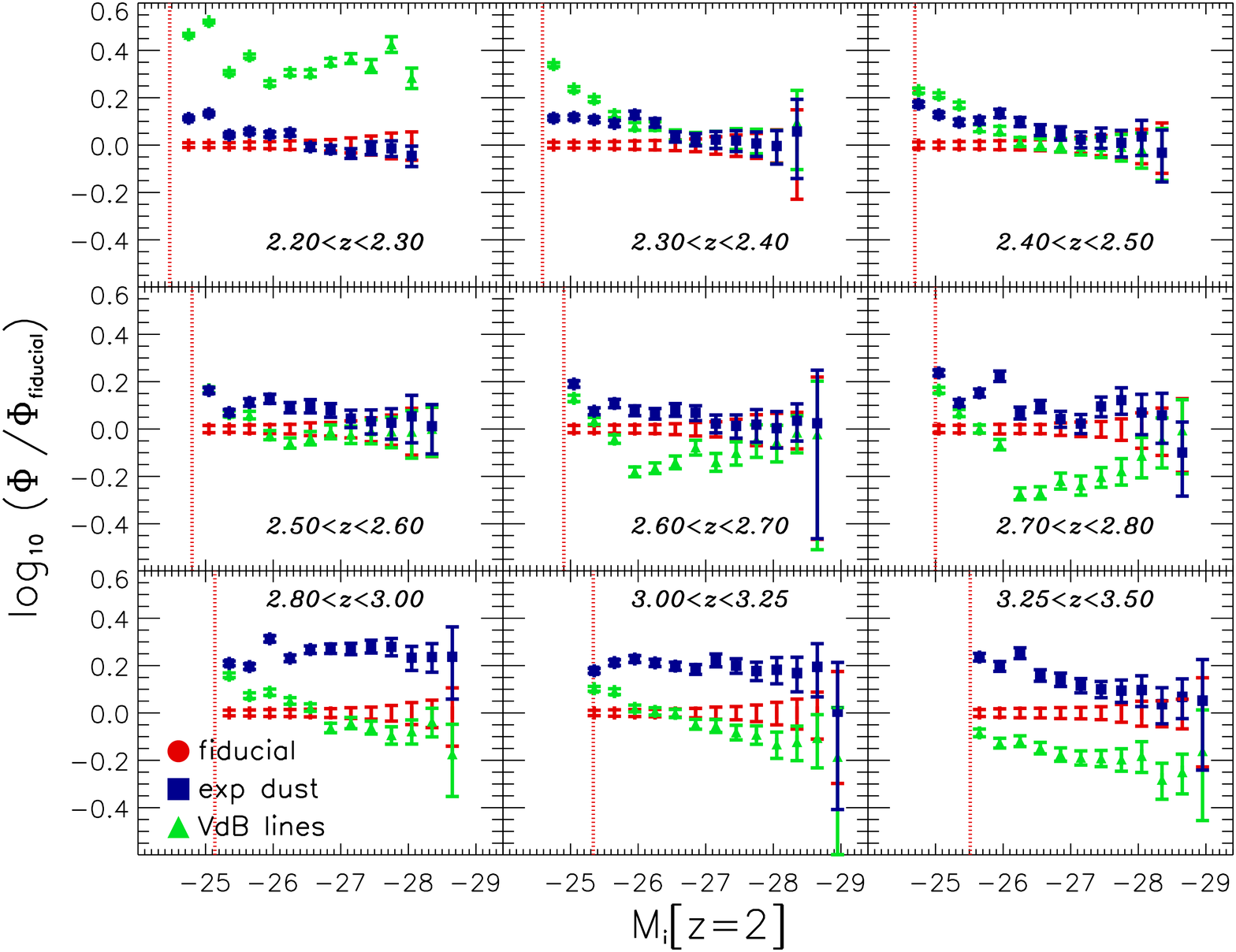}
        \vspace{-16pt}
         \caption{The sensitivity of the BOSS DR9 Quasar Luminosity Function
           to the selection function models used. The QLF is divided by the
           fiducial model. The red points show this fiducial model (at 0 by
           definition), with the extent of the points representing the
           statistical uncertainty range. The QLFs derived from the other two
           models are divided by the fiducial model QLF to highlight the effect
           of the choice of selection function on the derived QLF. The VdB lines
           model is shown as green points, and the exp dust model as blue
           points. In general there is good agreement between the three models,
           but in some redshift bins the disagreement can be 20\% or greater. The
           fiducial model provides the best fit to the observed color-redshift
           relation; however, the differences seen here quantify the systematic
           uncertainty inherent in not knowing the selection function exactly.
       }
      \label{fig:QLF_iband_narrow_selfunc_divmodel}
      \end{center}
    \end{figure*}

    Where the surveys overlap, we generally see very good agreement
    with the BOSS and SDSS data points, especially at $z\leq3$. Although
    there is overlap in $L-z$ coverage between the SDSS DR3 and BOSS
    measurements, since DR3 and DR9 cover different ares of the sky, there
    are only 304 quasars ($\lesssim2$\%) are common to both surveys,
    mostly in the $3<z<3.5$ redshift range.
    
    The limiting magnitude for BOSS DR9 quasar targets is $g< 22.00$
    or $r< 21.85$, and with $(r-i)\approx0.05$ for quasars at
    $z\approx2.5$, we show an $i$-band limiting of $i=21.8$ as a guide in
    Fig.~\ref{fig:QLF_iband}. There is strong evidence for a turn-over in
    the QLF, well before this limit, seen in all the redshift panels
    i.e. up to $z=3.5$. We shall see in Sec.~\ref{sec:compare}, that our
    results across $3.0<z<3.5$ are also consistent with a turn over seen
    in other experiments. This turn-over has been seen in the X-rays
    \citep{Miyaji01, Ueda03, Hasinger05, Aird10, Fiore12} and also in the
    optical \citep{Boyle88, Croom04, Croom09b}, with the BOSS DR9 now
    extending this evidence to redshifts $z=3.5$.
    
   \begin{figure*}
      \begin{center}
        \includegraphics[height=14.0cm,width=18.0cm]
        {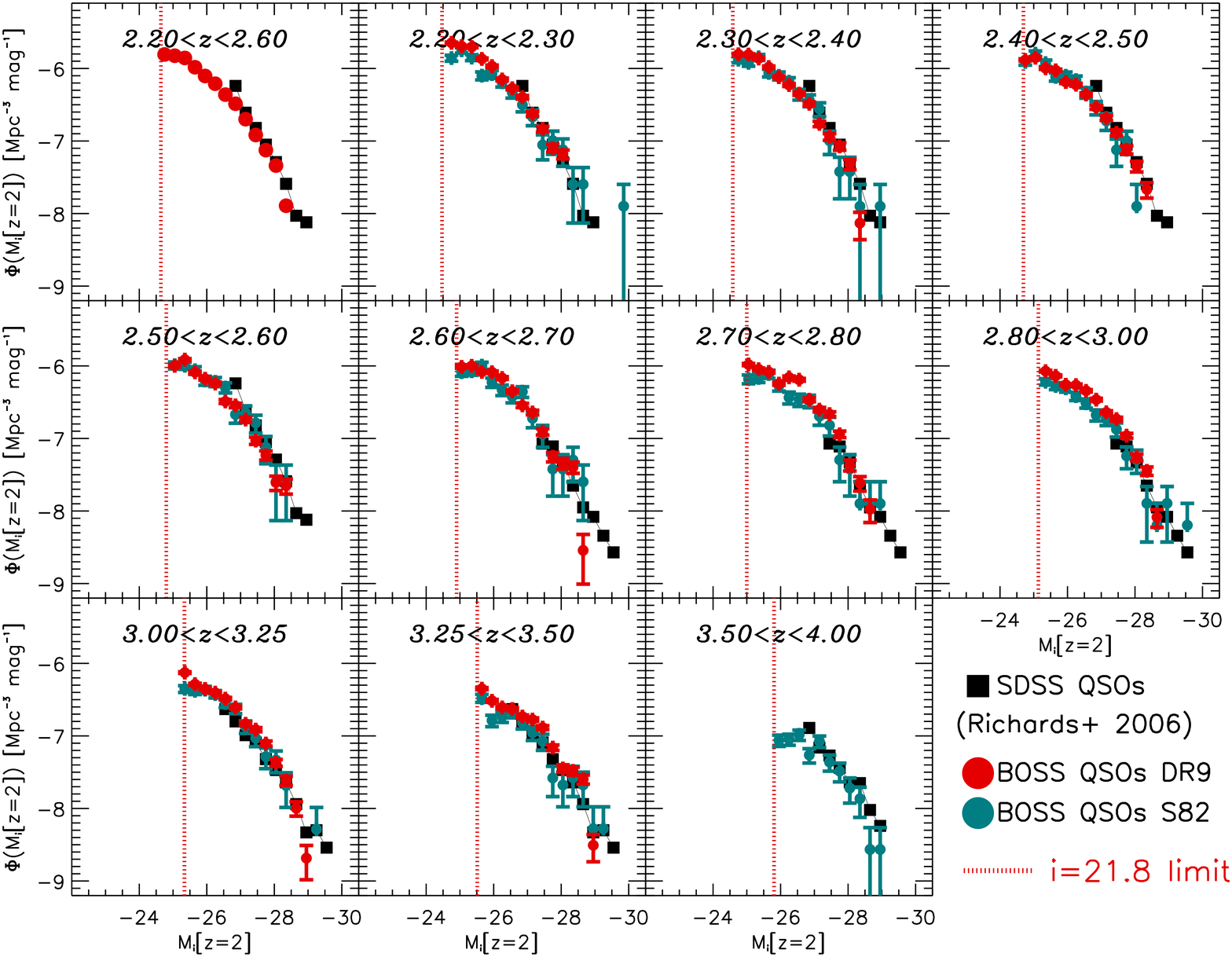}
        \vspace{-30pt}
        \caption{
          The {\it i}-band Quasar Luminosity Function. Black squares
          are from \citet{Richards06}, the filled red circles are our
          measurements from the full DR9 sample, while the teal points are from
          our analysis of the 5,476 BOSS quasars with $2.2<z<3.5$ located and
          observed on Stripe 82. The top-left panel is the 2.2$<z<$2.6
          measurement from Fig.~\ref{fig:QLF_iband} shown as a guide.} 
        \label{fig:QLF_iband_narrow}
      \end{center}
    \end{figure*}
    Our calculation of the errorbars given in Fig.~\ref{fig:QLF_iband}
    assumes the errors in each bin are independent and are dominated by
    Poisson statistics of the observed objects. For the DR9 sample overall
    this is reasonable, given the very large volume surveyed (which
    reduces fluctuations due to large-scale structure) and the low mean
    occupancy of quasars in halos (which reduces the impact of halo count
    fluctuations on the correlations). When comparing to surveys of
    smaller volume, sample variance may dominate over the Poisson
    errors. In some redshift ranges, however, BOSS is quite incomplete,
    and require a significant selection function correction (compare the
    open red circles to the filled red circles in the $2.6<z<3.0$ bin of
    Fig.~\ref{fig:QLF_iband} for example). In these bins the error is
    dominated not by Poisson statistics but by the uncertainty in our
    estimate of the selection function (see
    Fig.~\ref{fig:compare_stripe82_to_sims_selfn}). This uncertainty can
    reach 50\%, fractionally, for faint quasars in the most incomplete
    redshift range, leading to a similar fractional uncertainty in the
    QLF. However for most of the range plotted the uncertainty is
    significantly smaller.

    In Fig.~\ref{fig:QLF_iband_narrow_selfunc_divmodel} the effect of
    the selection function correction is investigated further. Here we
    plot the logarithm of the ratio of the QLF number densities for the
    two other selection function models, ``VdB lines'' and ``exp dust'',
    introduced in Sec.~\ref{sec:simqso}, compared to our fiducial model
    (that is used to calculate the QLF presented in
    Fig.~\ref{fig:QLF_iband}). We concentrate on the redshift range
    $2.20<z<3.50$. The errorbars for each model represent the Poisson
    uncertainties, and the differences between selection function models
    dominate over these statistical uncertainties, especially at the faint
    end. The corrections derived from the exp dust model generally augment
    the estimated luminosity function, particularly at low luminosities
    and higher redshifts. This is likely due to the fact that BOSS quasar
    selection is flux-limited in the $g$ and $r$ bands, so that fainter
    and higher redshift objects subjected to dust reddening will be
    extincted out of survey selection. The corrections derived from the
    VdB lines model show an even stronger trend with luminosity. The
    dependence of observed quasar colors on intrinsic luminosity resulting
    from the Baldwin Effect leads to a luminosity-dependent selection
    function. A QLF estimate that does not account for this effect will
    incur an artificial tilt as a function of luminosity, as highlighted
    by the figure. This tilt will further affect QLF parameters such as
    the power law slopes.
    
    \begin{figure*}
      \begin{center}
        \includegraphics[height=5.90cm,width=5.90cm,  trim =  0mm  0mm 5mm 0mm, clip]
        {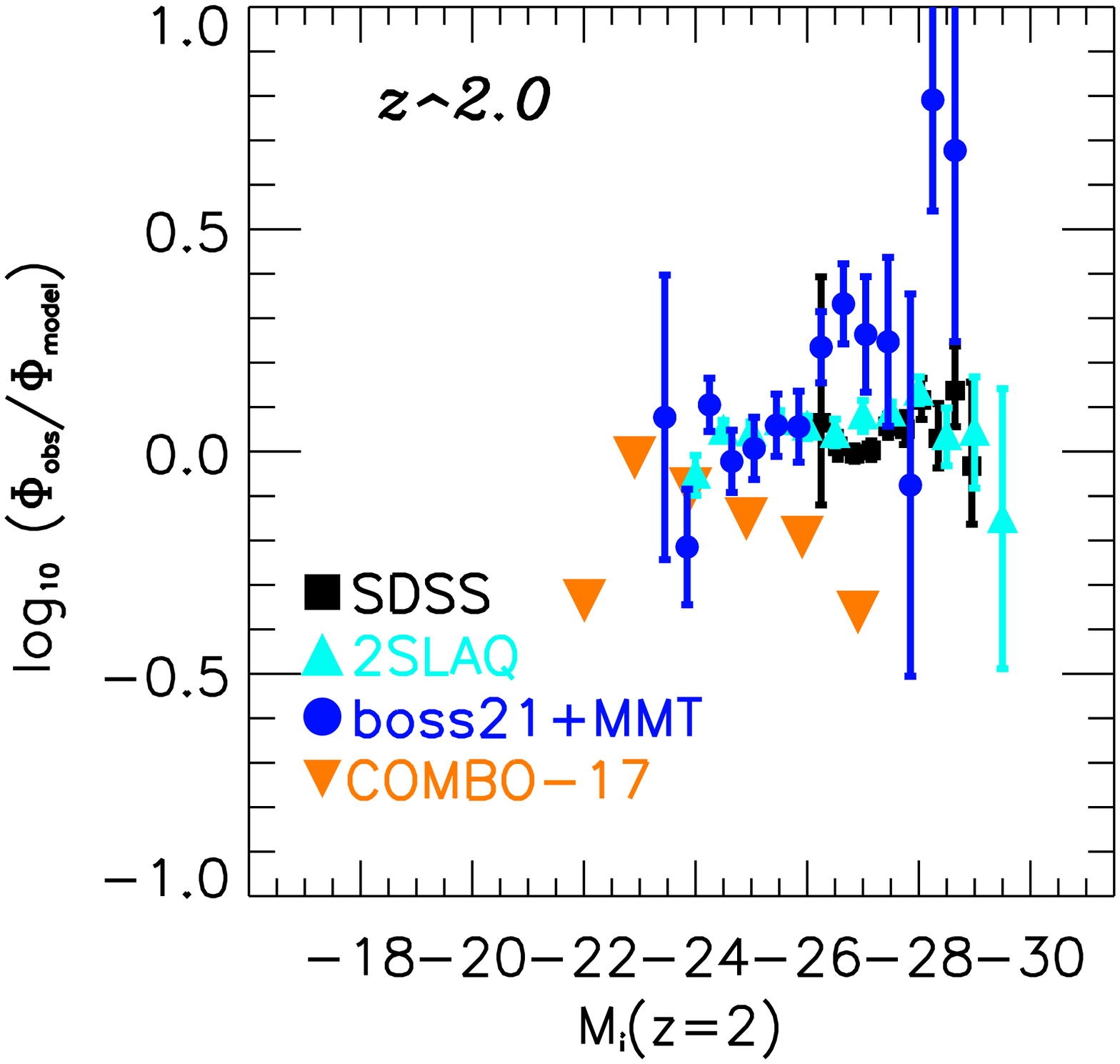}
        \includegraphics[height=5.90cm,width=5.90cm,  trim =  0mm  0mm 5mm 0mm, clip]
        {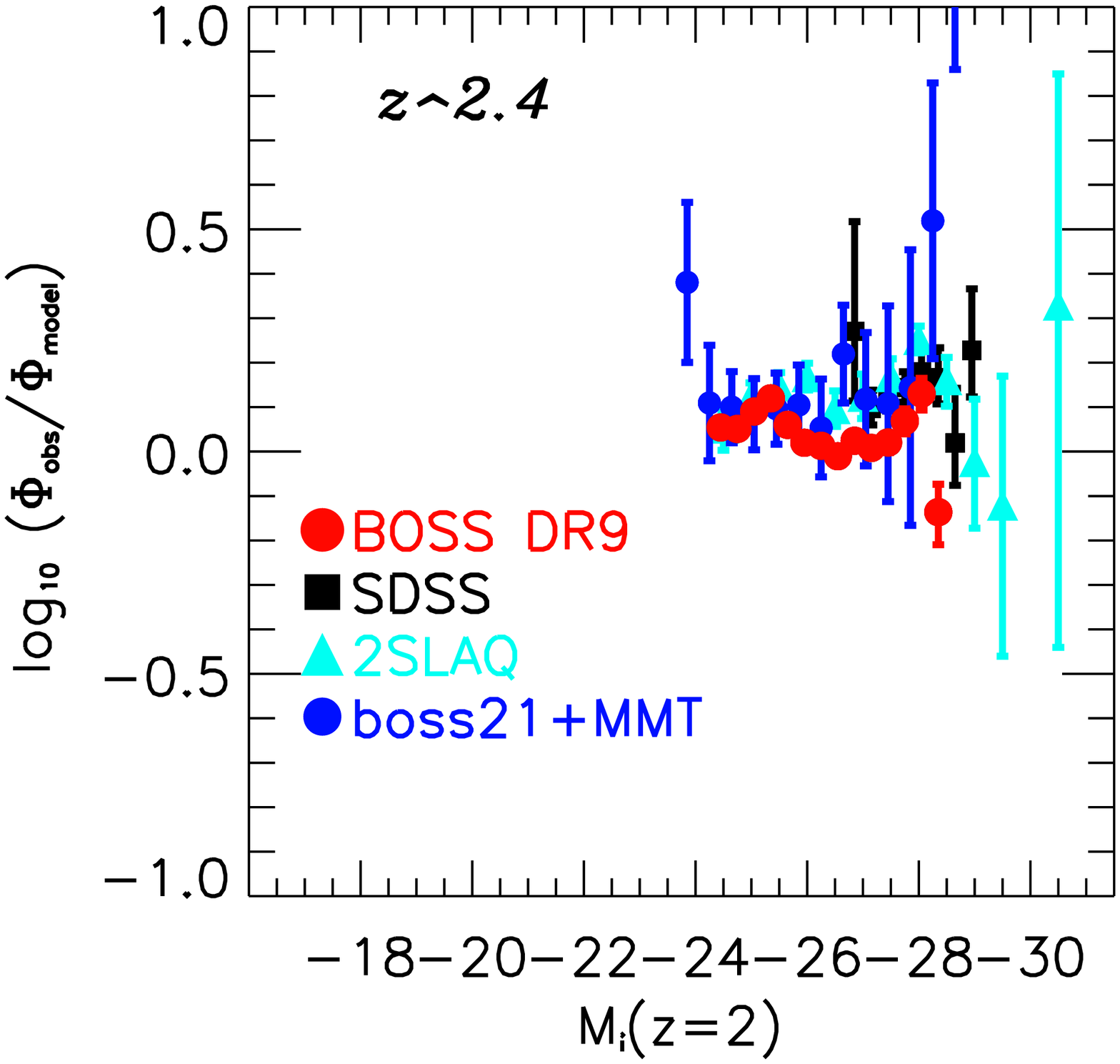}
        \includegraphics[height=5.90cm,width=5.90cm,  trim = 0mm 0mm  5mm 0mm, clip]
        {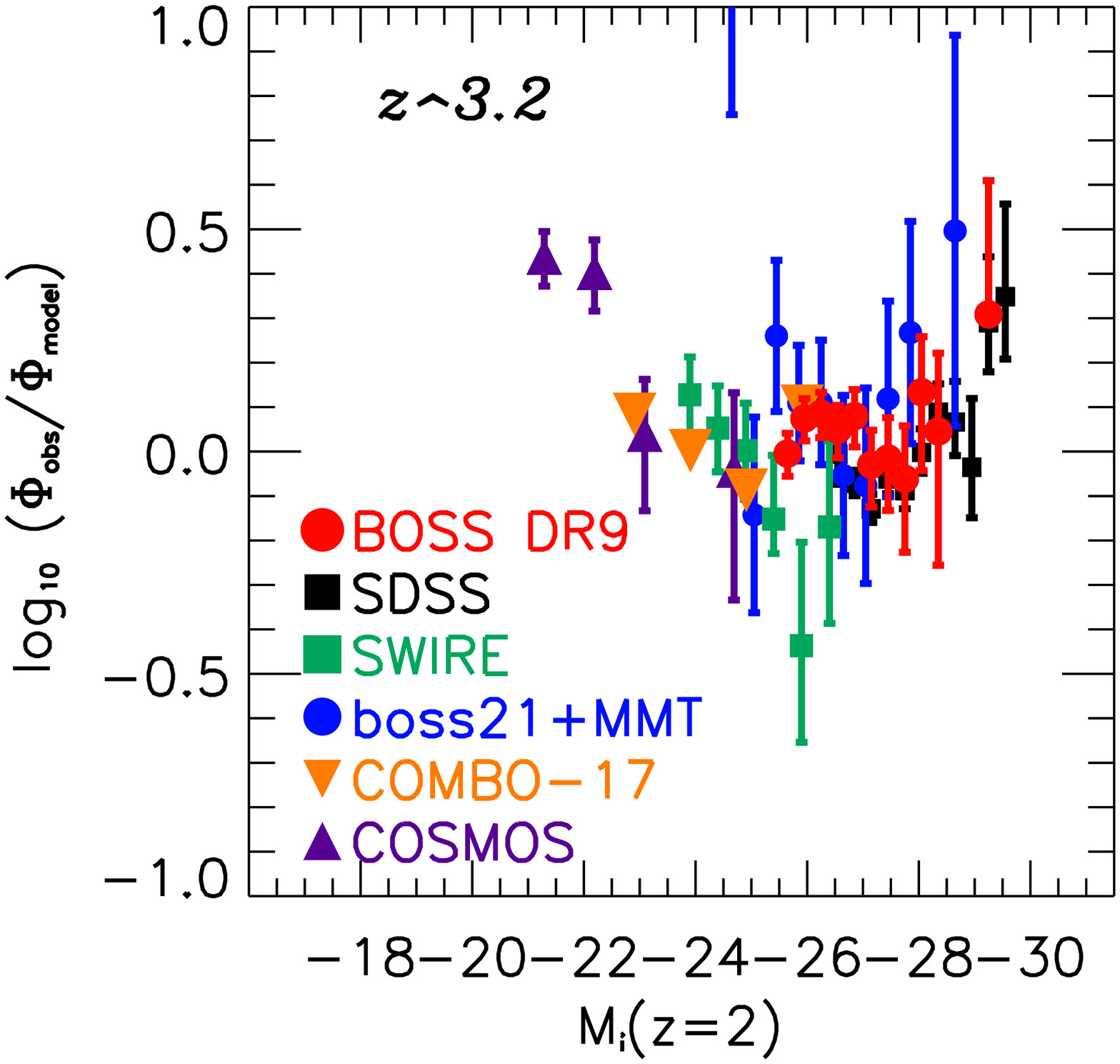}
        \caption{The BOSS DR9 Quasar Luminosity Function compared to other surveys. 
          {\it Left:} 
          Measurement in the $1.8<z<2.2$ range using data from BOSS
          (this paper; red circles), SDSS \citep[black squares; ][]{Richards06},
          2SLAQ QSO survey \citep[light blue up-triangles; ][]{Croom09b}, the
          ``boss21+MMT'' survey \citep[dark blue circles;
          ][]{Palanque-Delabrouille12} and the COMBO-17 survey \citep[orange
          down-triangles; ][]{Wolf03}.
          {\it Center:} 
          Measurement in the $2.2<z<2.6$ range;  
          {\it Right:} 
          Measurement at $z\sim3.2$, now adding data from the COSMOS
          survey \citep[purple diamonds; ][]{Masters12} and SWIRE \citep[dark
          green squares; ][]{Siana08}.  In each panel, we have divided the
          various QLFs by our best-fit ``log-linear'' LEDE model, which is
          described below in Sec.~\ref{sec:PLE_models} and
          Table~\ref{tab:PLE_fits}.}
       \label{fig:QLF_data_comparisons}
      \end{center}
    \end{figure*}

    In Fig.~\ref{fig:QLF_iband_narrow}, we continue to concentrate on
    the redshift range $2.2\leq z \leq3.5$, and divide it more finely in
    redshift than in Fig.~\ref{fig:QLF_iband}. The data displayed in
    Fig.~\ref{fig:QLF_iband_narrow} are presented in tabular form in
    Appendix~\ref{appndx:tables}, and it will be these data that we will
    fit models to in Sec.~\ref{sec:PLE_models}. We compare the BOSS DR9
    measurements to the \hbox{5 476} $2.2<z<3.5$ quasars on Stripe 82 that
    were selected via variability. The BOSS DR9 and Stripe 82 measurements
    are in very good agreement below $z\sim 2.7$, consistent with the
    selection function agreement in Fig.~\ref{fig:compare_stripe82_to_sims_selfn}.
    
    However, there are differences between the two datasets for
    $z\gtrsim2.7$, especially at the fainter end. The DR9 measurement
    implies a higher space density than the Stripe 82 variability
    measurements. One possible explanation could be that the selection
    function is underestimated (in the sense that it over corrects
    $N_{q}$) from Sec.~\ref{sec:simqso}. However, this would potentially
    lead to higher DR9 space densities at the faint end at all
    redshifts. Another possibility is that the variability selection is
    beginning to break down at the faint end, as the selection is based on
    light-curves taken from single-epoch imaging, and is susceptible to
    imaging incompleteness.

   \subsection{Comparison to Other Results}\label{sec:compare}
    In Fig.~\ref{fig:QLF_data_comparisons}, we compare our BOSS DR9
    QLF to other measurements of the QLF at $z\geq2$. In each panel, we
    divide the QLFs by our best-fit ``log-linear'' Luminosity Evolution
    and Density Evolution (LEDE) model, described in
    Section~\ref{sec:PLE_models}. We concentrate on the redshifts
    $z\approx2.0$ and $z\approx2.4$, and the results of \citet{Croom09b}
    from the 2SLAQ QSO survey.  We also compare our measurements at
    $z\approx3.2$ with recent results from \citet{Masters12} using
    observations from COSMOS \citep{Scoville07}. We additionally compare
    the results from our sister study, \citet{Palanque-Delabrouille12}, at
    all three epochs.
    
    The 2SLAQ results are presented in \citet{Croom09b} as a function
    of $M_{\rm g}(z=2)$, and the COSMOS results in \citet{Masters12} in
    $M_{\rm 1450}$, so in order to make a direct comparison, we convert
    these to $M_{\rm i}(z=2)$, with the transformation:
    \begin{eqnarray}
      M_{\rm i}(z=2) & = & M_{\rm g}(z=2) - 0.25 \\
                             & = & M_{\rm 1450}   -  0.29. 
%                             & = &M_{\rm b_{\rm J}} - 0.71.	\\
    \label{eqn:mag_conversions}
    \end{eqnarray}

    One underlying assumption in these conversions is that the 2SLAQ,
    and indeed the BOSS, quasars have a distribution in spectral power-law
    slopes ($\alpha_{\nu}$, where $F(\nu) \propto \nu^{\alpha_{\nu}}$) in
    the UV/Blue/Optical that is comparable to that of the SDSS quasar
    sample. Although the BOSS target selection avoids sources that would
    satisfy a UV Excess selection \citep[see ][]{Ross12, Paris12}, there
    is not strong {\it a priori} reason to suspect that these populations
    would deviate from a range of intrinsic slopes $-1<\alpha_{\nu}<0$,
    centered around $\alpha_{\nu} \sim -0.40$.
    
    Our comparison to the 2SLAQ result is shown in the left and center
    panels of Fig.~\ref{fig:QLF_data_comparisons}, for the redshift range
    $1.8\lesssim z \lesssim 2.2$ and $2.2 \lesssim z \lesssim 2.6$,
    respectively. We note that the 2SLAQ result is based on a combination
    of the 2SLAQ QSO survey (which dominates the signal at the faint end
    of the QLF) and the SDSS results from DR3 (which is responsible for
    the bright end measurement). Thus, the 2SLAQ points, represented as
    light blue upwards-pointing triangles in
    Fig.~\ref{fig:QLF_data_comparisons}, are not independent from the SDSS
    (black) squares.

    Concentrating on the $z\approx2.0$ panel, at the bright,
    $M_{i}(z=2) < -26$, end the boss21+MMT points, given by the blue
    filled points, seem $\sim$0.2-0.4 dex higher than e.g. the 2SLAQ and
    SDSS data, though are generally consistent within the quoted
    (statistical) error. \citet{Palanque-Delabrouille12} explore the
    variability selection in more detail than presented here, and
    resolution of this issue will be aided by new, forthcoming,
    variability-selected data, since sample variance uncertainties over
    the 14 deg$^{2}$ boss21+MMT field could well be an issue. At the faint
    end, boss21+MMT, 2SLAQ and SDSS are all consistent. All the displayed
    measurements are consistent with the COMBO-17 points (orange
    down-triangles), due to the large error associated with those points
    (not shown).
    
    At $z\sim2.4$ the BOSS DR9, SDSS, 2SLAQ and ``boss21+MMT'' are in
    excellent agreement, at both the bright and faint ends.
    
    We compare to the COSMOS result at $z\sim3.2$ \citep{Masters12},
    in the right panel of Fig.~\ref{fig:QLF_data_comparisons}. The BOSS
    QLF measurement is in good agreement with the COSMOS results, given by
    the purple upward-pointing triangles.  We also plot results from
    \citet[][green squares]{Siana08}, who use an optical/infra-red
    selection over 11.7 deg$^{2}$ from the {\it Spitzer} Wide-area
    Infrared Extragalactic \citep[SWIRE; ][]{Lonsdale03} Legacy
    Survey. The measurements from the $3<z<3.5$ bin from
    \citet{Palanque-Delabrouille12} is also given (blue circles). The BOSS
    DR9, SDSS, and boss21+MMT data are all in good agreement, and
    consisent given the errors. The faintest BOSS points are consistent
    with the brightest SWIRE, COMBO-17 and COSMOS points, again given the
    associated errors. There seems to be an inflection around $M_{i}(z=2)
    \approx -25.5$, suggesting that our best-fit model is under-predicting
    the QLF at the both the bright and faint end, i.e. the bright end
    slope of the model is too steep, while the faint end slope is too
    shallow. We discuss this further in Sec.~\ref{sec:PLE_models}.
    
    The R06 points lie below the other determinations, suggesting
    that they slightly underestimated the number density of $3.0<z<3.5$
    quasars. \citet{Worseck11} used UV data from the {\it GALEX} satellite
    \citep{Martin05, Morrissey07}, to show that the SDSS quasar target
    selection systematically misses quasars with blue $u-g\lesssim2$
    colors at $3\lesssim z \lesssim 3.5$ and preferentially selects
    quasars at these redshifts with intervening {H\,{\sc i}\ } Lyman limit
    systems, causing the QLF to be underestimated. Indeed, we specifically
    use the \citet{Worseck11} Monte Carlo model to describe the {H\,{\sc
        i}\ } Lyman series/forest and continuum absorption when creating our
    BOSS selection function, so we have corrected for this effect.

%%%%%%%%%%%%%%%%%%%%%%%%%%%%%%%%%%%%%%%%%%%%%%%%%%%%%%%%%%%%%%
%%%%%%%%%%%%%%%%%%%%%%%%%%%%%%%%%%%%%%%%%%%%%%%%%%%%%%%%%%%%%%
%%
%%   SECTION 6     SECTION 6     SECTION 6     SECTION 6     SECTION 6     SECTION 6     
%%   SECTION 6     SECTION 6     SECTION 6     SECTION 6     SECTION 6     SECTION 6     
%%   SECTION 6     SECTION 6     SECTION 6     SECTION 6     SECTION 6     SECTION 6     
%%
%%%%%%%%%%%%%%%%%%%%%%%%%%%%%%%%%%%%%%%%%%%%%%%%%%%%%%%%%%%%%%
%%%%%%%%%%%%%%%%%%%%%%%%%%%%%%%%%%%%%%%%%%%%%%%%%%%%%%%%%%%%%%
\section{QLF Fits, Models and Discussion}\label{sec:model_fits}
In this section, we fit parametric models to our binned QLF
and examine the evolution of the fitted parameters with 
redshift. We then compare our data to predictions based on more 
physical models of quasar evolution. Finally, we place our results in
a broader context regarding the AGN population and its link to galaxy
evolution.

    \subsection{QLF model fits}\label{sec:PLE_models} 
    The QLF is traditionally fit by a double power-law of the form 
    in Eq.~\ref{eq:double_powerlaw}. This functional form has four
    basic parameters, and various phenomenological models have been
    proposed to describe how those parameters evolve with redshift.
    In Pure Luminosity Evolution (PLE), only the break magnitude/luminosity
    evolves, leaving the overall number density constant. The opposite
    occurs in Pure Density Evolution: the shape of the QLF remains
    constant while the number density evolves. Various hybrid models
    allow both to vary but hold the bright- and faint-end slopes fixed.
    In Luminosity Evolution and Density Evolution (LEDE), $\mi^*(z)$ and
    $\Phi^{*}(z)$ evolve independently, while in Luminosity Dependent Density
    Evolution (LDDE), the evolution of $\Phi^{*}(z)$ is related to that of
    $\mi^*(z)$. Finally, extensions to these models allow the power law
    slopes to evolve as well.

    We begin with a simple PLE model for our data. In principle,
    $\mi^*(z)$ can take any functional form, but we follow 
    \citet{Boyle00} by fitting it with a second order polynomial: 
    \begin{equation}
      \mi^*(z)=\mi^*(z=0)-2.5(k_{1}z+k_{2}z^{2})~.
      \label{eq:mstarevol_poly}
    \end{equation}
    We note that this quadratic form for $\mi^*(z)$ requires symmetric
    evolution about the brightest $\mi^*$ value, and that this is known to
    break down at redshifts well above the peak
    \citep[e.g.][]{Richards06}. However, we are motivated to continue to
    use the quadratic PLE description as a historical reference and
    because over a limited redshift range the general form of our QLF is
    {\it qualitatively\/} consistent with a PLE model. For example, if the
    solid red line (representing the BOSS DR9 QLF at $2.2 <z< 2.6$) in
    Fig.~\ref{fig:QLF_iband} is compared to the measured QLF at
    $z\lesssim3$, one sees the broader trends in the data are encapsulated
    by a shift in $\mi^*$ with little change in normalization.
    
    We fit the PLE model with Eqn.~\ref{eq:mstarevol_poly} to our data
    over various redshift ranges. We use the combination of SDSS (R06),
    boss21+MMT and BOSS Stripe 82 dataset to perform the fits. These data
    span $0.30 < z < 4.75$ in redshift, $-29.55 \leq M_{\rm i}(z=2) \leq
    -22.96$ in magnitude and $\Phi=2.2 \times10^{-9} - 2.2 \times10^{-6}$
    Mpc$^{-3}$ mag$^{-1}$ in number density. We fit to the Stripe 82 data,
    since we expect that this data is less affected by systematics, and
    thus more meaningful $\chi^2$ values can be obtained from the
    statistical uncertainties\footnote{Note that while Stripe 82 does not
      have a correction applied for color selection effects, the
      $k$-correction still introduces uncertainty that may have systematic
      trends with redshift and luminosity.}. We have also found that the S82
    data is a fair representation of the DR9 data
    (Fig.~\ref{fig:QLF_iband_narrow}).

    We perform $\chi^{2}$ fits to the binned data with six total
    free parameters in the PLE model, using the Levenberg-Marquardt
    optimization method to find the best-fit parameters by minimizing
    the $\chi^2$. The parameter values for our 
    best-fit PLE models are given in Table~\ref{tab:PLE_fits}. We first
    restrict our fits to $z<2.2$, where previous work has generally
    found that PLE models provide a reasonably good fit. Fitting 
    over $0.30 < z < 2.20$ results in $\chi^{2}/\nu= 155 / 75$. Most of
    the disagreement with our data comes at $z<1$; by restricting to the 
    range $1.06 < z < 2.20$ the fit improves to $\chi^{2}/\nu= 83 / 52$. 
    Thus we find that PLE models do indeed provide a reasonable description
    of our low redshift data, though clearly there is room for improvement
    in the $\chi^2$. At higher redshift, the PLE model fails. Within the
    BOSS redshift range of $z=2.20-3.50$ we have $\chi^{2}/\nu= 286 / 113$;
    the result is even worse over the full redshift range of our data 
    ($z=0.30-3.50$), with $\chi^{2}/\nu= 662 / 195$.
    
\begin{figure}
      \begin{center}
        \hspace{-16pt}
        \includegraphics[width=9.0cm]
        {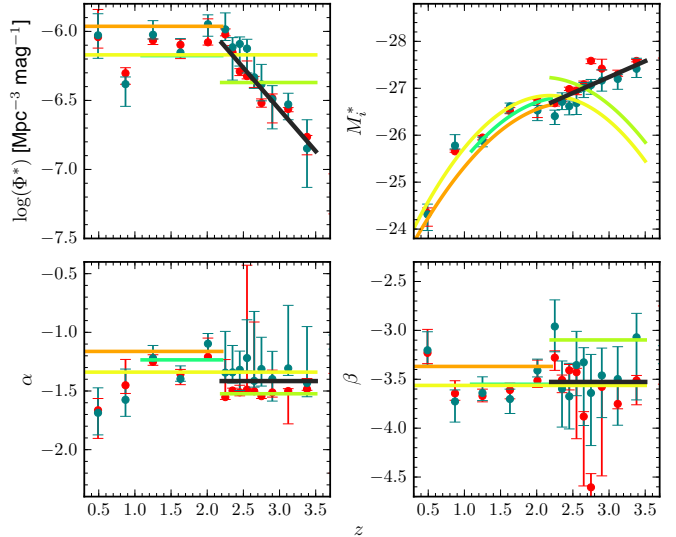}
        \caption{The best fit values for the parameters $\Phi^{*}$,
          $\mi^{*}$, $\alpha$ and $\beta$ as a function of redshift. A
          double-power law model is fit to the Stripe 82 data in each redshift bin. 
          The teal points
          are for the Stripe 82 data, while the red points are the BOSS DR9
          CORE data. The four colored lines represent the four best fitting PLE
          models in Table~\ref{tab:PLE_fits} over the respective redshift
          ranges, while the solid black line is the log-linear LEDE model
          (eqn.~\ref{eq:LEDE_linear}). }
        \label{fig:qlfparams}
      \end{center}
    \end{figure}

    Fig.~\ref{fig:qlfparams} demonstrates why this is the case, and 
    where the PLE model breaks down. Here we show the behavior with 
    redshift of the parameters $\Phi^{*}$, $\mi^{*}$, $\alpha$ and 
    $\beta$. The parameter values and uncertainties are determined by 
    $\chi^2$ minimization in each redshift bin independently.
    In the top right panel, we see that although at $z\la2.2$, a quadratic 
    description of the evolution of $\mi^*$ describes the general trend 
    of the data, at $z\ga2.2$, $\mi^*(z)$ continues to get brighter and 
    does not exhibit the turn over needed for the parameterization given 
    in equation~\ref{eq:mstarevol_poly} to work. However, even if a new 
    description of the evolution of $\mi^*$ could be found, the PLE model, 
    with no allowance for density evolution, is not suitable. This is 
    shown by the top left panel of Fig.~\ref{fig:qlfparams}; we can see 
    that there is essentially {\it no} evolution of $\Phi^{*}$ across 
    the range $0.5 < z \la 2.2$, but then $\log\Phi^{*}$ declines in a
    roughly linear fashion with redshift at $z\geq2.2$, corresponding
    to a drop in $\Phi^{*}$ by a factor of $\sim$6 between $z=2.2$ 
    and $z=3.5$. 
    
    Motivated by the evolution of $\log\Phi^{*}(z)$ and $\mi^{*}(z)$ 
    seen in Fig.~\ref{fig:qlfparams} across the range $z=2.2-3.5$, we 
    implement a form of the LEDE model where the normalization and
    break luminosity evolve in a log-linear manner; e.g.,
    \begin{eqnarray}
      \log[\Phi^{*}(z)]  & = & \log[\Phi^{*}(z=2.2)] + c_{1}(z-2.2)\\
                   \mi^{*}(z) & = &\mi^{*}(z=2.2) + c_{2}(z-2.2).
     \label{eq:LEDE_linear}
    \end{eqnarray} 
    For the BOSS Stripe 82 data across the redshift range $z=2.2-3.5$, 
    this returns a value of $\chi^{2}/\nu= 136 / 113$, indicating a 
    reasonable fit to the data. If we instead fit to the BOSS DR9 data,
    we find generally good agreement in the fitted parameters, but a
    dramatically worse $\chi^2$ value. While the binned QLF data from
    DR9 and Stripe 82 are in good agreement, the statistical uncertainties
    in the DR9 data are far smaller due to the much greater number of
    quasars. This inflates the $\chi^2$ for the same model fit; however,
    as explained in  Sections~\ref{sec:simqso} and~\ref{sec:qlf}, we 
    expect the true uncertainties of the DR9 data to be dominated by 
    systematics, in particular, in the need to correct for color selection 
    effects without knowing the true distribution of quasar colors. 
    The systematic effects associated with correcting for the selection
    function are obviously a general problem, and are especially
    problematic as the selection function affects the points in a 
    correlated way. Here we have taken advantage of two quasar samples
    selected by independent means; we leave this as a cautionary note
    for surveys relying on an unknown selection function, particularly
    where the data is dominated by objects found in regions where the
    selection efficiency is low.

    We do not extend our LEDE model below $z=2.2$, where it
    clearly would not describe the data. We also note that PDE models
    cannot capture the strong evolution in $\mi^{*}$ and are easily
    ruled out. Qualitatively, there is no clear relationship between
    the smooth evolution of $\mi^{*}$ and the disjoint behavior of
    $\log\Phi^*$, thus we also do not consider LDDE models. In summary,
    our data is best described as PLE evolution until $z\sim2.2$, at
    which point a transition to LEDE evolution occurs.
    
    Finally, we do not see evidence for evolution in the power law
    slopes, though these are not well constrained by our data. In
    particular, we do not find a strong evolution of the bright end slope
    (bottom right panel of Fig.~\ref{fig:qlfparams}) at $z>2.5$, in
    contrast to R06. This could be because the evolution in $M^{*}$
    affects the R06 results, or, very likely since we resolve the break in
    the QLF at $z=2.2-3.5$, and consequently fit a double-power law model
    \citep[cf. the single power-law in R06; see also the discussions
    in][]{Assef11, Shen_Kelly12}. However, comparing the points from
    Fig. 21 of R06, to our Fig.~\ref{fig:qlfparams}, the bright end slope
    measurements are consistent with each other, given the error bars.

    \begin{figure*}
      \begin{center}
        \includegraphics[height=5.90cm,width=5.90cm,  trim =  0mm  0mm 5mm 0mm, clip]
        {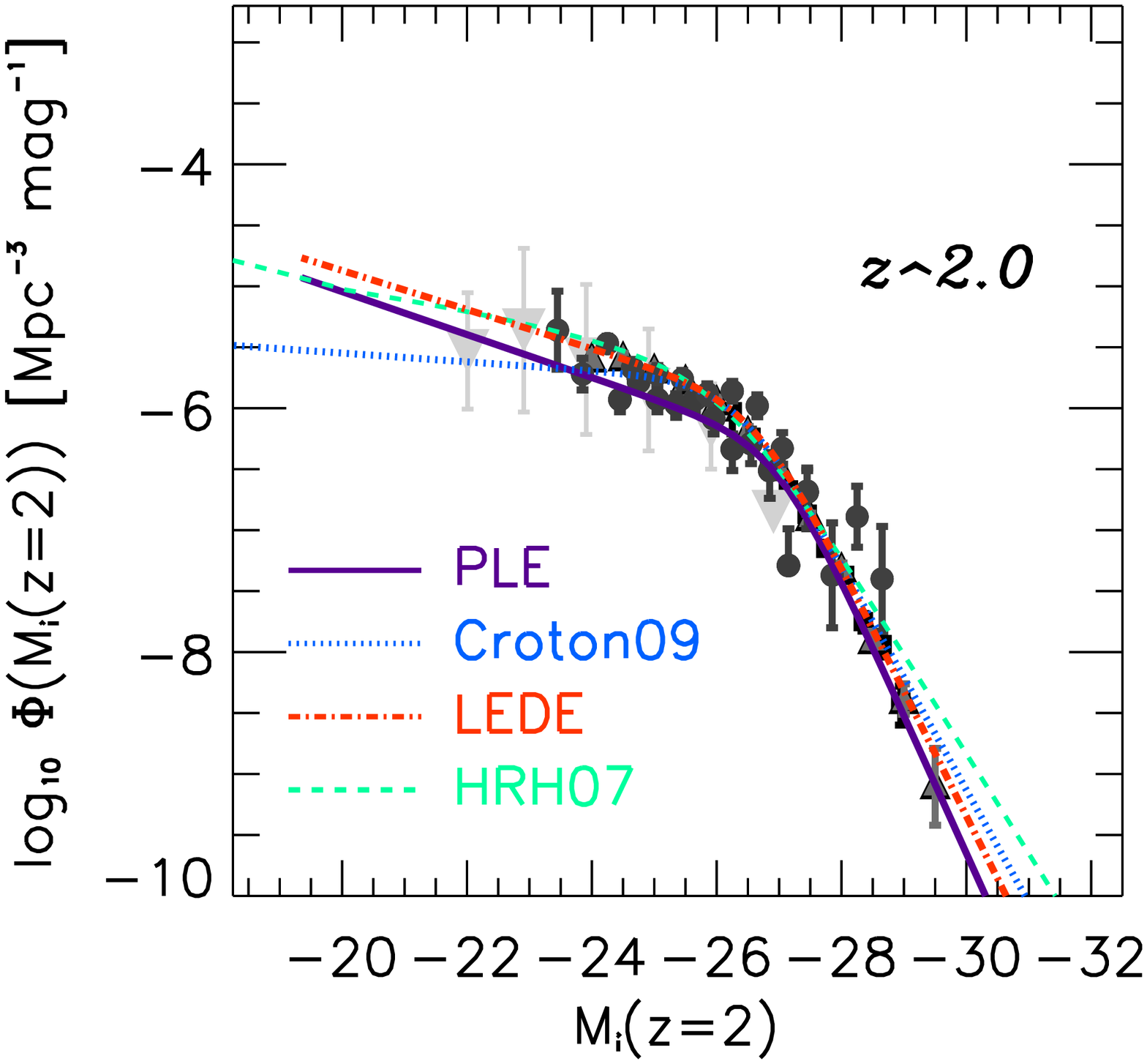}
        \includegraphics[height=5.90cm,width=5.90cm,  trim =  0mm  0mm 5mm 0mm, clip]
        {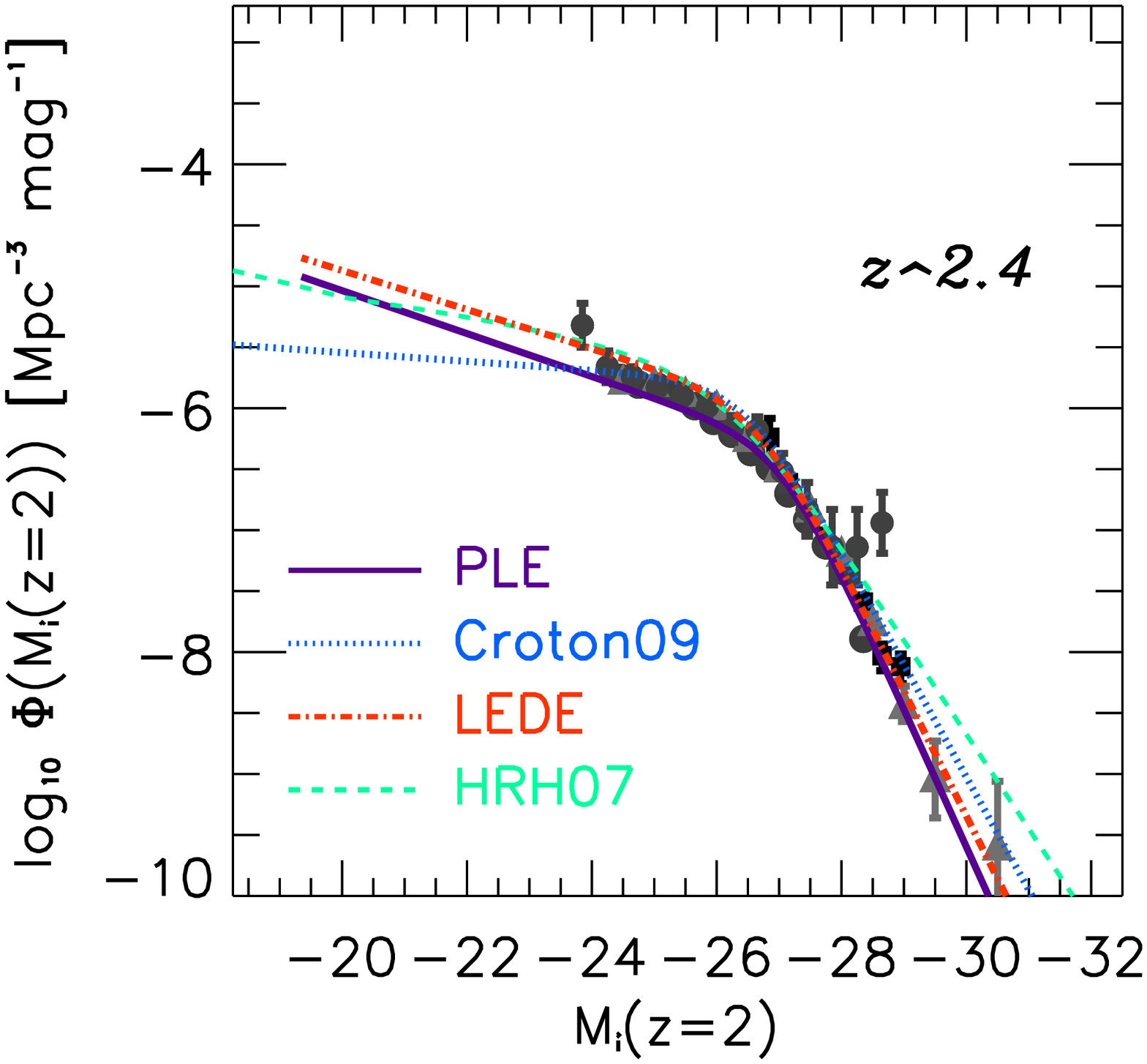}
        \includegraphics[height=5.90cm,width=5.90cm,  trim = 0mm 0mm  5mm 0mm, clip]
        {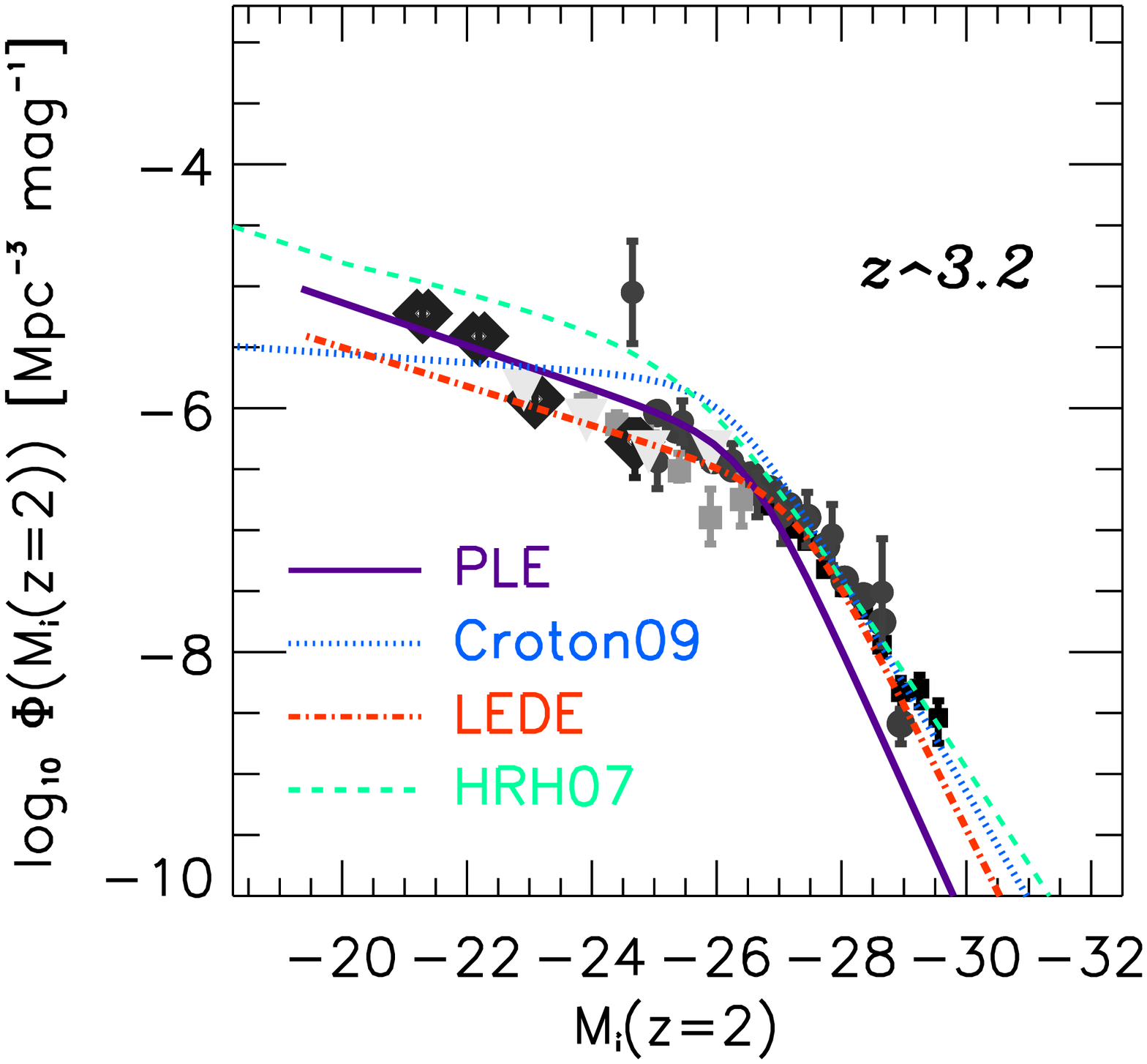}
        \caption{
          The BOSS DR9 Quasar Luminosity Function compared to a series
          of QLF fits. {\it Left:} Measurement in the $1.8<z<2.2$ range; {\it
            Center:} Measurement in the $2.2<z<2.6$ range and {\it Right:}
          Measurement at $z\sim3.2$.  In each panel we plot our best-fits PLE
          fit, given by the solid (purple) line, which is the fit over the
          redshift range $0.4<z<2.2$ (top line in Table~\ref{tab:PLE_fits}),
          Also shown is our best-fit log-linear LEDE model, given by
          the (orange) dot-dashed line, with the fitting parameters also in
          Table~\ref{tab:PLE_fits}. The extension to the 2QZ QLF as given in
          \citet{Croton09}, is shown by the dotted (light blue) line, while the
          ``Full'' model, of \citet{HRH07} is given by the (turquoise) dashed
          lines.}
        \label{fig:QLF_fits_comparisons}
      \end{center}
    \end{figure*}

    In Fig.~\ref{fig:QLF_fits_comparisons}, we show our best-fit
    PLE and LEDE models in three redshift bins, and compare our fits 
    to other models that have been presented in the literature.
    
    \citet{Croton09} presented a modification of the PLE fitting function
    of \citet{Croom04} in which the decline of $M_*$ with redshift is
    softened and the bright end power-law slope evolves above $z=3$. This
    was found to fit the higher redshift SDSS data better than the
    original fitting form, which was fit only to the 2QZ data. We
    reproduce this modified fit in Table~\ref{tab:PLE_fits} and
    Fig.~\ref{fig:QLF_fits_comparisons}, shown by the dotted (blue) line.
    We see that this model describes the data well at $z\sim2.0$ and 2.4, 
    but has a too high a normalization and (potentially) too flat a faint-end 
    slope at $z\sim3.2$.    
    
    \citet{HRH07} collected a large set of QLF measurements, from the
    rest-frame optical, soft and hard X-ray, and mid-IR bands, in order
    to obtain accurate bolometric corrections and thus determine the
    bolometric QLF in the redshift interval $z=0-6$.
    The observational dataset assembled by \citet{HRH07} is
    impressive, though most of the power in the $z>2$ dataset is from the
    (R06) optical measurements of the QLF. Taking the traditional
    double-power law approach, \citet{HRH07} then derive a series of
    best-fit models to the QLF, including a PLE and luminosity-dependent
    density evolution (LDDE) model. Their ``Full'' model, which is an
    LDDE-based model and includes a luminosity-dependent bolometric
    correction, is shown in Figure~\ref{fig:QLF_fits_comparisons} by the
    (turquoise) dashed lines. This model fits the data well until the highest
    redshift bin at $z\sim3.2$. In the \citet{HRH07} model, the break
    luminosity turns over at $z\sim2$ and becomes fainter at higher redshift,
    while the bright end slope flattens and the normalization is constant.
    This is apparent in Fig.~\ref{fig:QLF_fits_comparisons}, where the 
    break luminosity is clearly much fainter than in our data and the
    faint end number densities are overpredicted.
    
   \begin{figure}
      \includegraphics[width=8.8cm] 
     {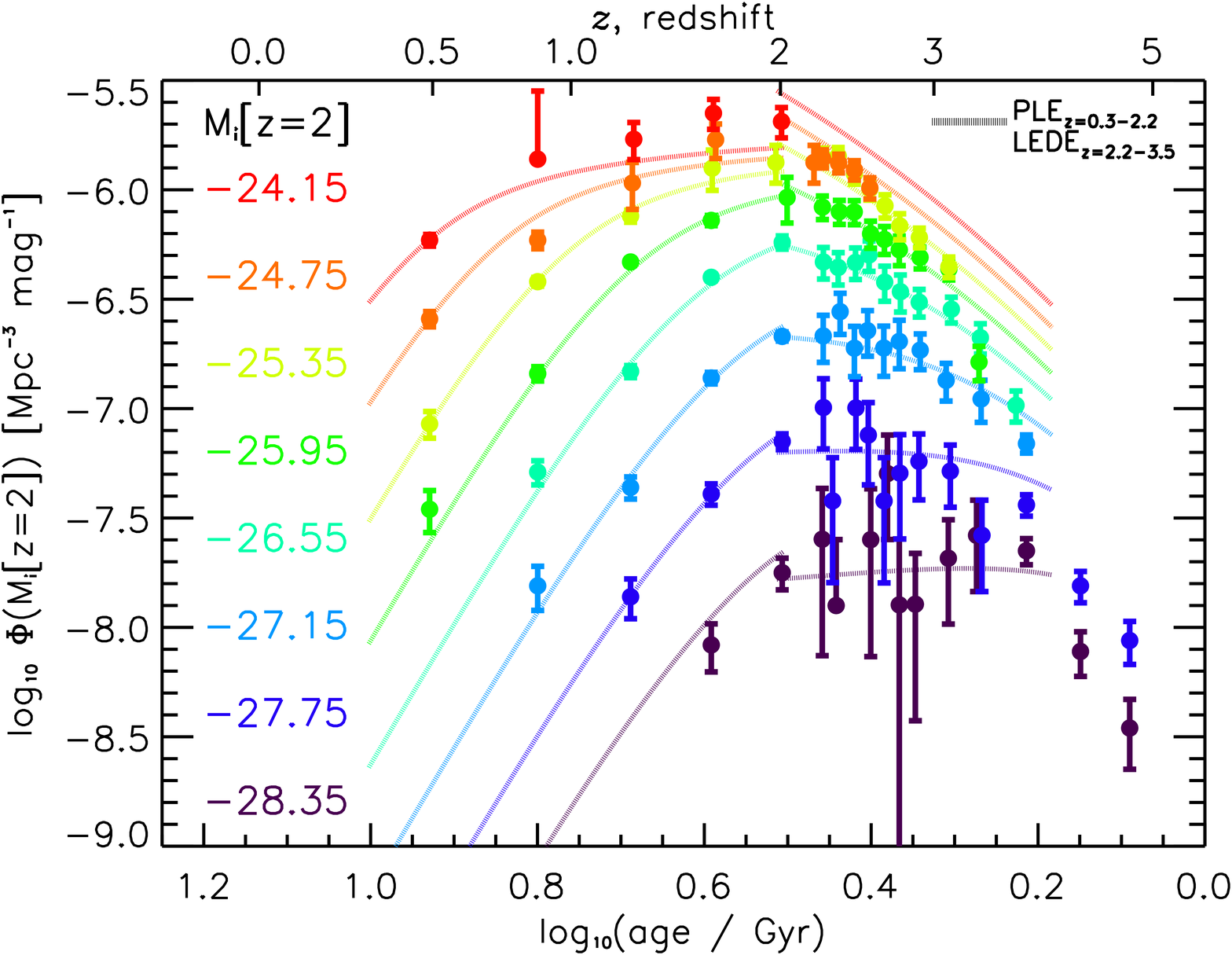}
      \caption{The comparison of our best fit phenomonological models, 
        dashed lines,  to the 
        SDSS, BOSS Stripe 82 and boss21+MMT QLF data (points). The number density of various
        magnitude bins are shown as a function of cosmological time 
        The best-fitting PLE model over $z=0.3-2.2$ and best-fitting LEDE model
        over $z=2.2-3.5$ from Table~\ref{tab:PLE_fits} are given by the dashed curves.
        A mismatch in number density at $z=2.2$ for the fainter magnitudes is 
        apparent, but since we do not require the fits to link, is not surprising and 
        within the uncertainties.}
     \label{fig:downsizing}
    \end{figure}
    Fig.~\ref{fig:downsizing} shows the redshift evolution of the QLF in
    a series of luminosity bins, including both our data and the best-fit
    PLE+LEDE model. 
    Previous measurements, especially in the deep X-ray studies 
    \citep{Miyaji01, Ueda03, Hasinger05}, and then in the optical by 
    the 2SLAQ QSO survey \citep{Richards05, Croom09b} see the trend 
    for ``AGN Downsizing'', with the number density of fainter AGN peaking 
    at lower-redshift than the luminous AGN. These studies, especially
    in the optical, have generally suggested that PLE works up to
    $z\approx2$, but not to higher $z$. Our BOSS results agree with this
    statement, but we use the longer redshift baseline of our data, and in
    particular the fact that we have resolved the break luminosity to
    $z\sim3.5$, to find a simple prescription for the evolution at
    $z>2$. Interestingly, we find that the {\it shape} of the QLF does not
    change (in terms of the power law slopes). Going from high to low
    redshift there is a build up of quasar activity (the log-linear trend
    in $\Phi^{*}$) until $z\approx2$, at which point the number density
    stalls. In this LEDE-to-PLE toy model scenario, AGN downsizing is then
    simply a trend in $L^{*}(z)$.

    Our optical QLF results are also in general agreement with the latest
    determination of the hard, 2-10 keV X-ray luminosity function
    \citep[XLF; ][]{Aird10}. These authors also find an LEDE model (which
    they name LADE) describe their XLF well, and that an XLF that also
    retains the same shape, but shifts in luminosity and density,
    describes the observed evolutionary behavior. We also agree with
    \citet{Aird10} in that the (QLF) LEDE model shows a much weaker
    signature of ``AGN Downsizing'' than previous studies
    \citep[][]{Hasinger05, Silverman08}. One caveat here is that the hard
    X-ray samples used in \citet{Aird10} are most secure at
    $z<1.2$. Overall, these trends of a simple log-linear LEDE model
    describing both the QLF and XLF lends weight to the theory that the
    X-ray selected AGN population at $z\sim1$ is a direct descendent of the
    optical quasar population at $z\sim2$; a scenario also suggested by
    quasar and X-ray AGN clustering results \citep{Hickox09, Ross09,
      Koutoulidis12}.

    \begin{table*}
      \caption{Values from a set of the best fit double-power law evolution
        models (e.g.  Eqns.~\ref{eq:double_powerlaw_mag}, \ref{eq:mstarevol_poly} and 
        \ref{eq:LEDE_linear}).  Listed are the redshift ranges of the
        data fitted, and the best fit values of the model parameters. We
        perform our fits using only the statistical error on the QLF. }
       \label{tab:PLE_fits}
      \centering
      \setlength{\tabcolsep}{2pt}
      \begin{tabular}{@{}l c ccc ccc ccc crr@{}}
        \hline
        Model & Redshift  & $\alpha$     & $\beta$ & $\mi^*(z=0)$ & $k_{1}$ & $k_{2}$ &  $\log(\Phi^{*})$                                  & $\chi^2$/$\nu$   \\
                   &  range     &   (faint end)  & (bright end)  &                   &            &            &    ${\rm Mpc}^{-3}{\rm mag}^{-1}$  & \\
        \hline 
        \hline 
       PLE & 0.3--2.2 &  $-1.16^{+0.02}_{-0.04}$ &  $-3.37^{+0.03}_{-0.05}$ &  $-22.85^{+0.05}_{-0.11}$ &  $1.241^{+0.010}_{-0.028}$ &  $-0.249^{+0.006}_{-0.017}$ &  $-5.96^{+0.02}_{-0.06}$ &   155/75 \\
       PLE & 1.06--2.2 &  $-1.23^{+0.06}_{-0.01}$ &  $-3.55^{+0.06}_{-0.05}$ &  $-22.92^{+0.24}_{-0.03}$ &  $1.293^{+0.061}_{-0.014}$ &  $-0.268^{+0.007}_{-0.031}$ &  $-6.04^{+0.07}_{-0.02}$ &   83/52 \\
       PLE & 2.2--3.5 &  $-1.52^{+0.05}_{-0.06}$ &  $-3.10^{+0.15}_{-0.07}$ &  $-24.29^{+0.26}_{-0.15}$ &  $1.134^{+0.041}_{-0.047}$ &  $-0.273^{+0.008}_{-0.006}$ &  $-6.37^{+0.10}_{-0.06}$ &   286/113 \\
       PLE & 0.3--3.5 &  $-1.34^{+0.06}_{-0.01}$ &  $-3.56^{+0.08}_{-0.05}$ &  $-23.04^{+0.15}_{-0.02}$ &  $1.396^{+0.032}_{-0.009}$ &  $-0.320^{+0.005}_{-0.004}$ &  $-6.17^{+0.03}_{-0.01}$ &   622/195 \\
\\
       Crot09  & $z<3$      & $-1.09$   &$-3.31$                 & $-22.32$ & $1.39$ & $-0.29$ & $-5.78$ \\ 
        Crot09  & $z\geq 3$ & $-1.09$ & $-3.33+0.5(z-3)$  & $-22.32$ & $1.22$ & $-0.23$ & $-5.78$ \\ 
      \hline
                                &               &     &  & $\mi^*(z=2.2)$ & $c_{1}$ & $c_{2}$ &    &    \\
       LEDE & 2.2--3.5 &  $-1.42^{+0.51}_{-0.01}$ &  $-3.53^{+0.09}_{-0.29}$ &  $-26.70^{+0.22}_{-0.06}$ &  $-0.604^{+0.005}_{-0.104}$ &  $-0.678^{+0.216}_{-0.037}$ &  $-6.08^{+0.39}_{-0.02}$ &   136/113 \\
       LEDE (DR9) & 2.2--3.5 &  $-1.46^{+0.03}_{-0.01}$ &  $-3.71^{+0.06}_{-0.02}$ &  $-26.70^{+0.02}_{-0.02}$ &  $-0.576^{+0.001}_{-0.039}$ &  $-0.774^{+0.034}_{-0.010}$ &  $-6.06^{+0.10}_{-0.01}$ &   1366/107 \\
       \hline
      \hline 
      \end{tabular}
    \end{table*}

    \subsection{Quasar model predictions}
    There are many models for quasar evolution in the literature, but
    the modern ones come in three basic flavors. The first implements some
    of the quasar physics directly into numerical hydrodynamic simulations
    of galaxy formation or interaction. The second follows much of the
    same physics semi-analytically. The third tries to relate the
    properties of quasars and black holes directly to those of dark matter
    halos or the galaxies which reside in them. We give recent examples
    from each of these classes of models here.

   \begin{figure*}
      \begin{center}
        \includegraphics[height=5.90cm,width=5.90cm,  trim =  0mm  0mm 5mm 0mm, clip]
        {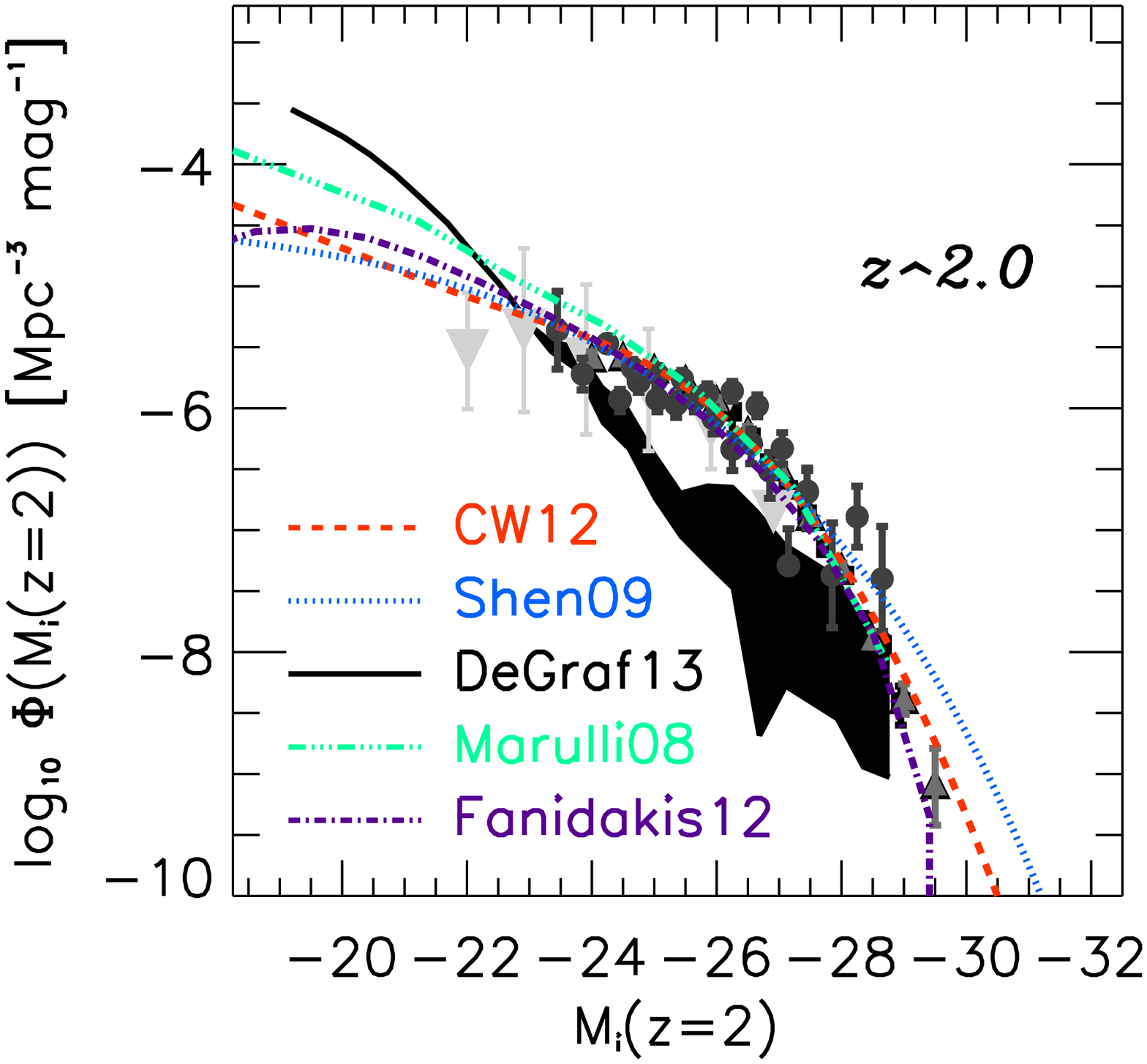}
        \includegraphics[height=5.90cm,width=5.90cm,  trim =  0mm  0mm 5mm 0mm, clip]
        {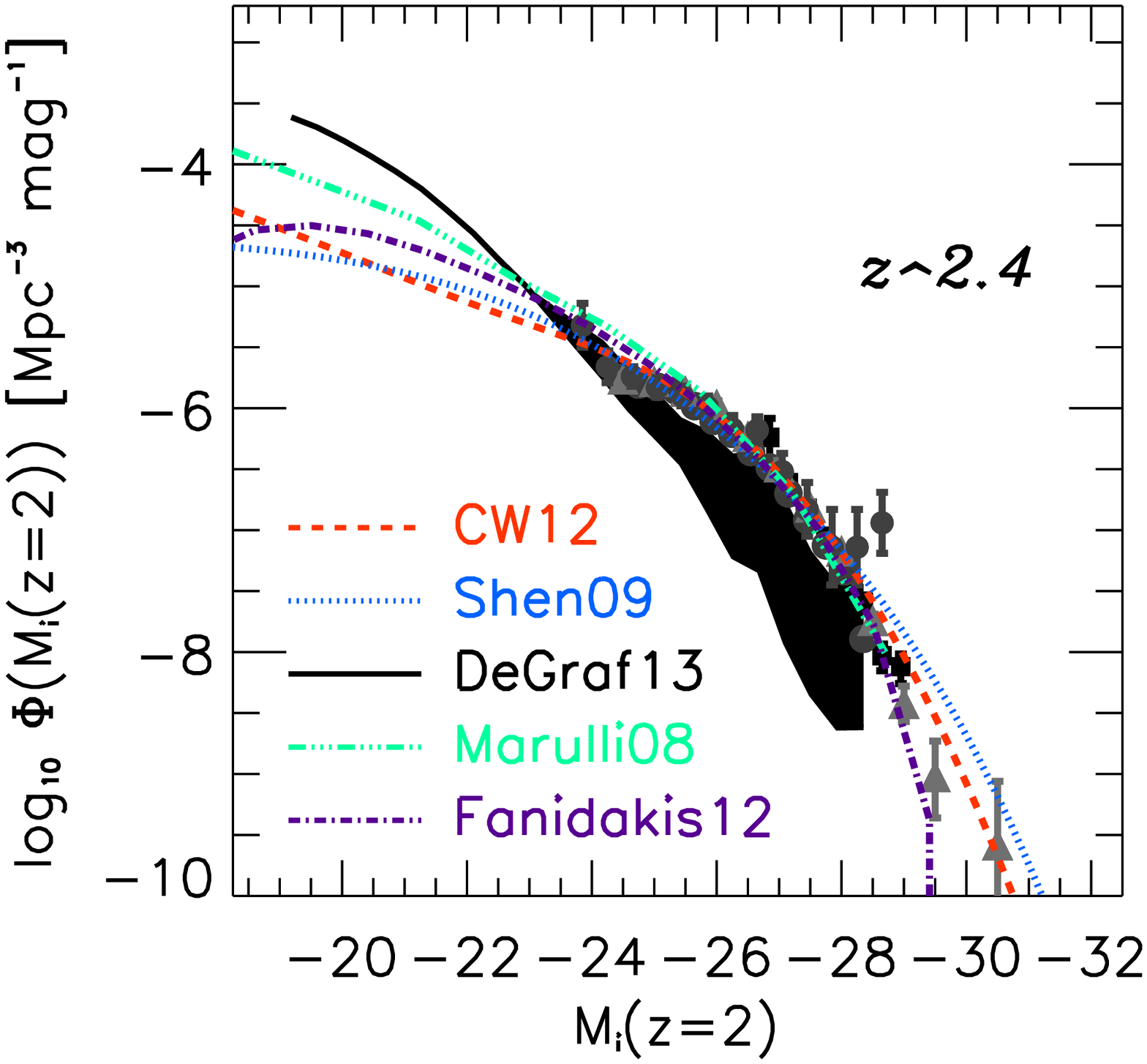}
        \includegraphics[height=5.90cm,width=5.90cm,  trim = 0mm 0mm  5mm 0mm, clip]
        {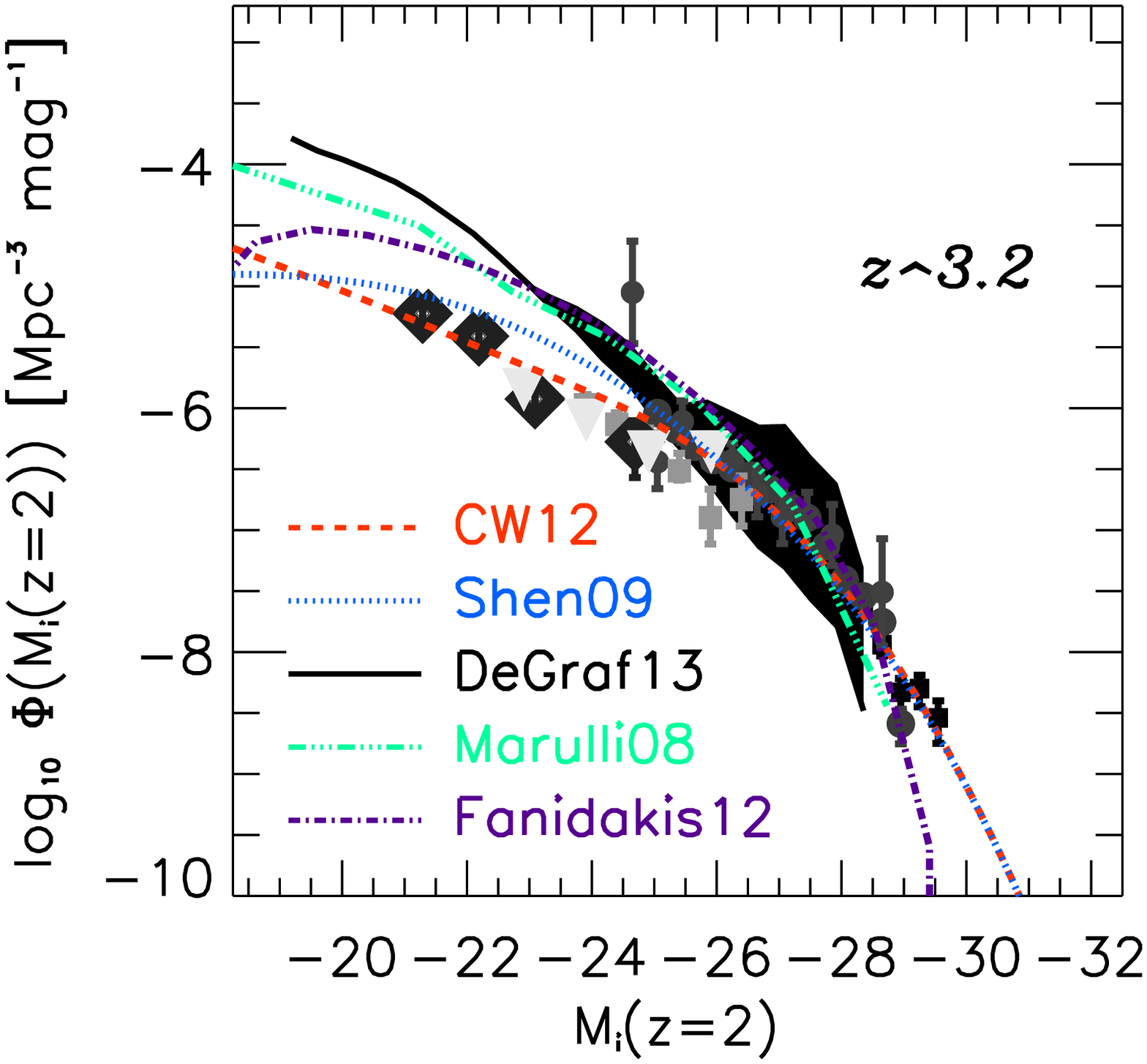}
        \caption{The BOSS DR9 Quasar Luminosity Function compared to a series 
          of QLF models from the literature. The model from \citet{CW12} is
          given by the dashed orange line, while the model from \citet{Shen09b}
          is given by the light-blue dotted line. The \citet{Marulli08} model is
          given by the solid turquoise line, the \citet{Fanidakis12} model by
          the dot-dashed purple line and the model from (DeGraf et al. 2013) is
          the shaded black region. We refer the interested reader to the given
          papers for presentation and discussion of the uncertainties associated
          with the published models.
          {\it Left:} Measurement in the $1.8<z<2.2$ range;      
          {\it Center:} Measurement in the $2.2<z<2.6$ range; 
          {\it Right:} Measurement at $z\sim3.2$. Note, the Shen09 and CW12 models 
          are on top of each other at $M_{i}<-26$. } 
        \label{fig:QLF_model_comparisons}
      \end{center}
    \end{figure*}

    DeGraf et al. (2013, in prep.) present models for the QLF using
    the new ``MassiveBlackII'' hydrodynamic simulation, which has a
    boxsize of 100 $h^{-1}$ Mpc, number of particles, $N_{\rm p}=2\times
    1792^3$ and a gravitational softening of $\epsilon=1.85$ h$^{-1}$ kpc,
    and employs a WMAP7 \citep{Komatsu11} cosmology. These simulations
    incorporate the physics of hydrodynamics, radiative cooling, star
    formation, black holes and associated feedback in order to make {\it
      ab initio} predictions for the observed properties of galaxies and
    quasars. The QLF for each redshift bin is computed using the complete
    luminosity history of every black hole, producing the best available
    statistics and extending the predictions to the brightest luminosity
    by catching rare objects that only occasionally reach very
    high-$L$. The predictions from these hydrodynamic simulations are
    given by the shaded black region in
    Fig.~\ref{fig:QLF_model_comparisons}. Note that they extend to
    luminosities fainter than BOSS generally probes.
    
    There are discrepancies between the simulations and the data,
    especially at $z\approx2.0$ and 2.4, which may be due to several
    effects. Previous work on smaller simulations
    \citep[e.g.,][]{DeGraf10} found that lower resolution simulations
    produce steeper faint end luminosity functions. Thus increased
    resolution should further flatten the faint end.  At the bright end,
    volume limitations become significant, with only several black holes
    reaching the brightest luminosities.  The shaded region in the bright
    end QLF represents an estimate for the cosmic variance using a larger
    volume simulation \citep[``MassiveBlack'', see][]{DeGraf12,
      DiMatteo12}. The larger simulation avoids the volume limitations
    resulting in the upper bound of this region, suggesting that within
    volume limitations the simulations are consistent with current data.

    \citet{Marulli08} model the cosmological co-evolution of galaxies
    and their central supermassive black holes within a semi-analytical
    framework developed on the outputs of the Millennium Simulation
    \citep{Springel05}. These authors use the galaxy formation model of
    \citet{Croton06} as updated by \citet{DeLucia07} as their starting
    point. Luminous quasars in this model occur when a BH accretes cold
    gas after a major merger of two gas rich galaxies. The accreted mass
    is proportional to the total cold gas mass present, but with an
    efficiency which is a function of the size of the system and the
    merger mass ratio, and chosen to reproduce the observed local $M_{\rm
      BH}-M_{\rm bulge}$ relation. \citet{Marulli08} then couple this
    accretion to various light curve models. The predictions for the
    luminosity function are shown in Fig.~\ref{fig:QLF_model_comparisons}
    by the triple-dotted-dashed (turquoise) line. We see that this
    model does well in the lower redshift bins at $z\sim2.0$ and 2.4 at
    reproducing the data, but perhaps over predicts the number of faint
    quasars at $z\sim3.2$.
    
    For comparison we also consider a second semi-analytic model
    \citep{Fanidakis12}. This model is embedded in the semi-analytical
    galaxy formation code GALFORM \citep[][see also \citet{Baugh05,
      Bower06}]{Cole00} and predicts the masses, spins \citep{Fanidakis11}
    and mass accretion histories of BHs in tandem with the formation of
    their host galaxies. In addition to merger-induced triggering they
    allow triggering when discs becoming dynamically unstable \citep[based
    on the arguments in][]{Efstathiou82}. As in \citet{Marulli08} they
    also follow quasi-hydrostatic hot gas accretion (known variously as
    ``hot halo mode'', ``radio mode'' or ``radiative mode'' accretion)
    with a rate orders of magnitude below the Eddington limit. The key
    aspect of the \citet{Fanidakis12} model in our comparison is that
    their starburst mode, and thus the BH mass growth, is mainly driven by
    disc instabilities. Comparison of \citet{Marulli08} and
    \citet{Fanidakis12} thus allows insights into how the triggering mode
    of quasar activity can potentially be tested by measurements such as ours.
    The number densities from the \citet{Fanidakis12} model are calculated
    considering the entire population of AGN (both obscured and
    unobscured) and include the empirical obscuration prescription from
    \citet{Hasinger08}. The QLFs for the unobscured population are shown
    as (purple) dot-dashed lines in Fig.~\ref{fig:QLF_model_comparisons}.
    
    \citet{Hirschmann12} also used semi-analytic models, based on those from
    \citet{Somerville08}, to examine the properties of accreting BHs and the
    evolution of the QLF.
    (We do not show the \citet{Hirschmann12} predictions in
    Fig.~\ref{fig:QLF_model_comparisons}, but their best fitting model
    fits our data well with potentially a slight overproduction of the
    faintest QSOs at $z>2.5$; see their Fig.~7.)
    These authors find that their best fitting model (which includes using
    ``heavy'' black hole seeds of $M_{\rm seed} \approx 10^{5-6} M_{\odot}$
    at very high $z$ and a varying {\it sub}-Eddington limit for the maximum
    accretion rate at $z\leq1$) suggests a scenario in which the disc
    instabilities are the main driver for moderately luminous Seyfert galaxies
    at low redshift, but major mergers remain the key trigger for luminous
    AGN/quasars, especially at high $z$.
    
    \citet{Shen09b} presents a phenomenological model for the growth
    and cosmic evolution of SMBHs, in which the quasar properties are tied
    to the properties of dark matter halos, rather than galaxies drawn
    from a semi-analytic model. This model assumes that quasar activity is
    triggered by major mergers of host halos, and that the resulting light
    curve follows a universal form, in which its peak luminosity is
    correlated with the (post)merger halo mass. Quasar activity is
    quenched at low $z$ and in lower mass halos with phenomenological
    rules. In particular, the quasar triggering rate depends on a
    ``quasar-on'' factor \citep[called $f_{\rm QSO}$ in][]{Shen09b} which
    has exponential cut-offs both at the low and high mass ends which are
    adjusted to fit the data.  These cut-offs ensure that halos with too
    small a (postmerger) halo mass cannot trigger any quasar activity, while
    those above a (redshift dependent) maximum mass cannot cool gas efficiently
    and BH growth halts. With these assumptions, the quasar LF and SMBH
    growth are tracked self-consistently across cosmic time.
    The QLF predicted by this model is shown in
    Fig.~\ref{fig:QLF_model_comparisons} by the dotted (blue) line. This
    model does well at reproducing the data in all three redshift slices,
    though with a slight over-production of bright quasars at $z\simeq 2.0$.
    
    Recently \citet{CW12} presented a model for quasar demographics in
    which quasars populate galaxies in a simple manner and many of the
    properties of the quasar population follow naturally from the known,
    evolving properties of galaxies. A simple ``scattered lightbulb''
    model is adopted, with BHs shining at a fixed fraction of the
    Eddington luminosity during accretion episodes with Eddington ratios
    drawn from a lognormal distribution. The quasar duty cycle is
    explicitly independent of galaxy and BH mass and luminosity, in
    contrast to the strong dependence invoked in \citet{Shen09b} when
    connecting quasars to halos. The QLF predictions for that model are
    shown in Fig.~\ref{fig:QLF_model_comparisons} as the (red) dashed
    lines.
    
    While the models we have highlighted agree with the existing data
    relatively well, they explain the qualitative behaviors we see in
    different ways. For example, it is well known that the abundance of
    bright quasars drops rapidly to low $z$ and that lower mass black hole
    growth peaks at lower redshift than higher mass black holes
    \citep[][]{Hasinger05, Croom09b}. In the model of
    \citet{CW12} this is explained through a combination of slow growth of
    massive galaxies and evolution in the Eddington ratio. In the model of
    \citet{Shen09b}, it involves a suppressing function which simulates the
    effects of cold gas consumption with time. In the model of
    \citet{Fanidakis12} it arises due to a combination of factors,
    including obscuration evolution.

    The models differ significantly in the mass and redshift dependence of
    the duty cycle, and predict subtle differences in the width of the halo
    mass distribution at any redshift. 
    In almost all models the characteristic halo mass associated with
    existing quasar samples is almost independent of redshift.  This arises
    largely due to a chance cancellation of trends in the absolute magnitude
    limit, the relation between galaxy and halo properties and galaxies,
    black holes and Eddington ratios.
    
    In the model of \citet{CW12}, the evolution of the characteristic
    luminosity is driven by the evolution in the $L-M_{\rm gal}$ and
    $M_{\rm gal}-M_h$ relations, while the break in the LF arises
    primarily due to the shape of the $M_{\rm gal}-M_h$ relation.
    The \citet{Shen09b} model adjusts the typical host halo of luminous
    quasars to fit the observed evolution of the break luminosity.
    In the semi-analytic models, the starburst/quasar mode is powered
    by one or a combination of major galaxy mergers and disk instabilities
    with the relative contributions possibly evolving with time.  The
    evolution of the characteristic luminosity thus arises from a complex
    interplay of factors.
    \citet{CW12} predict that the faint end slope of the LF does not vary
    significantly, while the bright end slope appears shallower at higher
    $z$. In the model of \citet{Shen09b}, the LF is predicted to turn down
    at sufficiently low luminosities and high redshifts, since at $z>2$
    the minimum Eddington ratio is constrained by the age of the Universe.
    The hydrodynamic simulations predict a steep faint-end slope at $z\sim 2$.
    Most of the models have considerable scatter between quasar luminosity
    and galaxy or halo mass, and thus predict a power-law tail to high
    luminosity, as observed.  Further measurements of this tail at higher $z$
    may provide better constraints on this aspect of the models.

    In summary, all the models reproduce the QLF and quasar demographics
    overall reasonably well.  We agree with \citet{Hirschmann12} when they
    state that further progress on these issues will require data beyond
    just the luminosity function.

    \subsection{Discussion}

    In this final section, we tie our QLF results (and comparisons to
    models) into the broader context of the link between SMBH growth (see
    as seen AGN activity) and the properties of galaxies. We
    take as our starting point the QLF reported here and the clustering
    measurement and discussion of the BOSS DR9 uniform quasar sample
    reported in \citet{White12}. Using the same arguments as in
    \citet{White12}, and the conversions of \citet{Croom05} and
    \citet{Shen09}, we place the median BOSS quasar with a bolometric
    luminosity of $L_{\rm bol} = 2 - 4 \times 10^{46}$~erg s$^{-1}$, in
    dark matter haloes of characteristic mass of $\sim 2 \times 10^{12}\
    h^{-1} M_{\odot}$ at $z\approx2.5$. Either making the assumption that
    the BOSS quasars are consistent with the $M_{\rm BH} - M_{\rm halo}$
    relation \citep{Ferrarese02, Fine06}, or, that the quasars radiate at
    close to the Eddington Limit, $L_{\rm Edd} = 10^{47.1} (M_{\rm
      BH}/10^{9} M_{\odot})$ erg s$^{-1}$, suggests that the median $M_{\rm
      BH}$ in our sample is $\sim 2 \times 10^{8} M_{\odot}$. As a guide, a
    typical $\sim 2 \times 10^{8} M_{\odot}$ BH, accreting continuously
    since $z\sim2.5$, with an accretion efficiency of $\epsilon=0.1$, and
    not merging, would have a mass at redshift $z\sim0$ of $M_{\rm BH}
    \sim 6\times10^{10} M_{\odot}$. This would place these objects at the
    very highest BH masses observed, but also inline with recent results
    \citep{McConnell11}. A more realistic scenario, where the duty cycle
    is $1$\%, would lead to $M_{\rm BH} \sim 6\times10^{8} M_{\odot}$,
    placing these objects in bulges with $\sigma\sim 250-300$ km s$^{-1}$,
    and thus in early-type galaxies from the relations in
    e.g.~\citet{Gultekin09}. 
     
    From the observed clustering (and indeed essentially any of the
    models quoted above) the {\it typical\/} halo for a BOSS quasar at
    $z\approx2.5$ would grow to host a small group by $z\sim0$.  The
    most likely host galaxy is the central galaxy of the group, since at
    higher-$z$, any satellites would not be massive enough to host a
    SMBH. Thus, the typical BOSS quasar host descendant would be the
    central galaxy of a small group - though we caution that including
    e.g. the diversity of growth histories of DM halos and scatter in any
    of the given relations, can easily lead to an order of magnitude
    dispersion in the above statements \citep{White12}.
    Placing these quasars at the centers of groups at $z=0$ is consistent
    with the suggested velocity dispersions given above.
    This potentially also suggests that BOSS
    quasars today are very likely not on the ``SF Main Sequence'' any more
    (i.e.~they are quenched) even if they were initially.
    This is also consistent with the recent work by \citet{Kelly_Shen12b}.
    
    Leaving the properties of the median BOSS quasar, we now focus on
    the ``extremes'' of our population. Taking the most luminous quasars,
    we find these objects to have close to $\log (L_{\rm bol}) = 46.0$,
    and thus black holes in the mass range $\sim3\times 10^{9} M_{\odot}$
    (assuming an Eddington luminosity). At the bright end, the QLF is
    described by a power-law fall-off, while the massive end of the
    stellar mass function, the abundance declines exponentially. With a
    relationship known to exist between $M_{\rm BH}/M_{\rm bulge}$ (and
    where $M_{\rm bulge} \sim M_{\rm gal}$ for these compact massive
    galaxies), this argues that there is scatter in $L_{\rm Q}$ at fixed
    $M_{\rm gal}$. This is perhaps not surprising: at low $z$, $M_{\rm
      BH}/M_{\rm bulge}$ is measured to have $\sim$0.3 dex in scatter and
    Eddington ratios are also measured to have $\sim$0.3 dex scatter, so a
    scatter of at least 0.4 dex overall could be expected. However, this
    leads to the situation that at high-$L$, scatter is increasingly
    important, and that bright quasars are ``overbright'', and it is
    currently unclear what underlying physical mechanisms would lead to
    this enhanced up-scatter. We leave further investigation into the
    potential evolution of $M_{\rm BH} / M_{\rm gal}$, and the different
    channels that drives the growth of black holes, the evolution of the
    number density of quasars, and that of AGN activity in general for
    future study.

\section{Conclusions}\label{sec:concs}

The quasar luminosity function is one of the most fundamental observables
of this class of important cosmological objects.  The shape and evolution
of the QLF provides constraints on models of quasar fueling, feedback and
galaxy evolution and the ionization history of the inter-galactic gas.
Despite its importance, it has proven difficult observationally to probe
the quasar luminosity function at magnitudes below the break at the peak
of the quasar epoch. 

Here we measure the QLF using data from the SDSS-III: Baryon Oscillation
Spectroscopic Survey (BOSS) using a uniformly selected sample of
\hbox{23 301} quasars, and fill in the $L-z$ plane with published results from
the SDSS-I/II.
We probe the faint end of the QLF to $M_{i}= -24.5$ at $z= 2.2$ and complement
our uniform color-selection with a sample of variability-selected quasars from
the ``Stripe 82'' field.  We also provide a cross-check of our selection
function using new, simulated, model, quasar spectra. 
Amongst our findings are:

\begin{itemize}
\item{That down to a magnitude limit of $i=21.5$, there are 26.2 and
    48.0 quasars deg$^{-2}$ across the redshift ranges $2.2<z<3.5$ and
    $1.0<z<2.2$ respectively. Using the deeper boss21+MMT data, for the
    unobscured $1.0<z<2.2$ quasar population, there are 78 objects
    deg$^{-2}$ brighter than $i\approx 23.0$, a surface density similar to
    that selected by a shallow mid-infrared selection \citep{Stern12}.  }
\item{Our combined SDSS+BOSS QLF is reasonably well described by 
    a double power-law, quadratic, pure luminosity evolution (PLE) model 
    across the redshift range $0.3 < z < 2.2$, with a bright end slope 
    $-3.37^{+0.03}_{-0.05}$, a faint end slope $-1.16^{+0.02}_{-0.04}$,
    $M^{*}_{i}(z=0)=-22.85^{+0.05}_{-0.11}$, 
    $k_{1}=1.24^{+0.01}_{-0.03}$, $k_{2}=-0.25^{+0.01}_{-0.02}$ and
    $\log\Phi^{*}= -5.96^{+0.02}_{-0.05}$.}
\item{The simple PLE model breaks down at $z\ga2.2$. We replace it with
    a luminosity evolution and density evolution (LEDE) model that has
    a log-linear trend in both $\Phi^*$ and $L^*$. This simple form
    provides a good fit to the data at $2.2<z<3.5$, capturing both the
    steep decline in number density and the rise in the break luminosity.
    The data are consistent with no evolution in the power law slopes,
    though do not strongly constrain the lack of evolution. }
\item{We compare our measured QLF to theoretical models and find a wide variety
    of models describe our data reasonably well.  While the latest hydrodynamic
    simulations do not fit as well, semi-analytic models in which luminous
    quasar activity is triggered by major mergers, disk instabilities or a
    combination of channels can fit our data over a wide range of redshifts.
    Models based on directly populating halos with quasars can fit the shape
    of our QLF by assuming a mass and redshift-dependent duty-cycle which is
    sharply peaked around a characteristic mass.
    We also find that models which relate black hole mass linearly to galaxy
    mass and assume a mass-independent duty-cycle match our QLF well.}
\end{itemize}

The results presented here are from the first two, of five, years of BOSS
spectroscopy.  The upcoming Data Release Ten dataset will cover
$\sim 7000\,$deg$^{2}$, include $\sim150,000$ quasars and will more than
double the number in our uniform selection.
Future investigations will be able to use this enhanced dataset in order to
further quantify, and refine, the selection function for the $2.2<z<3.5$
quasar sample and thus reduce the errors further.
This release will include quasars that were observed by BOSS because of their
near- and mid-infrared colors, and with these samples we will be able to
infer further key properties of quasars at the height of the quasar epoch.

\acknowledgments
The JavaScript Cosmology Calculator was used whilst preparing this
paper \citep{Wright06}. This research made use of the NASA
Astrophysics Data System. Heavy use was made of the
\href{http://code.google.com/p/red-idl-cosmology/}{RED} IDL cosmology
routines written by L. and J. Moustakas, and based on \citet{Hogg99}.

Funding for SDSS-III has been provided by the Alfred P. Sloan
Foundation, the Participating Institutions, the National Science
Foundation, and the U.S. Department of Energy Office of Science. The
SDSS-III web site is
\href{http://www.sdss3.org/}{http://www.sdss3.org/}.

SDSS-III is managed by the Astrophysical Research Consortium for the
Participating Institutions of the SDSS-III Collaboration including the
University of Arizona, the Brazilian Participation Group, Brookhaven
National Laboratory, University of Cambridge, Carnegie Mellon
University, University of Florida, the French Participation Group, the
German Participation Group, Harvard University, the Instituto de
Astrofisica de Canarias, the Michigan State/Notre Dame/JINA
Participation Group, Johns Hopkins University, Lawrence Berkeley
National Laboratory, Max Planck Institute for Astrophysics, Max Planck
Institute for Extraterrestrial Physics, New Mexico State University,
New York University, Ohio State University, Pennsylvania State
University, University of Portsmouth, Princeton University, the
Spanish Participation Group, University of Tokyo, University of Utah,
Vanderbilt University, University of Virginia, University of
Washington, and Yale University. 

NPR warmly thanks Silvia Bonoli, Federico Marulli, Nikos Fanidakis and
Phil Hopkins for providing their model QLF data in a prompt manner.
Matt George, Genevieve Graves, Tom Shanks, Julie Wardlow and Gabor
Worseck, also provided very useful discussions.
%%
%% Die Zauberflaute, Daft Punk and Hans Zimmer, also provided much 
%% inspiration. 
%%
IDM and XF acknowledge support from a David and Lucile Packard
Fellowship, and NSF Grants AST 08-06861 and AST 11-07682.

{\it Facilities: SDSS}

\appendix

%\clearpage
\section{Appendix A. Comparison of Quasar Spectral Models}\label{appndx:spec_models}

In Section~\ref{sec:simqso} we introduced three models for quasar
spectral features. In brief, the fiducial model (adopted for our
primary analysis) includes a luminosity-dependent emission line
template derived from fitting of composite quasar spectra. The
composite spectra are created within narrow bins of luminosity, so
that mean trends of emission line features with luminosity are
reproduced; in particular, the anti-correlation of line equivalent
width with continuum luminosity \citep[the ``Baldwin
Effect'',][]{Baldwin77}. Introducing this feature accounts for the
luminosity dependence of quasar colors and the effect this has on
quasar selection.

For comparison, we include two additional models. The first is based
on a fixed emission line template with no luminosity dependence. As
this is most similar to models used in previous work
\citep[e.g.,][]{Richards06,Croom09b}, it provides a reference point
for comparison to QLF estimates that did not include a Baldwin Effect
in the selection function estimation. Finally, we also include a model
with dust extinction (``exp dust''), motivated by observations of SDSS
quasars with mild dust reddening
\citep[e.g.,][]{Richards03,Hopkins04}.
 
In Figure~\ref{fig:QLF_iband_narrow_selfunc_divmodel} we compare the
estimated QLFs derived from each of the three selection function
models. The systematic effects resulting from imperfect knowledge of
the true selection function (or by corollary, the intrinsic
distribution of quasar spectral features) is greater than the
statistical uncertainties resulting from Poisson variations.

Here, we motivate our choice of selection function (the fiducial
model). Our method is a simple qualitative comparison of the observed
color-redshift relation for the three models as compared to the
data. The method for constructing the color redshift relation is
described in Section~\ref{sec:simqso}.

\begin{figure*}[!th]
  \begin{center}
    \includegraphics[scale=1.0]{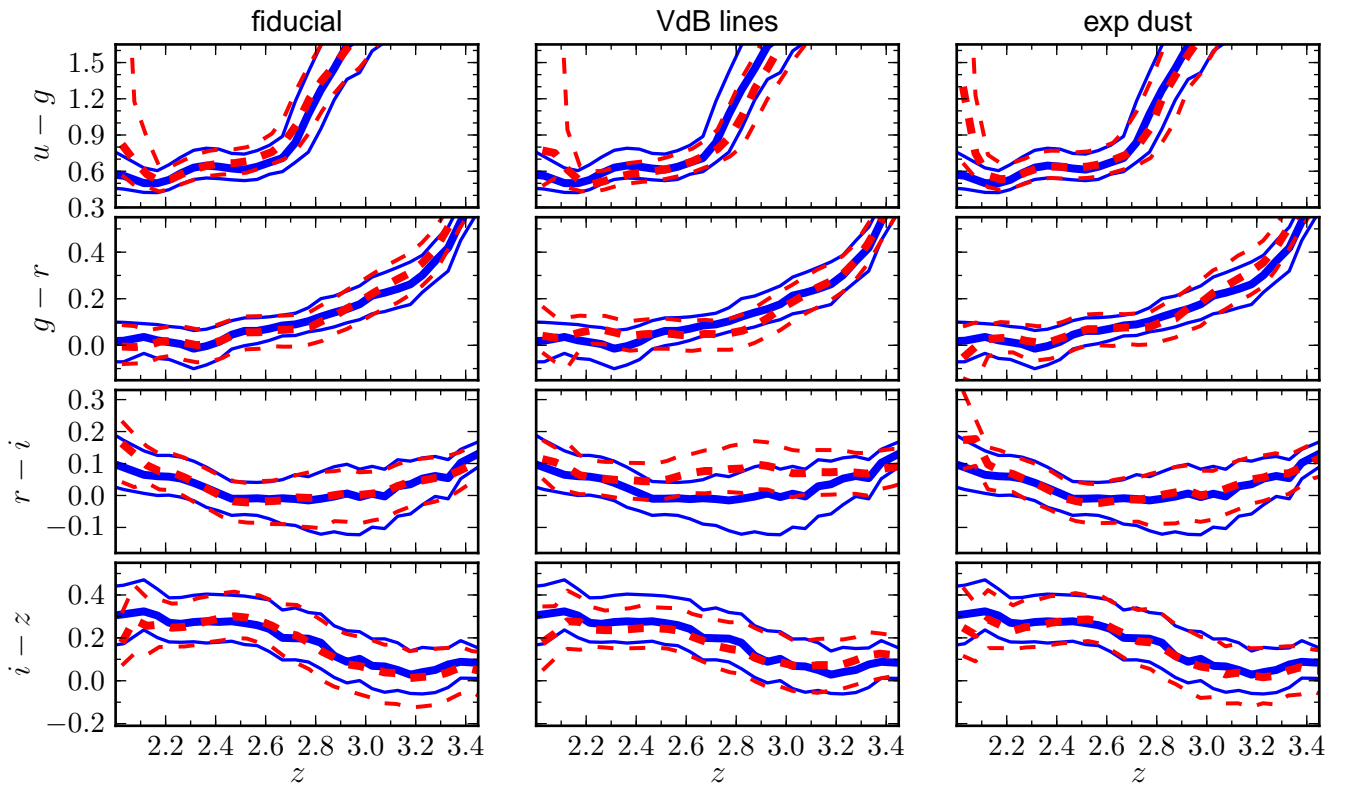}
    \caption{Color-redshift relations for the three quasar
      spectral models. Each column of panels represents one
      of the quasar models named above the top panel. As in
      Fig.~\ref{fig:colorz_simcompare}, the solid blue lines
      are the mean and $\pm1\sigma$ scatter of the colors in
      redshift bins of $\Delta z = 0.05$. The dashed red lines
      are the same color relations derived from the simulated
      quasars.
    }
   \label{fig:colorz_comparemodels}
  \end{center}
\end{figure*}

The resulting relations are shown in
Figure~\ref{fig:colorz_comparemodels}. The ``VdB lines'' model does
poorly at reproducing the observed colors in the range $2.4 \la z \la
3.3$, when Ly$\alpha$ and \ion{C}{4} are in the $g$ and $r$ bands,
respectively. The only difference between this model and the fiducial
model is the emission line template, thus a model that does not
account for the Baldwin Effect will have difficulty reproducing quasar
colors at these redshifts. Note that this effect is likely less
pronounced in the SDSS data \cite[e.g.,][]{Richards06}, as it covers
less dynamic range in luminosity, and thus most quasars are closer to
the mean luminosity represented by a single composite spectrum. The
exp dust model appears to do as well as the fiducial model. We
chose not to include dust in order to remain more consistent with
previous work, and since the focus of this work is on the unobscured
quasar population. Unsurprisingly, the exp dust model results in a
lower overall completeness, so that the estimated luminosity function
is higher overall
(Figure~\ref{fig:QLF_iband_narrow_selfunc_divmodel}). We will consider
these issues further in subsequent work.

%\clearpage
\section{Appendix B. Additional BOSS QLF Tables}\label{appndx:tables}
Here we present additional tables reporting the BOSS QLF for the
various samples given in the main text. 
Table~\ref{tab:QLF_iband_narrow} gives the BOSS DR9 QLF as shown in
Fig.~\ref {fig:QLF_iband_narrow}, while
Table~\ref{tab:qlf_iband_narrow_S82} gives the calculated QLF for
the Stripe 82 dataset over the same redshift range and binning
(teal points in Fig.~\ref {fig:QLF_iband_narrow}). 

\begin{table}
  \begin{center}
    \begin{tabular}{ccccrr}
      \hline
      \hline
      $<z>$ & $<M_{i}(z=2)>$ & $M_{i}$ bin  & $N_{\rm Q}$ & $\log(\Phi)$ & $\sigma\times10^{-9}$  \\
      \hline
      2.260 &   -28.015 &   -28.050 &      25 &   -7.181 &    9.054 \\
      2.256 &   -27.743 &   -27.750 &      29 &   -7.096 &    9.985 \\
      2.258 &   -27.431 &   -27.450 &      70 &   -6.836 &   13.472 \\
      2.256 &   -27.130 &   -27.150 &     114 &   -6.625 &   17.167 \\
      2.257 &   -26.853 &   -26.850 &     203 &   -6.396 &   22.356 \\
      2.255 &   -26.544 &   -26.550 &     254 &   -6.273 &   25.761 \\
      2.253 &   -26.248 &   -26.250 &     346 &   -6.152 &   29.598 \\
      2.255 &   -25.947 &   -25.950 &     473 &   -5.968 &   36.587 \\
      2.254 &   -25.653 &   -25.650 &     490 &   -5.868 &   41.044 \\
      2.255 &   -25.352 &   -25.350 &     570 &   -5.703 &   49.647 \\
     \hline
      \hline
    \end{tabular}
    \caption{
      The narrowly binned BOSS DR9 Quasar Luminosity
      Function. The columns are the same as
      Table~\ref{tab:qlf_iband}. The full table appears in the
      electronic edition of {\it The Astrophysical Journal}.}. 
    \label{tab:QLF_iband_narrow}
  \end{center}
\end{table}

\begin{table}
  \begin{center}
    \begin{tabular}{ccccrr}
      \hline
      \hline
      $<z>$ & $<M_{i}(z=2)>$ & $M_{i}$ bin  & $N_{\rm Q}$ & $\log(\Phi)$ & $\sigma\times10^{-9}$  \\
      2.268 &   -29.762 &   -29.850 &       1 &   -7.898 &   12.636 \\
      2.266 &   -28.668 &   -28.650 &       2 &   -7.597 &   17.870 \\
      2.246 &   -28.334 &   -28.350 &       2 &   -7.597 &   17.870 \\
      2.252 &   -28.080 &   -28.050 &       6 &   -7.120 &   30.952 \\
      2.253 &   -27.713 &   -27.750 &       8 &   -6.995 &   35.740 \\
      2.244 &   -27.426 &   -27.450 &       7 &   -7.053 &   33.432 \\
      2.252 &   -27.150 &   -27.150 &      17 &   -6.668 &   52.100 \\
      2.251 &   -26.833 &   -26.850 &      25 &   -6.500 &   63.181 \\
      2.253 &   -26.534 &   -26.550 &      37 &   -6.330 &   76.863 \\
      2.250 &   -26.247 &   -26.250 &      51 &   -6.191 &   90.240 \\
      \hline
     \hline
      \hline
    \end{tabular}
    \caption{
      The narrowly binned BOSS Quasar Luminosity Function using
      data from 5731 (5476) $2.20<z<4.00\ (3.50)$ quasars selected 
      via their variability signature on Stripe 82 (Sec.~\ref{sec:variability}). 
      The columns are the same as
      Table~\ref{tab:qlf_iband}. The full table appears in the
      electronic edition of {\it The Astrophysical Journal}.}. 
    \label{tab:qlf_iband_narrow_S82}
  \end{center}
\end{table}

\bibliographystyle{mn2e} 
%\bibliography{../../biblio/tester_mnras}
\bibliography{tester_mnras}

\end{document}